\documentclass[submission,copyright,creativecommons]{eptcs}
\usepackage{underscore}           

\usepackage[capitalize,noabbrev]{cleveref}
\usepackage{graphicx}
\usepackage{lineno}

\title{SeCaV: A Sequent Calculus Verifier in Isabelle/HOL} 

\author{Asta Halkjær From \qquad Frederik Krogsdal Jacobsen \qquad Jørgen Villadsen
\institute{DTU Compute - Department of Applied Mathematics and Computer Science - Technical University of Denmark}
}

\def\authorrunning{A.\,H. From, F.\,K. Jacobsen and J. Villadsen}
\def\titlerunning{SeCaV: A Sequent Calculus Verifier in Isabelle/HOL}

\usepackage{isabelle, isabellesym}
\isabellestyle{it} 

\newcommand{\DefineSnippet}[2]{\expandafter\newcommand\csname snippet--#1\endcsname{#2}}
\DefineSnippet{Sequent-Calculus-Verifier:section:142272d4bfb0206-0}{%
\isadelimdocument
\endisadelimdocument
\isatagdocument
\isamarkupsection{Appendix: Completeness%
}
\isamarkuptrue%
\endisatagdocument
{\isafolddocument}%
\isadelimdocument
\endisadelimdocument
}
\DefineSnippet{Sequent-Calculus-Verifier:theory:Sequent-Calculus-Verifier-0}{%
\isadelimtheory
\endisadelimtheory
\isatagtheory
\isacommand{theory}\isamarkupfalse%
\ Sequent{\isacharunderscore}Calculus{\isacharunderscore}Verifier\ \isakeyword{imports}\ Sequent\ SeCaV\ \isakeyword{begin}%
\endisatagtheory
{\isafoldtheory}%
\isadelimtheory
\endisadelimtheory
}
\DefineSnippet{Sequent-Calculus-Verifier:primrec:from-tm-0}{%
\isacommand{primrec}\isamarkupfalse%
\ from{\isacharunderscore}tm\ \isakeyword{and}\ from{\isacharunderscore}tm{\isacharunderscore}list\ \isakeyword{where}\isanewline
}
\DefineSnippet{Sequent-Calculus-Verifier:primrec:from-tm-1}{%
\ \ {\isacartoucheopen}from{\isacharunderscore}tm\ {\isacharparenleft}Var\ n{\isacharparenright}\ {\isacharequal}\ FOL{\isacharunderscore}Fitting{\isachardot}Var\ n{\isacartoucheclose}\ {\isacharbar}\isanewline
}
\DefineSnippet{Sequent-Calculus-Verifier:primrec:from-tm-2}{%
\ \ {\isacartoucheopen}from{\isacharunderscore}tm\ {\isacharparenleft}Fun\ a\ ts{\isacharparenright}\ {\isacharequal}\ App\ a\ {\isacharparenleft}from{\isacharunderscore}tm{\isacharunderscore}list\ ts{\isacharparenright}{\isacartoucheclose}\ {\isacharbar}\isanewline
}
\DefineSnippet{Sequent-Calculus-Verifier:primrec:from-tm-3}{%
\ \ {\isacartoucheopen}from{\isacharunderscore}tm{\isacharunderscore}list\ {\isacharbrackleft}{\isacharbrackright}\ {\isacharequal}\ {\isacharbrackleft}{\isacharbrackright}{\isacartoucheclose}\ {\isacharbar}\isanewline
}
\DefineSnippet{Sequent-Calculus-Verifier:primrec:from-tm-4}{%
\ \ {\isacartoucheopen}from{\isacharunderscore}tm{\isacharunderscore}list\ {\isacharparenleft}t\ {\isacharhash}\ ts{\isacharparenright}\ {\isacharequal}\ from{\isacharunderscore}tm\ t\ {\isacharhash}\ from{\isacharunderscore}tm{\isacharunderscore}list\ ts{\isacartoucheclose}%
}
\DefineSnippet{Sequent-Calculus-Verifier:primrec:from-fm-0}{%
\isacommand{primrec}\isamarkupfalse%
\ from{\isacharunderscore}fm\ \isakeyword{where}\isanewline
}
\DefineSnippet{Sequent-Calculus-Verifier:primrec:from-fm-1}{%
\ \ {\isacartoucheopen}from{\isacharunderscore}fm\ {\isacharparenleft}Pre\ b\ ts{\isacharparenright}\ {\isacharequal}\ Pred\ b\ {\isacharparenleft}from{\isacharunderscore}tm{\isacharunderscore}list\ ts{\isacharparenright}{\isacartoucheclose}\ {\isacharbar}\isanewline
}
\DefineSnippet{Sequent-Calculus-Verifier:primrec:from-fm-2}{%
\ \ {\isacartoucheopen}from{\isacharunderscore}fm\ {\isacharparenleft}Con\ p\ q{\isacharparenright}\ {\isacharequal}\ And\ {\isacharparenleft}from{\isacharunderscore}fm\ p{\isacharparenright}\ {\isacharparenleft}from{\isacharunderscore}fm\ q{\isacharparenright}{\isacartoucheclose}\ {\isacharbar}\isanewline
}
\DefineSnippet{Sequent-Calculus-Verifier:primrec:from-fm-3}{%
\ \ {\isacartoucheopen}from{\isacharunderscore}fm\ {\isacharparenleft}Dis\ p\ q{\isacharparenright}\ {\isacharequal}\ Or\ {\isacharparenleft}from{\isacharunderscore}fm\ p{\isacharparenright}\ {\isacharparenleft}from{\isacharunderscore}fm\ q{\isacharparenright}{\isacartoucheclose}\ {\isacharbar}\isanewline
}
\DefineSnippet{Sequent-Calculus-Verifier:primrec:from-fm-4}{%
\ \ {\isacartoucheopen}from{\isacharunderscore}fm\ {\isacharparenleft}Imp\ p\ q{\isacharparenright}\ {\isacharequal}\ Impl\ {\isacharparenleft}from{\isacharunderscore}fm\ p{\isacharparenright}\ {\isacharparenleft}from{\isacharunderscore}fm\ q{\isacharparenright}{\isacartoucheclose}\ {\isacharbar}\isanewline
}
\DefineSnippet{Sequent-Calculus-Verifier:primrec:from-fm-5}{%
\ \ {\isacartoucheopen}from{\isacharunderscore}fm\ {\isacharparenleft}Neg\ p{\isacharparenright}\ {\isacharequal}\ FOL{\isacharunderscore}Fitting{\isachardot}Neg\ {\isacharparenleft}from{\isacharunderscore}fm\ p{\isacharparenright}{\isacartoucheclose}\ {\isacharbar}\isanewline
}
\DefineSnippet{Sequent-Calculus-Verifier:primrec:from-fm-6}{%
\ \ {\isacartoucheopen}from{\isacharunderscore}fm\ {\isacharparenleft}Uni\ p{\isacharparenright}\ {\isacharequal}\ Forall\ {\isacharparenleft}from{\isacharunderscore}fm\ p{\isacharparenright}{\isacartoucheclose}\ {\isacharbar}\isanewline
}
\DefineSnippet{Sequent-Calculus-Verifier:primrec:from-fm-7}{%
\ \ {\isacartoucheopen}from{\isacharunderscore}fm\ {\isacharparenleft}Exi\ p{\isacharparenright}\ {\isacharequal}\ Exists\ {\isacharparenleft}from{\isacharunderscore}fm\ p{\isacharparenright}{\isacartoucheclose}%
}
\DefineSnippet{Sequent-Calculus-Verifier:primrec:to-tm-0}{%
\isacommand{primrec}\isamarkupfalse%
\ to{\isacharunderscore}tm\ \isakeyword{and}\ to{\isacharunderscore}tm{\isacharunderscore}list\ \isakeyword{where}\isanewline
}
\DefineSnippet{Sequent-Calculus-Verifier:primrec:to-tm-1}{%
\ \ {\isacartoucheopen}to{\isacharunderscore}tm\ {\isacharparenleft}FOL{\isacharunderscore}Fitting{\isachardot}Var\ n{\isacharparenright}\ {\isacharequal}\ Var\ n{\isacartoucheclose}\ {\isacharbar}\isanewline
}
\DefineSnippet{Sequent-Calculus-Verifier:primrec:to-tm-2}{%
\ \ {\isacartoucheopen}to{\isacharunderscore}tm\ {\isacharparenleft}App\ a\ ts{\isacharparenright}\ {\isacharequal}\ Fun\ a\ {\isacharparenleft}to{\isacharunderscore}tm{\isacharunderscore}list\ ts{\isacharparenright}{\isacartoucheclose}\ {\isacharbar}\isanewline
}
\DefineSnippet{Sequent-Calculus-Verifier:primrec:to-tm-3}{%
\ \ {\isacartoucheopen}to{\isacharunderscore}tm{\isacharunderscore}list\ {\isacharbrackleft}{\isacharbrackright}\ {\isacharequal}\ {\isacharbrackleft}{\isacharbrackright}{\isacartoucheclose}\ {\isacharbar}\isanewline
}
\DefineSnippet{Sequent-Calculus-Verifier:primrec:to-tm-4}{%
\ \ {\isacartoucheopen}to{\isacharunderscore}tm{\isacharunderscore}list\ {\isacharparenleft}t\ {\isacharhash}\ ts{\isacharparenright}\ {\isacharequal}\ to{\isacharunderscore}tm\ t\ {\isacharhash}\ to{\isacharunderscore}tm{\isacharunderscore}list\ ts{\isacartoucheclose}%
}
\DefineSnippet{Sequent-Calculus-Verifier:primrec:to-fm-0}{%
\isacommand{primrec}\isamarkupfalse%
\ to{\isacharunderscore}fm\ \isakeyword{where}\isanewline
}
\DefineSnippet{Sequent-Calculus-Verifier:primrec:to-fm-1}{%
\ \ {\isacartoucheopen}to{\isacharunderscore}fm\ {\isasymbottom}\ {\isacharequal}\ Neg\ {\isacharparenleft}Imp\ {\isacharparenleft}Pre\ {\isadigit{0}}\ {\isacharbrackleft}{\isacharbrackright}{\isacharparenright}\ {\isacharparenleft}Pre\ {\isadigit{0}}\ {\isacharbrackleft}{\isacharbrackright}{\isacharparenright}{\isacharparenright}{\isacartoucheclose}\ {\isacharbar}\isanewline
}
\DefineSnippet{Sequent-Calculus-Verifier:primrec:to-fm-2}{%
\ \ {\isacartoucheopen}to{\isacharunderscore}fm\ {\isasymtop}\ {\isacharequal}\ Imp\ {\isacharparenleft}Pre\ {\isadigit{0}}\ {\isacharbrackleft}{\isacharbrackright}{\isacharparenright}\ {\isacharparenleft}Pre\ {\isadigit{0}}\ {\isacharbrackleft}{\isacharbrackright}{\isacharparenright}{\isacartoucheclose}\ {\isacharbar}\isanewline
}
\DefineSnippet{Sequent-Calculus-Verifier:primrec:to-fm-3}{%
\ \ {\isacartoucheopen}to{\isacharunderscore}fm\ {\isacharparenleft}Pred\ b\ ts{\isacharparenright}\ {\isacharequal}\ Pre\ b\ {\isacharparenleft}to{\isacharunderscore}tm{\isacharunderscore}list\ ts{\isacharparenright}{\isacartoucheclose}\ {\isacharbar}\isanewline
}
\DefineSnippet{Sequent-Calculus-Verifier:primrec:to-fm-4}{%
\ \ {\isacartoucheopen}to{\isacharunderscore}fm\ {\isacharparenleft}And\ p\ q{\isacharparenright}\ {\isacharequal}\ Con\ {\isacharparenleft}to{\isacharunderscore}fm\ p{\isacharparenright}\ {\isacharparenleft}to{\isacharunderscore}fm\ q{\isacharparenright}{\isacartoucheclose}\ {\isacharbar}\isanewline
}
\DefineSnippet{Sequent-Calculus-Verifier:primrec:to-fm-5}{%
\ \ {\isacartoucheopen}to{\isacharunderscore}fm\ {\isacharparenleft}Or\ p\ q{\isacharparenright}\ {\isacharequal}\ Dis\ {\isacharparenleft}to{\isacharunderscore}fm\ p{\isacharparenright}\ {\isacharparenleft}to{\isacharunderscore}fm\ q{\isacharparenright}{\isacartoucheclose}\ {\isacharbar}\isanewline
}
\DefineSnippet{Sequent-Calculus-Verifier:primrec:to-fm-6}{%
\ \ {\isacartoucheopen}to{\isacharunderscore}fm\ {\isacharparenleft}Impl\ p\ q{\isacharparenright}\ {\isacharequal}\ Imp\ {\isacharparenleft}to{\isacharunderscore}fm\ p{\isacharparenright}\ {\isacharparenleft}to{\isacharunderscore}fm\ q{\isacharparenright}{\isacartoucheclose}\ {\isacharbar}\isanewline
}
\DefineSnippet{Sequent-Calculus-Verifier:primrec:to-fm-7}{%
\ \ {\isacartoucheopen}to{\isacharunderscore}fm\ {\isacharparenleft}FOL{\isacharunderscore}Fitting{\isachardot}Neg\ p{\isacharparenright}\ {\isacharequal}\ Neg\ {\isacharparenleft}to{\isacharunderscore}fm\ p{\isacharparenright}{\isacartoucheclose}\ {\isacharbar}\isanewline
}
\DefineSnippet{Sequent-Calculus-Verifier:primrec:to-fm-8}{%
\ \ {\isacartoucheopen}to{\isacharunderscore}fm\ {\isacharparenleft}Forall\ p{\isacharparenright}\ {\isacharequal}\ Uni\ {\isacharparenleft}to{\isacharunderscore}fm\ p{\isacharparenright}{\isacartoucheclose}\ {\isacharbar}\isanewline
}
\DefineSnippet{Sequent-Calculus-Verifier:primrec:to-fm-9}{%
\ \ {\isacartoucheopen}to{\isacharunderscore}fm\ {\isacharparenleft}Exists\ p{\isacharparenright}\ {\isacharequal}\ Exi\ {\isacharparenleft}to{\isacharunderscore}fm\ p{\isacharparenright}{\isacartoucheclose}%
}
\DefineSnippet{Sequent-Calculus-Verifier:theorem:to-from-tm-0}{%
\isacommand{theorem}\isamarkupfalse%
\ to{\isacharunderscore}from{\isacharunderscore}tm\ {\isacharbrackleft}simp{\isacharbrackright}{\isacharcolon}\ {\isacartoucheopen}to{\isacharunderscore}tm\ {\isacharparenleft}from{\isacharunderscore}tm\ t{\isacharparenright}\ {\isacharequal}\ t{\isacartoucheclose}\ {\isacartoucheopen}to{\isacharunderscore}tm{\isacharunderscore}list\ {\isacharparenleft}from{\isacharunderscore}tm{\isacharunderscore}list\ ts{\isacharparenright}\ {\isacharequal}\ ts{\isacartoucheclose}\isanewline
}
\DefineSnippet{Sequent-Calculus-Verifier:theorem:to-from-tm-1}{%
\isadelimproof
\ \ %
\endisadelimproof
\isatagproof
\isacommand{by}\isamarkupfalse%
\ {\isacharparenleft}induct\ t\ \isakeyword{and}\ ts\ rule{\isacharcolon}\ from{\isacharunderscore}tm{\isachardot}induct\ from{\isacharunderscore}tm{\isacharunderscore}list{\isachardot}induct{\isacharparenright}\ simp{\isacharunderscore}all%
\endisatagproof
{\isafoldproof}%
\isadelimproof
\endisadelimproof
}
\DefineSnippet{Sequent-Calculus-Verifier:theorem:to-from-fm-0}{%
\isacommand{theorem}\isamarkupfalse%
\ to{\isacharunderscore}from{\isacharunderscore}fm\ {\isacharbrackleft}simp{\isacharbrackright}{\isacharcolon}\ {\isacartoucheopen}to{\isacharunderscore}fm\ {\isacharparenleft}from{\isacharunderscore}fm\ p{\isacharparenright}\ {\isacharequal}\ p{\isacartoucheclose}\isanewline
}
\DefineSnippet{Sequent-Calculus-Verifier:theorem:to-from-fm-1}{%
\isadelimproof
\ \ %
\endisadelimproof
\isatagproof
\isacommand{by}\isamarkupfalse%
\ {\isacharparenleft}induct\ p{\isacharparenright}\ simp{\isacharunderscore}all%
\endisatagproof
{\isafoldproof}%
\isadelimproof
\endisadelimproof
}
\DefineSnippet{Sequent-Calculus-Verifier:lemma:Truth-0}{%
\isacommand{lemma}\isamarkupfalse%
\ Truth\ {\isacharbrackleft}simp{\isacharbrackright}{\isacharcolon}\ {\isacartoucheopen}{\isasymtturnstile}\ Imp\ {\isacharparenleft}Pre\ {\isadigit{0}}\ {\isacharbrackleft}{\isacharbrackright}{\isacharparenright}\ {\isacharparenleft}Pre\ {\isadigit{0}}\ {\isacharbrackleft}{\isacharbrackright}{\isacharparenright}\ {\isacharhash}\ z{\isacartoucheclose}\isanewline
}
\DefineSnippet{Sequent-Calculus-Verifier:lemma:Truth-1}{%
\isadelimproof
\ \ %
\endisadelimproof
\isatagproof
\isacommand{using}\isamarkupfalse%
\ AlphaImp\ Basic\ Ext\ ext{\isachardot}simps\ member{\isachardot}simps{\isacharparenleft}{\isadigit{2}}{\isacharparenright}\ \isacommand{by}\isamarkupfalse%
\ metis%
\endisatagproof
{\isafoldproof}%
\isadelimproof
\endisadelimproof
}
\DefineSnippet{Sequent-Calculus-Verifier:lemma:paramst-0}{%
\isacommand{lemma}\isamarkupfalse%
\ paramst\ {\isacharbrackleft}simp{\isacharbrackright}{\isacharcolon}\isanewline
}
\DefineSnippet{Sequent-Calculus-Verifier:lemma:paramst-1}{%
\ \ {\isacartoucheopen}FOL{\isacharunderscore}Fitting{\isachardot}new{\isacharunderscore}term\ c\ t\ {\isacharequal}\ new{\isacharunderscore}term\ c\ {\isacharparenleft}to{\isacharunderscore}tm\ t{\isacharparenright}{\isacartoucheclose}\isanewline
}
\DefineSnippet{Sequent-Calculus-Verifier:lemma:paramst-2}{%
\ \ {\isacartoucheopen}FOL{\isacharunderscore}Fitting{\isachardot}new{\isacharunderscore}list\ c\ l\ {\isacharequal}\ new{\isacharunderscore}list\ c\ {\isacharparenleft}to{\isacharunderscore}tm{\isacharunderscore}list\ l{\isacharparenright}{\isacartoucheclose}\isanewline
}
\DefineSnippet{Sequent-Calculus-Verifier:lemma:paramst-3}{%
\isadelimproof
\ \ %
\endisadelimproof
\isatagproof
\isacommand{by}\isamarkupfalse%
\ {\isacharparenleft}induct\ t\ \isakeyword{and}\ l\ rule{\isacharcolon}\ FOL{\isacharunderscore}Fitting{\isachardot}paramst{\isachardot}induct\ FOL{\isacharunderscore}Fitting{\isachardot}paramsts{\isachardot}induct{\isacharparenright}\ simp{\isacharunderscore}all%
\endisatagproof
{\isafoldproof}%
\isadelimproof
\endisadelimproof
}
\DefineSnippet{Sequent-Calculus-Verifier:lemma:params-0}{%
\isacommand{lemma}\isamarkupfalse%
\ params\ {\isacharbrackleft}iff{\isacharbrackright}{\isacharcolon}\ {\isacartoucheopen}FOL{\isacharunderscore}Fitting{\isachardot}new\ c\ p\ {\isasymlongleftrightarrow}\ new\ c\ {\isacharparenleft}to{\isacharunderscore}fm\ p{\isacharparenright}{\isacartoucheclose}\isanewline
}
\DefineSnippet{Sequent-Calculus-Verifier:lemma:params-1}{%
\isadelimproof
\ \ %
\endisadelimproof
\isatagproof
\isacommand{by}\isamarkupfalse%
\ {\isacharparenleft}induct\ p{\isacharparenright}\ simp{\isacharunderscore}all%
\endisatagproof
{\isafoldproof}%
\isadelimproof
\endisadelimproof
}
\DefineSnippet{Sequent-Calculus-Verifier:lemma:list-params-0}{%
\isacommand{lemma}\isamarkupfalse%
\ list{\isacharunderscore}params\ {\isacharbrackleft}iff{\isacharbrackright}{\isacharcolon}\ {\isacartoucheopen}FOL{\isacharunderscore}Fitting{\isachardot}news\ c\ z\ {\isasymlongleftrightarrow}\ news\ c\ {\isacharparenleft}map\ to{\isacharunderscore}fm\ z{\isacharparenright}{\isacartoucheclose}\isanewline
}
\DefineSnippet{Sequent-Calculus-Verifier:lemma:list-params-1}{%
\isadelimproof
\ \ %
\endisadelimproof
\isatagproof
\isacommand{by}\isamarkupfalse%
\ {\isacharparenleft}induct\ z{\isacharparenright}\ simp{\isacharunderscore}all%
\endisatagproof
{\isafoldproof}%
\isadelimproof
\endisadelimproof
}
\DefineSnippet{Sequent-Calculus-Verifier:lemma:liftt-0}{%
\isacommand{lemma}\isamarkupfalse%
\ liftt\ {\isacharbrackleft}simp{\isacharbrackright}{\isacharcolon}\isanewline
}
\DefineSnippet{Sequent-Calculus-Verifier:lemma:liftt-1}{%
\ \ {\isacartoucheopen}to{\isacharunderscore}tm\ {\isacharparenleft}FOL{\isacharunderscore}Fitting{\isachardot}liftt\ t{\isacharparenright}\ {\isacharequal}\ inc{\isacharunderscore}term\ {\isacharparenleft}to{\isacharunderscore}tm\ t{\isacharparenright}{\isacartoucheclose}\isanewline
}
\DefineSnippet{Sequent-Calculus-Verifier:lemma:liftt-2}{%
\ \ {\isacartoucheopen}to{\isacharunderscore}tm{\isacharunderscore}list\ {\isacharparenleft}FOL{\isacharunderscore}Fitting{\isachardot}liftts\ l{\isacharparenright}\ {\isacharequal}\ inc{\isacharunderscore}list\ {\isacharparenleft}to{\isacharunderscore}tm{\isacharunderscore}list\ l{\isacharparenright}{\isacartoucheclose}\isanewline
}
\DefineSnippet{Sequent-Calculus-Verifier:lemma:liftt-3}{%
\isadelimproof
\ \ %
\endisadelimproof
\isatagproof
\isacommand{by}\isamarkupfalse%
\ {\isacharparenleft}induct\ t\ \isakeyword{and}\ l\ rule{\isacharcolon}\ FOL{\isacharunderscore}Fitting{\isachardot}liftt{\isachardot}induct\ FOL{\isacharunderscore}Fitting{\isachardot}liftts{\isachardot}induct{\isacharparenright}\ simp{\isacharunderscore}all%
\endisatagproof
{\isafoldproof}%
\isadelimproof
\endisadelimproof
}
\DefineSnippet{Sequent-Calculus-Verifier:lemma:substt-0}{%
\isacommand{lemma}\isamarkupfalse%
\ substt\ {\isacharbrackleft}simp{\isacharbrackright}{\isacharcolon}\isanewline
}
\DefineSnippet{Sequent-Calculus-Verifier:lemma:substt-1}{%
\ \ {\isacartoucheopen}to{\isacharunderscore}tm\ {\isacharparenleft}FOL{\isacharunderscore}Fitting{\isachardot}substt\ t\ s\ v{\isacharparenright}\ {\isacharequal}\ sub{\isacharunderscore}term\ v\ {\isacharparenleft}to{\isacharunderscore}tm\ s{\isacharparenright}\ {\isacharparenleft}to{\isacharunderscore}tm\ t{\isacharparenright}{\isacartoucheclose}\isanewline
}
\DefineSnippet{Sequent-Calculus-Verifier:lemma:substt-2}{%
\ \ {\isacartoucheopen}to{\isacharunderscore}tm{\isacharunderscore}list\ {\isacharparenleft}FOL{\isacharunderscore}Fitting{\isachardot}substts\ l\ s\ v{\isacharparenright}\ {\isacharequal}\ sub{\isacharunderscore}list\ v\ {\isacharparenleft}to{\isacharunderscore}tm\ s{\isacharparenright}\ {\isacharparenleft}to{\isacharunderscore}tm{\isacharunderscore}list\ l{\isacharparenright}{\isacartoucheclose}\isanewline
}
\DefineSnippet{Sequent-Calculus-Verifier:lemma:substt-3}{%
\isadelimproof
\ \ %
\endisadelimproof
\isatagproof
\isacommand{by}\isamarkupfalse%
\ {\isacharparenleft}induct\ t\ \isakeyword{and}\ l\ rule{\isacharcolon}\ FOL{\isacharunderscore}Fitting{\isachardot}substt{\isachardot}induct\ FOL{\isacharunderscore}Fitting{\isachardot}substts{\isachardot}induct{\isacharparenright}\ simp{\isacharunderscore}all%
\endisatagproof
{\isafoldproof}%
\isadelimproof
\endisadelimproof
}
\DefineSnippet{Sequent-Calculus-Verifier:lemma:subst-0}{%
\isacommand{lemma}\isamarkupfalse%
\ subst\ {\isacharbrackleft}simp{\isacharbrackright}{\isacharcolon}\ {\isacartoucheopen}to{\isacharunderscore}fm\ {\isacharparenleft}FOL{\isacharunderscore}Fitting{\isachardot}subst\ A\ t\ s{\isacharparenright}\ {\isacharequal}\ sub\ s\ {\isacharparenleft}to{\isacharunderscore}tm\ t{\isacharparenright}\ {\isacharparenleft}to{\isacharunderscore}fm\ A{\isacharparenright}{\isacartoucheclose}\isanewline
}
\DefineSnippet{Sequent-Calculus-Verifier:lemma:subst-1}{%
\isadelimproof
\ \ %
\endisadelimproof
\isatagproof
\isacommand{by}\isamarkupfalse%
\ {\isacharparenleft}induct\ A\ arbitrary{\isacharcolon}\ t\ s{\isacharparenright}\ simp{\isacharunderscore}all%
\endisatagproof
{\isafoldproof}%
\isadelimproof
\endisadelimproof
}
\DefineSnippet{Sequent-Calculus-Verifier:lemma:sim-0}{%
\isacommand{lemma}\isamarkupfalse%
\ sim{\isacharcolon}\ {\isacartoucheopen}{\isacharparenleft}{\isasymturnstile}\ x{\isacharparenright}\ {\isasymLongrightarrow}\ {\isacharparenleft}{\isasymtturnstile}\ {\isacharparenleft}map\ to{\isacharunderscore}fm\ x{\isacharparenright}{\isacharparenright}{\isacartoucheclose}\isanewline
}
\DefineSnippet{Sequent-Calculus-Verifier:lemma:sim-1}{%
\isadelimproof
\ \ %
\endisadelimproof
\isatagproof
\isacommand{by}\isamarkupfalse%
\ {\isacharparenleft}induct\ rule{\isacharcolon}\ SC{\isachardot}induct{\isacharparenright}\ {\isacharparenleft}force\ intro{\isacharcolon}\ sequent{\isacharunderscore}calculus{\isachardot}intros{\isacharparenright}{\isacharplus}%
\endisatagproof
{\isafoldproof}%
\isadelimproof
\endisadelimproof
}
\DefineSnippet{Sequent-Calculus-Verifier:lemma:evalt-0}{%
\isacommand{lemma}\isamarkupfalse%
\ evalt\ {\isacharbrackleft}simp{\isacharbrackright}{\isacharcolon}\isanewline
}
\DefineSnippet{Sequent-Calculus-Verifier:lemma:evalt-1}{%
\ \ {\isacartoucheopen}semantics{\isacharunderscore}term\ e\ f\ t\ {\isacharequal}\ evalt\ e\ f\ {\isacharparenleft}from{\isacharunderscore}tm\ t{\isacharparenright}{\isacartoucheclose}\isanewline
}
\DefineSnippet{Sequent-Calculus-Verifier:lemma:evalt-2}{%
\ \ {\isacartoucheopen}semantics{\isacharunderscore}list\ e\ f\ ts\ {\isacharequal}\ evalts\ e\ f\ {\isacharparenleft}from{\isacharunderscore}tm{\isacharunderscore}list\ ts{\isacharparenright}{\isacartoucheclose}\isanewline
}
\DefineSnippet{Sequent-Calculus-Verifier:lemma:evalt-3}{%
\isadelimproof
\ \ %
\endisadelimproof
\isatagproof
\isacommand{by}\isamarkupfalse%
\ {\isacharparenleft}induct\ t\ \isakeyword{and}\ ts\ rule{\isacharcolon}\ from{\isacharunderscore}tm{\isachardot}induct\ from{\isacharunderscore}tm{\isacharunderscore}list{\isachardot}induct{\isacharparenright}\ simp{\isacharunderscore}all%
\endisatagproof
{\isafoldproof}%
\isadelimproof
\endisadelimproof
}
\DefineSnippet{Sequent-Calculus-Verifier:lemma:shift-0}{%
\isacommand{lemma}\isamarkupfalse%
\ shift\ {\isacharbrackleft}simp{\isacharbrackright}{\isacharcolon}\ {\isacartoucheopen}shift\ e\ {\isadigit{0}}\ x\ {\isacharequal}\ e{\isasymlangle}{\isadigit{0}}{\isacharcolon}x{\isasymrangle}{\isacartoucheclose}\isanewline
}
\DefineSnippet{Sequent-Calculus-Verifier:lemma:shift-1}{%
\isadelimproof
\ \ %
\endisadelimproof
\isatagproof
\isacommand{unfolding}\isamarkupfalse%
\ shift{\isacharunderscore}def\ FOL{\isacharunderscore}Fitting{\isachardot}shift{\isacharunderscore}def\ \isacommand{by}\isamarkupfalse%
\ simp%
\endisatagproof
{\isafoldproof}%
\isadelimproof
\endisadelimproof
}
\DefineSnippet{Sequent-Calculus-Verifier:lemma:semantics-0}{%
\isacommand{lemma}\isamarkupfalse%
\ semantics\ {\isacharbrackleft}iff{\isacharbrackright}{\isacharcolon}\ {\isacartoucheopen}semantics\ e\ f\ g\ p\ {\isasymlongleftrightarrow}\ eval\ e\ f\ g\ {\isacharparenleft}from{\isacharunderscore}fm\ p{\isacharparenright}{\isacartoucheclose}\isanewline
}
\DefineSnippet{Sequent-Calculus-Verifier:lemma:semantics-1}{%
\isadelimproof
\ \ %
\endisadelimproof
\isatagproof
\isacommand{by}\isamarkupfalse%
\ {\isacharparenleft}induct\ p\ arbitrary{\isacharcolon}\ e{\isacharparenright}\ simp{\isacharunderscore}all%
\endisatagproof
{\isafoldproof}%
\isadelimproof
\endisadelimproof
}
\DefineSnippet{Sequent-Calculus-Verifier:abbreviation:valid-0}{%
\isacommand{abbreviation}\isamarkupfalse%
\ valid\ {\isacharparenleft}{\isachardoublequoteopen}{\isasymthen}\ {\isacharunderscore}{\isachardoublequoteclose}\ {\isadigit{0}}{\isacharparenright}\ \isakeyword{where}\isanewline
}
\DefineSnippet{Sequent-Calculus-Verifier:abbreviation:valid-1}{%
\ \ {\isacartoucheopen}{\isacharparenleft}{\isasymthen}\ p{\isacharparenright}\ {\isasymequiv}\ {\isasymforall}{\isacharparenleft}e\ {\isacharcolon}{\isacharcolon}\ {\isacharunderscore}\ {\isasymRightarrow}\ nat\ hterm{\isacharparenright}\ f\ g{\isachardot}\ semantics\ e\ f\ g\ p{\isacartoucheclose}%
}
\DefineSnippet{Sequent-Calculus-Verifier:theorem:complete-sound-0}{%
\isacommand{theorem}\isamarkupfalse%
\ complete{\isacharunderscore}sound{\isacharcolon}\ {\isacartoucheopen}{\isasymthen}\ p\ {\isasymLongrightarrow}\ {\isasymtturnstile}\ {\isacharbrackleft}p{\isacharbrackright}{\isacartoucheclose}\ {\isacartoucheopen}{\isasymtturnstile}\ {\isacharbrackleft}q{\isacharbrackright}\ {\isasymLongrightarrow}\ semantics\ e\ f\ g\ q{\isacartoucheclose}\isanewline
}
\DefineSnippet{Sequent-Calculus-Verifier:theorem:complete-sound-1}{%
\isadelimproof
\ \ %
\endisadelimproof
\isatagproof
\isacommand{by}\isamarkupfalse%
\ {\isacharparenleft}metis\ to{\isacharunderscore}from{\isacharunderscore}fm\ sim\ semantics\ list{\isachardot}map\ SC{\isacharunderscore}completeness{\isacharparenright}\ {\isacharparenleft}use\ sound\ \isakeyword{in}\ force{\isacharparenright}%
\endisatagproof
{\isafoldproof}%
\isadelimproof
\endisadelimproof
}
\DefineSnippet{Sequent-Calculus-Verifier:corollary:c0ba05d8f2cce949-0}{%
\isacommand{corollary}\isamarkupfalse%
\ {\isacartoucheopen}{\isacharparenleft}{\isasymthen}\ p{\isacharparenright}\ {\isasymlongleftrightarrow}\ {\isacharparenleft}{\isasymtturnstile}\ {\isacharbrackleft}p{\isacharbrackright}{\isacharparenright}{\isacartoucheclose}\isanewline
}
\DefineSnippet{Sequent-Calculus-Verifier:corollary:c0ba05d8f2cce949-1}{%
\isadelimproof
\ \ %
\endisadelimproof
\isatagproof
\isacommand{using}\isamarkupfalse%
\ complete{\isacharunderscore}sound\ \isacommand{by}\isamarkupfalse%
\ fast%
\endisatagproof
{\isafoldproof}%
\isadelimproof
\endisadelimproof
}
\DefineSnippet{Sequent-Calculus-Verifier:section:333af9ea13740c84-0}{%
\isadelimdocument
\endisadelimdocument
\isatagdocument
\isamarkupsection{Reference%
}
\isamarkuptrue%
\endisatagdocument
{\isafolddocument}%
\isadelimdocument
\endisadelimdocument
}
\DefineSnippet{Sequent-Calculus-Verifier:end:7158f6523c28304a-0}{%
\isadelimtheory
\endisadelimtheory
\isatagtheory
\isacommand{end}\isamarkupfalse%
\endisatagtheory
{\isafoldtheory}%
\isadelimtheory
\endisadelimtheory
}
\DefineSnippet{Examples:theory:Examples-0}{%
\isadelimtheory
\endisadelimtheory
\isatagtheory
\isacommand{theory}\isamarkupfalse%
\ Examples\ \isakeyword{imports}\ SeCaV\ \isakeyword{begin}%
\endisatagtheory
{\isafoldtheory}%
\isadelimtheory
\endisadelimtheory
}
\DefineSnippet{Examples:section:4cfee1ef64a0f4a8-0}{%
\isadelimdocument
\endisadelimdocument
\isatagdocument
\isamarkupsection{Example 1%
}
\isamarkuptrue%
\endisatagdocument
{\isafolddocument}%
\isadelimdocument
\endisadelimdocument
}
\DefineSnippet{Examples:proposition:4716e8e0a97c6fd4-0}{%
\isacommand{proposition}\isamarkupfalse%
\ {\isacartoucheopen}{\isacharparenleft}p\ a\ b{\isacharparenright}\ {\isasymor}\ {\isacharparenleft}{\isasymnot}{\isacharparenleft}p\ a\ b{\isacharparenright}{\isacharparenright}{\isacartoucheclose}%
\isadelimproof
\ %
\endisadelimproof
\isatagproof
\isacommand{by}\isamarkupfalse%
\ metis%
\endisatagproof
{\isafoldproof}%
\isadelimproof
\endisadelimproof
}
\DefineSnippet{Examples:text:c0610c4e750da3eb-0}{%
\begin{isamarkuptext}%
Predicate numbers
    0 = p
  Function numbers
    0 = a
    1 = b%
\end{isamarkuptext}\isamarkuptrue%
}
\DefineSnippet{Examples:lemma:67c91b7751fada70-0}{%
\isacommand{lemma}\isamarkupfalse%
\ {\isacartoucheopen}{\isasymtturnstile}\isanewline
}
\DefineSnippet{Examples:lemma:67c91b7751fada70-1}{%
\ \ {\isacharbrackleft}\isanewline
}
\DefineSnippet{Examples:lemma:67c91b7751fada70-2}{%
\ \ \ \ Dis\ {\isacharparenleft}Pre\ {\isadigit{0}}\ {\isacharbrackleft}Fun\ {\isadigit{0}}\ {\isacharbrackleft}{\isacharbrackright}{\isacharcomma}\ Fun\ {\isadigit{1}}\ {\isacharbrackleft}{\isacharbrackright}{\isacharbrackright}{\isacharparenright}\ {\isacharparenleft}Neg\ {\isacharparenleft}Pre\ {\isadigit{0}}\ {\isacharbrackleft}Fun\ {\isadigit{0}}\ {\isacharbrackleft}{\isacharbrackright}{\isacharcomma}\ Fun\ {\isadigit{1}}\ {\isacharbrackleft}{\isacharbrackright}{\isacharbrackright}{\isacharparenright}{\isacharparenright}\isanewline
}
\DefineSnippet{Examples:lemma:67c91b7751fada70-3}{%
\ \ {\isacharbrackright}\isanewline
}
\DefineSnippet{Examples:lemma:67c91b7751fada70-4}{%
\ \ {\isacartoucheclose}\isanewline
}
\DefineSnippet{Examples:lemma:67c91b7751fada70-5}{%
\isadelimproof
\endisadelimproof
\isatagproof
\isacommand{proof}\isamarkupfalse%
\ {\isacharminus}\isanewline
}
\DefineSnippet{Examples:lemma:67c91b7751fada70-6}{%
\ \ \isacommand{from}\isamarkupfalse%
\ AlphaDis\ \isacommand{have}\isamarkupfalse%
\ {\isacharquery}thesis\ \isakeyword{if}\ {\isacartoucheopen}{\isasymtturnstile}\isanewline
}
\DefineSnippet{Examples:lemma:67c91b7751fada70-7}{%
\ \ \ \ {\isacharbrackleft}\isanewline
}
\DefineSnippet{Examples:lemma:67c91b7751fada70-8}{%
\ \ \ \ \ \ Pre\ {\isadigit{0}}\ {\isacharbrackleft}Fun\ {\isadigit{0}}\ {\isacharbrackleft}{\isacharbrackright}{\isacharcomma}\ Fun\ {\isadigit{1}}\ {\isacharbrackleft}{\isacharbrackright}{\isacharbrackright}{\isacharcomma}\isanewline
}
\DefineSnippet{Examples:lemma:67c91b7751fada70-9}{%
\ \ \ \ \ \ Neg\ {\isacharparenleft}Pre\ {\isadigit{0}}\ {\isacharbrackleft}Fun\ {\isadigit{0}}\ {\isacharbrackleft}{\isacharbrackright}{\isacharcomma}\ Fun\ {\isadigit{1}}\ {\isacharbrackleft}{\isacharbrackright}{\isacharbrackright}{\isacharparenright}\isanewline
}
\DefineSnippet{Examples:lemma:67c91b7751fada70-10}{%
\ \ \ \ {\isacharbrackright}\isanewline
}
\DefineSnippet{Examples:lemma:67c91b7751fada70-11}{%
\ \ \ \ {\isacartoucheclose}\isanewline
}
\DefineSnippet{Examples:lemma:67c91b7751fada70-12}{%
\ \ \ \ \isacommand{using}\isamarkupfalse%
\ that\ \isacommand{by}\isamarkupfalse%
\ simp\isanewline
}
\DefineSnippet{Examples:lemma:67c91b7751fada70-13}{%
\ \ \isacommand{with}\isamarkupfalse%
\ Basic\ \isacommand{show}\isamarkupfalse%
\ {\isacharquery}thesis\isanewline
}
\DefineSnippet{Examples:lemma:67c91b7751fada70-14}{%
\ \ \ \ \isacommand{by}\isamarkupfalse%
\ simp\isanewline
}
\DefineSnippet{Examples:lemma:67c91b7751fada70-15}{%
\isacommand{qed}\isamarkupfalse%
\endisatagproof
{\isafoldproof}%
\isadelimproof
\endisadelimproof
}
\DefineSnippet{Examples:section:d802fa1e9e10a169-0}{%
\isadelimdocument
\endisadelimdocument
\isatagdocument
\isamarkupsection{Example 2%
}
\isamarkuptrue%
\endisatagdocument
{\isafolddocument}%
\isadelimdocument
\endisadelimdocument
}
\DefineSnippet{Examples:proposition:5892606a27dd15b0-0}{%
\isacommand{proposition}\isamarkupfalse%
\ {\isacartoucheopen}{\isacharparenleft}{\isasymforall}x{\isachardot}\ {\isacharparenleft}{\isasymforall}y{\isachardot}\ {\isacharparenleft}p\ x\ y{\isacharparenright}{\isacharparenright}{\isacharparenright}\ {\isasymlongrightarrow}\ {\isacharparenleft}p\ a\ a{\isacharparenright}{\isacartoucheclose}%
\isadelimproof
\ %
\endisadelimproof
\isatagproof
\isacommand{by}\isamarkupfalse%
\ metis%
\endisatagproof
{\isafoldproof}%
\isadelimproof
\endisadelimproof
}
\DefineSnippet{Examples:text:7594e6754e8f0b80-0}{%
\begin{isamarkuptext}%
Predicate numbers
    0 = p
  Function numbers
    0 = a%
\end{isamarkuptext}\isamarkuptrue%
}
\DefineSnippet{Examples:lemma:43b39acd50cf3f31-0}{%
\isacommand{lemma}\isamarkupfalse%
\ {\isacartoucheopen}{\isasymtturnstile}\isanewline
}
\DefineSnippet{Examples:lemma:43b39acd50cf3f31-1}{%
\ \ {\isacharbrackleft}\isanewline
}
\DefineSnippet{Examples:lemma:43b39acd50cf3f31-2}{%
\ \ \ \ Imp\ {\isacharparenleft}Uni\ {\isacharparenleft}Uni\ {\isacharparenleft}Pre\ {\isadigit{0}}\ {\isacharbrackleft}Var\ {\isadigit{1}}{\isacharcomma}\ Var\ {\isadigit{0}}{\isacharbrackright}{\isacharparenright}{\isacharparenright}{\isacharparenright}\ {\isacharparenleft}Pre\ {\isadigit{0}}\ {\isacharbrackleft}Fun\ {\isadigit{0}}\ {\isacharbrackleft}{\isacharbrackright}{\isacharcomma}\ Fun\ {\isadigit{0}}\ {\isacharbrackleft}{\isacharbrackright}{\isacharbrackright}{\isacharparenright}\isanewline
}
\DefineSnippet{Examples:lemma:43b39acd50cf3f31-3}{%
\ \ {\isacharbrackright}\isanewline
}
\DefineSnippet{Examples:lemma:43b39acd50cf3f31-4}{%
\ \ {\isacartoucheclose}\isanewline
}
\DefineSnippet{Examples:lemma:43b39acd50cf3f31-5}{%
\isadelimproof
\endisadelimproof
\isatagproof
\isacommand{proof}\isamarkupfalse%
\ {\isacharminus}\isanewline
}
\DefineSnippet{Examples:lemma:43b39acd50cf3f31-6}{%
\ \ \isacommand{from}\isamarkupfalse%
\ AlphaImp\ \isacommand{have}\isamarkupfalse%
\ {\isacharquery}thesis\ \isakeyword{if}\ {\isacartoucheopen}{\isasymtturnstile}\isanewline
}
\DefineSnippet{Examples:lemma:43b39acd50cf3f31-7}{%
\ \ \ \ {\isacharbrackleft}\isanewline
}
\DefineSnippet{Examples:lemma:43b39acd50cf3f31-8}{%
\ \ \ \ \ \ Neg\ {\isacharparenleft}Uni\ {\isacharparenleft}Uni\ {\isacharparenleft}Pre\ {\isadigit{0}}\ {\isacharbrackleft}Var\ {\isadigit{1}}{\isacharcomma}\ Var\ {\isadigit{0}}{\isacharbrackright}{\isacharparenright}{\isacharparenright}{\isacharparenright}{\isacharcomma}\isanewline
}
\DefineSnippet{Examples:lemma:43b39acd50cf3f31-9}{%
\ \ \ \ \ \ Pre\ {\isadigit{0}}\ {\isacharbrackleft}Fun\ {\isadigit{0}}\ {\isacharbrackleft}{\isacharbrackright}{\isacharcomma}\ Fun\ {\isadigit{0}}\ {\isacharbrackleft}{\isacharbrackright}{\isacharbrackright}\isanewline
}
\DefineSnippet{Examples:lemma:43b39acd50cf3f31-10}{%
\ \ \ \ {\isacharbrackright}\isanewline
}
\DefineSnippet{Examples:lemma:43b39acd50cf3f31-11}{%
\ \ \ \ {\isacartoucheclose}\isanewline
}
\DefineSnippet{Examples:lemma:43b39acd50cf3f31-12}{%
\ \ \ \ \isacommand{using}\isamarkupfalse%
\ that\ \isacommand{by}\isamarkupfalse%
\ simp\isanewline
}
\DefineSnippet{Examples:lemma:43b39acd50cf3f31-13}{%
\ \ \isacommand{with}\isamarkupfalse%
\ GammaUni{\isacharbrackleft}\isakeyword{where}\ t{\isacharequal}{\isacartoucheopen}Fun\ {\isadigit{0}}\ {\isacharbrackleft}{\isacharbrackright}{\isacartoucheclose}{\isacharbrackright}\ \isacommand{have}\isamarkupfalse%
\ {\isacharquery}thesis\ \isakeyword{if}\ {\isacartoucheopen}{\isasymtturnstile}\isanewline
}
\DefineSnippet{Examples:lemma:43b39acd50cf3f31-14}{%
\ \ \ \ {\isacharbrackleft}\isanewline
}
\DefineSnippet{Examples:lemma:43b39acd50cf3f31-15}{%
\ \ \ \ \ \ Neg\ {\isacharparenleft}Pre\ {\isadigit{0}}\ {\isacharbrackleft}Fun\ {\isadigit{0}}\ {\isacharbrackleft}{\isacharbrackright}{\isacharcomma}\ Fun\ {\isadigit{0}}\ {\isacharbrackleft}{\isacharbrackright}{\isacharbrackright}{\isacharparenright}{\isacharcomma}\isanewline
}
\DefineSnippet{Examples:lemma:43b39acd50cf3f31-16}{%
\ \ \ \ \ \ Pre\ {\isadigit{0}}\ {\isacharbrackleft}Fun\ {\isadigit{0}}\ {\isacharbrackleft}{\isacharbrackright}{\isacharcomma}\ Fun\ {\isadigit{0}}\ {\isacharbrackleft}{\isacharbrackright}{\isacharbrackright}\isanewline
}
\DefineSnippet{Examples:lemma:43b39acd50cf3f31-17}{%
\ \ \ \ {\isacharbrackright}\isanewline
}
\DefineSnippet{Examples:lemma:43b39acd50cf3f31-18}{%
\ \ \ \ {\isacartoucheclose}\isanewline
}
\DefineSnippet{Examples:lemma:43b39acd50cf3f31-19}{%
\ \ \ \ \isacommand{using}\isamarkupfalse%
\ that\ \isacommand{by}\isamarkupfalse%
\ simp\isanewline
}
\DefineSnippet{Examples:lemma:43b39acd50cf3f31-20}{%
\ \ \isacommand{with}\isamarkupfalse%
\ Ext\ \isacommand{have}\isamarkupfalse%
\ {\isacharquery}thesis\ \isakeyword{if}\ {\isacartoucheopen}{\isasymtturnstile}\isanewline
}
\DefineSnippet{Examples:lemma:43b39acd50cf3f31-21}{%
\ \ \ \ {\isacharbrackleft}\isanewline
}
\DefineSnippet{Examples:lemma:43b39acd50cf3f31-22}{%
\ \ \ \ \ \ Pre\ {\isadigit{0}}\ {\isacharbrackleft}Fun\ {\isadigit{0}}\ {\isacharbrackleft}{\isacharbrackright}{\isacharcomma}\ Fun\ {\isadigit{0}}\ {\isacharbrackleft}{\isacharbrackright}{\isacharbrackright}{\isacharcomma}\isanewline
}
\DefineSnippet{Examples:lemma:43b39acd50cf3f31-23}{%
\ \ \ \ \ \ Neg\ {\isacharparenleft}Pre\ {\isadigit{0}}\ {\isacharbrackleft}Fun\ {\isadigit{0}}\ {\isacharbrackleft}{\isacharbrackright}{\isacharcomma}\ Fun\ {\isadigit{0}}\ {\isacharbrackleft}{\isacharbrackright}{\isacharbrackright}{\isacharparenright}\isanewline
}
\DefineSnippet{Examples:lemma:43b39acd50cf3f31-24}{%
\ \ \ \ {\isacharbrackright}\isanewline
}
\DefineSnippet{Examples:lemma:43b39acd50cf3f31-25}{%
\ \ \ \ {\isacartoucheclose}\isanewline
}
\DefineSnippet{Examples:lemma:43b39acd50cf3f31-26}{%
\ \ \ \ \isacommand{using}\isamarkupfalse%
\ that\ \isacommand{by}\isamarkupfalse%
\ simp\isanewline
}
\DefineSnippet{Examples:lemma:43b39acd50cf3f31-27}{%
\ \ \isacommand{with}\isamarkupfalse%
\ Basic\ \isacommand{show}\isamarkupfalse%
\ {\isacharquery}thesis\isanewline
}
\DefineSnippet{Examples:lemma:43b39acd50cf3f31-28}{%
\ \ \ \ \isacommand{by}\isamarkupfalse%
\ simp\isanewline
}
\DefineSnippet{Examples:lemma:43b39acd50cf3f31-29}{%
\isacommand{qed}\isamarkupfalse%
\endisatagproof
{\isafoldproof}%
\isadelimproof
\endisadelimproof
}
\DefineSnippet{Examples:section:58094403e30661df-0}{%
\isadelimdocument
\endisadelimdocument
\isatagdocument
\isamarkupsection{Example 3%
}
\isamarkuptrue%
\endisatagdocument
{\isafolddocument}%
\isadelimdocument
\endisadelimdocument
}
\DefineSnippet{Examples:proposition:fc77485f97a2360a-0}{%
\isacommand{proposition}\isamarkupfalse%
\ {\isacartoucheopen}{\isacharparenleft}{\isasymforall}x{\isachardot}\ {\isacharparenleft}{\isacharparenleft}p\ x{\isacharparenright}\ {\isasymlongrightarrow}\ {\isacharparenleft}q\ x{\isacharparenright}{\isacharparenright}{\isacharparenright}\ {\isasymlongrightarrow}\ {\isacharparenleft}{\isacharparenleft}{\isasymexists}x{\isachardot}\ {\isacharparenleft}p\ x{\isacharparenright}{\isacharparenright}\ {\isasymlongrightarrow}\ {\isacharparenleft}{\isasymexists}x{\isachardot}\ {\isacharparenleft}q\ x{\isacharparenright}{\isacharparenright}{\isacharparenright}{\isacartoucheclose}%
\isadelimproof
\ %
\endisadelimproof
\isatagproof
\isacommand{by}\isamarkupfalse%
\ metis%
\endisatagproof
{\isafoldproof}%
\isadelimproof
\endisadelimproof
}
\DefineSnippet{Examples:text:a34726466f79db0b-0}{%
\begin{isamarkuptext}%
Predicate numbers
    0 = p
    1 = q
  Function numbers
    0 = a%
\end{isamarkuptext}\isamarkuptrue%
}
\DefineSnippet{Examples:lemma:b6645243b11899ae-0}{%
\isacommand{lemma}\isamarkupfalse%
\ {\isacartoucheopen}{\isasymtturnstile}\isanewline
}
\DefineSnippet{Examples:lemma:b6645243b11899ae-1}{%
\ \ {\isacharbrackleft}\isanewline
}
\DefineSnippet{Examples:lemma:b6645243b11899ae-2}{%
\ \ \ \ Imp\ {\isacharparenleft}Uni\ {\isacharparenleft}Imp\ {\isacharparenleft}Pre\ {\isadigit{0}}\ {\isacharbrackleft}Var\ {\isadigit{0}}{\isacharbrackright}{\isacharparenright}\ {\isacharparenleft}Pre\ {\isadigit{1}}\ {\isacharbrackleft}Var\ {\isadigit{0}}{\isacharbrackright}{\isacharparenright}{\isacharparenright}{\isacharparenright}\ {\isacharparenleft}Imp\ {\isacharparenleft}Exi\ {\isacharparenleft}Pre\ {\isadigit{0}}\ {\isacharbrackleft}Var\ {\isadigit{0}}{\isacharbrackright}{\isacharparenright}{\isacharparenright}\ {\isacharparenleft}Exi\ {\isacharparenleft}Pre\ {\isadigit{1}}\ {\isacharbrackleft}Var\ {\isadigit{0}}{\isacharbrackright}{\isacharparenright}{\isacharparenright}{\isacharparenright}\isanewline
}
\DefineSnippet{Examples:lemma:b6645243b11899ae-3}{%
\ \ {\isacharbrackright}\isanewline
}
\DefineSnippet{Examples:lemma:b6645243b11899ae-4}{%
\ \ {\isacartoucheclose}\isanewline
}
\DefineSnippet{Examples:lemma:b6645243b11899ae-5}{%
\isadelimproof
\endisadelimproof
\isatagproof
\isacommand{proof}\isamarkupfalse%
\ {\isacharminus}\isanewline
}
\DefineSnippet{Examples:lemma:b6645243b11899ae-6}{%
\ \ \isacommand{from}\isamarkupfalse%
\ AlphaImp\ \isacommand{have}\isamarkupfalse%
\ {\isacharquery}thesis\ \isakeyword{if}\ {\isacartoucheopen}{\isasymtturnstile}\isanewline
}
\DefineSnippet{Examples:lemma:b6645243b11899ae-7}{%
\ \ \ \ {\isacharbrackleft}\isanewline
}
\DefineSnippet{Examples:lemma:b6645243b11899ae-8}{%
\ \ \ \ \ \ Neg\ {\isacharparenleft}Uni\ {\isacharparenleft}Imp\ {\isacharparenleft}Pre\ {\isadigit{0}}\ {\isacharbrackleft}Var\ {\isadigit{0}}{\isacharbrackright}{\isacharparenright}\ {\isacharparenleft}Pre\ {\isadigit{1}}\ {\isacharbrackleft}Var\ {\isadigit{0}}{\isacharbrackright}{\isacharparenright}{\isacharparenright}{\isacharparenright}{\isacharcomma}\isanewline
}
\DefineSnippet{Examples:lemma:b6645243b11899ae-9}{%
\ \ \ \ \ \ Imp\ {\isacharparenleft}Exi\ {\isacharparenleft}Pre\ {\isadigit{0}}\ {\isacharbrackleft}Var\ {\isadigit{0}}{\isacharbrackright}{\isacharparenright}{\isacharparenright}\ {\isacharparenleft}Exi\ {\isacharparenleft}Pre\ {\isadigit{1}}\ {\isacharbrackleft}Var\ {\isadigit{0}}{\isacharbrackright}{\isacharparenright}{\isacharparenright}\isanewline
}
\DefineSnippet{Examples:lemma:b6645243b11899ae-10}{%
\ \ \ \ {\isacharbrackright}\isanewline
}
\DefineSnippet{Examples:lemma:b6645243b11899ae-11}{%
\ \ \ \ {\isacartoucheclose}\isanewline
}
\DefineSnippet{Examples:lemma:b6645243b11899ae-12}{%
\ \ \ \ \isacommand{using}\isamarkupfalse%
\ that\ \isacommand{by}\isamarkupfalse%
\ simp\isanewline
}
\DefineSnippet{Examples:lemma:b6645243b11899ae-13}{%
\ \ \isacommand{with}\isamarkupfalse%
\ Ext\ \isacommand{have}\isamarkupfalse%
\ {\isacharquery}thesis\ \isakeyword{if}\ {\isacartoucheopen}{\isasymtturnstile}\isanewline
}
\DefineSnippet{Examples:lemma:b6645243b11899ae-14}{%
\ \ \ \ {\isacharbrackleft}\isanewline
}
\DefineSnippet{Examples:lemma:b6645243b11899ae-15}{%
\ \ \ \ \ \ Imp\ {\isacharparenleft}Exi\ {\isacharparenleft}Pre\ {\isadigit{0}}\ {\isacharbrackleft}Var\ {\isadigit{0}}{\isacharbrackright}{\isacharparenright}{\isacharparenright}\ {\isacharparenleft}Exi\ {\isacharparenleft}Pre\ {\isadigit{1}}\ {\isacharbrackleft}Var\ {\isadigit{0}}{\isacharbrackright}{\isacharparenright}{\isacharparenright}{\isacharcomma}\isanewline
}
\DefineSnippet{Examples:lemma:b6645243b11899ae-16}{%
\ \ \ \ \ \ Neg\ {\isacharparenleft}Uni\ {\isacharparenleft}Imp\ {\isacharparenleft}Pre\ {\isadigit{0}}\ {\isacharbrackleft}Var\ {\isadigit{0}}{\isacharbrackright}{\isacharparenright}\ {\isacharparenleft}Pre\ {\isadigit{1}}\ {\isacharbrackleft}Var\ {\isadigit{0}}{\isacharbrackright}{\isacharparenright}{\isacharparenright}{\isacharparenright}\isanewline
}
\DefineSnippet{Examples:lemma:b6645243b11899ae-17}{%
\ \ \ \ {\isacharbrackright}\isanewline
}
\DefineSnippet{Examples:lemma:b6645243b11899ae-18}{%
\ \ \ \ {\isacartoucheclose}\isanewline
}
\DefineSnippet{Examples:lemma:b6645243b11899ae-19}{%
\ \ \ \ \isacommand{using}\isamarkupfalse%
\ that\ \isacommand{by}\isamarkupfalse%
\ simp\isanewline
}
\DefineSnippet{Examples:lemma:b6645243b11899ae-20}{%
\ \ \isacommand{with}\isamarkupfalse%
\ AlphaImp\ \isacommand{have}\isamarkupfalse%
\ {\isacharquery}thesis\ \isakeyword{if}\ {\isacartoucheopen}{\isasymtturnstile}\isanewline
}
\DefineSnippet{Examples:lemma:b6645243b11899ae-21}{%
\ \ \ \ {\isacharbrackleft}\isanewline
}
\DefineSnippet{Examples:lemma:b6645243b11899ae-22}{%
\ \ \ \ \ \ Neg\ {\isacharparenleft}Exi\ {\isacharparenleft}Pre\ {\isadigit{0}}\ {\isacharbrackleft}Var\ {\isadigit{0}}{\isacharbrackright}{\isacharparenright}{\isacharparenright}{\isacharcomma}\isanewline
}
\DefineSnippet{Examples:lemma:b6645243b11899ae-23}{%
\ \ \ \ \ \ Exi\ {\isacharparenleft}Pre\ {\isadigit{1}}\ {\isacharbrackleft}Var\ {\isadigit{0}}{\isacharbrackright}{\isacharparenright}{\isacharcomma}\isanewline
}
\DefineSnippet{Examples:lemma:b6645243b11899ae-24}{%
\ \ \ \ \ \ Neg\ {\isacharparenleft}Uni\ {\isacharparenleft}Imp\ {\isacharparenleft}Pre\ {\isadigit{0}}\ {\isacharbrackleft}Var\ {\isadigit{0}}{\isacharbrackright}{\isacharparenright}\ {\isacharparenleft}Pre\ {\isadigit{1}}\ {\isacharbrackleft}Var\ {\isadigit{0}}{\isacharbrackright}{\isacharparenright}{\isacharparenright}{\isacharparenright}\isanewline
}
\DefineSnippet{Examples:lemma:b6645243b11899ae-25}{%
\ \ \ \ {\isacharbrackright}\isanewline
}
\DefineSnippet{Examples:lemma:b6645243b11899ae-26}{%
\ \ \ \ {\isacartoucheclose}\isanewline
}
\DefineSnippet{Examples:lemma:b6645243b11899ae-27}{%
\ \ \ \ \isacommand{using}\isamarkupfalse%
\ that\ \isacommand{by}\isamarkupfalse%
\ simp\isanewline
}
\DefineSnippet{Examples:lemma:b6645243b11899ae-28}{%
\ \ \isacommand{with}\isamarkupfalse%
\ DeltaExi\ \isacommand{have}\isamarkupfalse%
\ {\isacharquery}thesis\ \isakeyword{if}\ {\isacartoucheopen}{\isasymtturnstile}\isanewline
}
\DefineSnippet{Examples:lemma:b6645243b11899ae-29}{%
\ \ \ \ {\isacharbrackleft}\isanewline
}
\DefineSnippet{Examples:lemma:b6645243b11899ae-30}{%
\ \ \ \ \ \ Neg\ {\isacharparenleft}Pre\ {\isadigit{0}}\ {\isacharbrackleft}Fun\ {\isadigit{0}}\ {\isacharbrackleft}{\isacharbrackright}{\isacharbrackright}{\isacharparenright}{\isacharcomma}\isanewline
}
\DefineSnippet{Examples:lemma:b6645243b11899ae-31}{%
\ \ \ \ \ \ Exi\ {\isacharparenleft}Pre\ {\isadigit{1}}\ {\isacharbrackleft}Var\ {\isadigit{0}}{\isacharbrackright}{\isacharparenright}{\isacharcomma}\isanewline
}
\DefineSnippet{Examples:lemma:b6645243b11899ae-32}{%
\ \ \ \ \ \ Neg\ {\isacharparenleft}Uni\ {\isacharparenleft}Imp\ {\isacharparenleft}Pre\ {\isadigit{0}}\ {\isacharbrackleft}Var\ {\isadigit{0}}{\isacharbrackright}{\isacharparenright}\ {\isacharparenleft}Pre\ {\isadigit{1}}\ {\isacharbrackleft}Var\ {\isadigit{0}}{\isacharbrackright}{\isacharparenright}{\isacharparenright}{\isacharparenright}\isanewline
}
\DefineSnippet{Examples:lemma:b6645243b11899ae-33}{%
\ \ \ \ {\isacharbrackright}\isanewline
}
\DefineSnippet{Examples:lemma:b6645243b11899ae-34}{%
\ \ \ \ {\isacartoucheclose}\isanewline
}
\DefineSnippet{Examples:lemma:b6645243b11899ae-35}{%
\ \ \ \ \isacommand{using}\isamarkupfalse%
\ that\ \isacommand{by}\isamarkupfalse%
\ simp\isanewline
}
\DefineSnippet{Examples:lemma:b6645243b11899ae-36}{%
\ \ \isacommand{with}\isamarkupfalse%
\ Ext\ \isacommand{have}\isamarkupfalse%
\ {\isacharquery}thesis\ \isakeyword{if}\ {\isacartoucheopen}{\isasymtturnstile}\isanewline
}
\DefineSnippet{Examples:lemma:b6645243b11899ae-37}{%
\ \ \ \ {\isacharbrackleft}\isanewline
}
\DefineSnippet{Examples:lemma:b6645243b11899ae-38}{%
\ \ \ \ \ \ Neg\ {\isacharparenleft}Uni\ {\isacharparenleft}Imp\ {\isacharparenleft}Pre\ {\isadigit{0}}\ {\isacharbrackleft}Var\ {\isadigit{0}}{\isacharbrackright}{\isacharparenright}\ {\isacharparenleft}Pre\ {\isadigit{1}}\ {\isacharbrackleft}Var\ {\isadigit{0}}{\isacharbrackright}{\isacharparenright}{\isacharparenright}{\isacharparenright}{\isacharcomma}\isanewline
}
\DefineSnippet{Examples:lemma:b6645243b11899ae-39}{%
\ \ \ \ \ \ Neg\ {\isacharparenleft}Pre\ {\isadigit{0}}\ {\isacharbrackleft}Fun\ {\isadigit{0}}\ {\isacharbrackleft}{\isacharbrackright}{\isacharbrackright}{\isacharparenright}{\isacharcomma}\isanewline
}
\DefineSnippet{Examples:lemma:b6645243b11899ae-40}{%
\ \ \ \ \ \ Exi\ {\isacharparenleft}Pre\ {\isadigit{1}}\ {\isacharbrackleft}Var\ {\isadigit{0}}{\isacharbrackright}{\isacharparenright}\isanewline
}
\DefineSnippet{Examples:lemma:b6645243b11899ae-41}{%
\ \ \ \ {\isacharbrackright}\isanewline
}
\DefineSnippet{Examples:lemma:b6645243b11899ae-42}{%
\ \ \ \ {\isacartoucheclose}\isanewline
}
\DefineSnippet{Examples:lemma:b6645243b11899ae-43}{%
\ \ \ \ \isacommand{using}\isamarkupfalse%
\ that\ \isacommand{by}\isamarkupfalse%
\ simp\isanewline
}
\DefineSnippet{Examples:lemma:b6645243b11899ae-44}{%
\ \ \isacommand{with}\isamarkupfalse%
\ GammaUni\ \isacommand{have}\isamarkupfalse%
\ {\isacharquery}thesis\ \isakeyword{if}\ {\isacartoucheopen}{\isasymtturnstile}\isanewline
}
\DefineSnippet{Examples:lemma:b6645243b11899ae-45}{%
\ \ \ \ {\isacharbrackleft}\isanewline
}
\DefineSnippet{Examples:lemma:b6645243b11899ae-46}{%
\ \ \ \ \ \ Neg\ {\isacharparenleft}Imp\ {\isacharparenleft}Pre\ {\isadigit{0}}\ {\isacharbrackleft}Fun\ {\isadigit{0}}\ {\isacharbrackleft}{\isacharbrackright}{\isacharbrackright}{\isacharparenright}\ {\isacharparenleft}Pre\ {\isadigit{1}}\ {\isacharbrackleft}Fun\ {\isadigit{0}}\ {\isacharbrackleft}{\isacharbrackright}{\isacharbrackright}{\isacharparenright}{\isacharparenright}{\isacharcomma}\isanewline
}
\DefineSnippet{Examples:lemma:b6645243b11899ae-47}{%
\ \ \ \ \ \ Neg\ {\isacharparenleft}Pre\ {\isadigit{0}}\ {\isacharbrackleft}Fun\ {\isadigit{0}}\ {\isacharbrackleft}{\isacharbrackright}{\isacharbrackright}{\isacharparenright}{\isacharcomma}\isanewline
}
\DefineSnippet{Examples:lemma:b6645243b11899ae-48}{%
\ \ \ \ \ \ Exi\ {\isacharparenleft}Pre\ {\isadigit{1}}\ {\isacharbrackleft}Var\ {\isadigit{0}}{\isacharbrackright}{\isacharparenright}\isanewline
}
\DefineSnippet{Examples:lemma:b6645243b11899ae-49}{%
\ \ \ \ {\isacharbrackright}\isanewline
}
\DefineSnippet{Examples:lemma:b6645243b11899ae-50}{%
\ \ \ \ {\isacartoucheclose}\isanewline
}
\DefineSnippet{Examples:lemma:b6645243b11899ae-51}{%
\ \ \ \ \isacommand{using}\isamarkupfalse%
\ that\ \isacommand{by}\isamarkupfalse%
\ simp\isanewline
}
\DefineSnippet{Examples:lemma:b6645243b11899ae-52}{%
\ \ \isacommand{with}\isamarkupfalse%
\ BetaImp\ \isacommand{have}\isamarkupfalse%
\ {\isacharquery}thesis\ \isakeyword{if}\ {\isacartoucheopen}{\isasymtturnstile}\isanewline
}
\DefineSnippet{Examples:lemma:b6645243b11899ae-53}{%
\ \ \ \ {\isacharbrackleft}\isanewline
}
\DefineSnippet{Examples:lemma:b6645243b11899ae-54}{%
\ \ \ \ \ \ Pre\ {\isadigit{0}}\ {\isacharbrackleft}Fun\ {\isadigit{0}}\ {\isacharbrackleft}{\isacharbrackright}{\isacharbrackright}{\isacharcomma}\isanewline
}
\DefineSnippet{Examples:lemma:b6645243b11899ae-55}{%
\ \ \ \ \ \ Neg\ {\isacharparenleft}Pre\ {\isadigit{0}}\ {\isacharbrackleft}Fun\ {\isadigit{0}}\ {\isacharbrackleft}{\isacharbrackright}{\isacharbrackright}{\isacharparenright}{\isacharcomma}\isanewline
}
\DefineSnippet{Examples:lemma:b6645243b11899ae-56}{%
\ \ \ \ \ \ Exi\ {\isacharparenleft}Pre\ {\isadigit{1}}\ {\isacharbrackleft}Var\ {\isadigit{0}}{\isacharbrackright}{\isacharparenright}\isanewline
}
\DefineSnippet{Examples:lemma:b6645243b11899ae-57}{%
\ \ \ \ {\isacharbrackright}\isanewline
}
\DefineSnippet{Examples:lemma:b6645243b11899ae-58}{%
\ \ \ \ {\isacartoucheclose}\ \isakeyword{and}\ {\isacartoucheopen}{\isasymtturnstile}\isanewline
}
\DefineSnippet{Examples:lemma:b6645243b11899ae-59}{%
\ \ \ \ {\isacharbrackleft}\isanewline
}
\DefineSnippet{Examples:lemma:b6645243b11899ae-60}{%
\ \ \ \ \ \ Neg\ {\isacharparenleft}Pre\ {\isadigit{1}}\ {\isacharbrackleft}Fun\ {\isadigit{0}}\ {\isacharbrackleft}{\isacharbrackright}{\isacharbrackright}{\isacharparenright}{\isacharcomma}\isanewline
}
\DefineSnippet{Examples:lemma:b6645243b11899ae-61}{%
\ \ \ \ \ \ Neg\ {\isacharparenleft}Pre\ {\isadigit{0}}\ {\isacharbrackleft}Fun\ {\isadigit{0}}\ {\isacharbrackleft}{\isacharbrackright}{\isacharbrackright}{\isacharparenright}{\isacharcomma}\isanewline
}
\DefineSnippet{Examples:lemma:b6645243b11899ae-62}{%
\ \ \ \ \ \ Exi\ {\isacharparenleft}Pre\ {\isadigit{1}}\ {\isacharbrackleft}Var\ {\isadigit{0}}{\isacharbrackright}{\isacharparenright}\isanewline
}
\DefineSnippet{Examples:lemma:b6645243b11899ae-63}{%
\ \ \ \ {\isacharbrackright}\isanewline
}
\DefineSnippet{Examples:lemma:b6645243b11899ae-64}{%
\ \ \ \ {\isacartoucheclose}\isanewline
}
\DefineSnippet{Examples:lemma:b6645243b11899ae-65}{%
\ \ \ \ \isacommand{using}\isamarkupfalse%
\ that\ \isacommand{by}\isamarkupfalse%
\ simp\isanewline
}
\DefineSnippet{Examples:lemma:b6645243b11899ae-66}{%
\ \ \isacommand{with}\isamarkupfalse%
\ Basic\ \isacommand{have}\isamarkupfalse%
\ {\isacharquery}thesis\ \isakeyword{if}\ {\isacartoucheopen}{\isasymtturnstile}\isanewline
}
\DefineSnippet{Examples:lemma:b6645243b11899ae-67}{%
\ \ \ \ {\isacharbrackleft}\isanewline
}
\DefineSnippet{Examples:lemma:b6645243b11899ae-68}{%
\ \ \ \ \ \ Neg\ {\isacharparenleft}Pre\ {\isadigit{1}}\ {\isacharbrackleft}Fun\ {\isadigit{0}}\ {\isacharbrackleft}{\isacharbrackright}{\isacharbrackright}{\isacharparenright}{\isacharcomma}\isanewline
}
\DefineSnippet{Examples:lemma:b6645243b11899ae-69}{%
\ \ \ \ \ \ Neg\ {\isacharparenleft}Pre\ {\isadigit{0}}\ {\isacharbrackleft}Fun\ {\isadigit{0}}\ {\isacharbrackleft}{\isacharbrackright}{\isacharbrackright}{\isacharparenright}{\isacharcomma}\isanewline
}
\DefineSnippet{Examples:lemma:b6645243b11899ae-70}{%
\ \ \ \ \ \ Exi\ {\isacharparenleft}Pre\ {\isadigit{1}}\ {\isacharbrackleft}Var\ {\isadigit{0}}{\isacharbrackright}{\isacharparenright}\isanewline
}
\DefineSnippet{Examples:lemma:b6645243b11899ae-71}{%
\ \ \ \ {\isacharbrackright}\isanewline
}
\DefineSnippet{Examples:lemma:b6645243b11899ae-72}{%
\ \ \ \ {\isacartoucheclose}\isanewline
}
\DefineSnippet{Examples:lemma:b6645243b11899ae-73}{%
\ \ \ \ \isacommand{using}\isamarkupfalse%
\ that\ \isacommand{by}\isamarkupfalse%
\ simp\isanewline
}
\DefineSnippet{Examples:lemma:b6645243b11899ae-74}{%
\ \ \isacommand{with}\isamarkupfalse%
\ Ext\ \isacommand{have}\isamarkupfalse%
\ {\isacharquery}thesis\ \isakeyword{if}\ {\isacartoucheopen}{\isasymtturnstile}\isanewline
}
\DefineSnippet{Examples:lemma:b6645243b11899ae-75}{%
\ \ \ \ {\isacharbrackleft}\isanewline
}
\DefineSnippet{Examples:lemma:b6645243b11899ae-76}{%
\ \ \ \ \ \ Exi\ {\isacharparenleft}Pre\ {\isadigit{1}}\ {\isacharbrackleft}Var\ {\isadigit{0}}{\isacharbrackright}{\isacharparenright}{\isacharcomma}\isanewline
}
\DefineSnippet{Examples:lemma:b6645243b11899ae-77}{%
\ \ \ \ \ \ Neg\ {\isacharparenleft}Pre\ {\isadigit{1}}\ {\isacharbrackleft}Fun\ {\isadigit{0}}\ {\isacharbrackleft}{\isacharbrackright}{\isacharbrackright}{\isacharparenright}\isanewline
}
\DefineSnippet{Examples:lemma:b6645243b11899ae-78}{%
\ \ \ \ {\isacharbrackright}\isanewline
}
\DefineSnippet{Examples:lemma:b6645243b11899ae-79}{%
\ \ \ \ {\isacartoucheclose}\isanewline
}
\DefineSnippet{Examples:lemma:b6645243b11899ae-80}{%
\ \ \ \ \isacommand{using}\isamarkupfalse%
\ that\ \isacommand{by}\isamarkupfalse%
\ simp\isanewline
}
\DefineSnippet{Examples:lemma:b6645243b11899ae-81}{%
\ \ \isacommand{with}\isamarkupfalse%
\ GammaExi\ \isacommand{have}\isamarkupfalse%
\ {\isacharquery}thesis\ \isakeyword{if}\ {\isacartoucheopen}{\isasymtturnstile}\isanewline
}
\DefineSnippet{Examples:lemma:b6645243b11899ae-82}{%
\ \ \ \ {\isacharbrackleft}\isanewline
}
\DefineSnippet{Examples:lemma:b6645243b11899ae-83}{%
\ \ \ \ \ \ Pre\ {\isadigit{1}}\ {\isacharbrackleft}Fun\ {\isadigit{0}}\ {\isacharbrackleft}{\isacharbrackright}{\isacharbrackright}{\isacharcomma}\isanewline
}
\DefineSnippet{Examples:lemma:b6645243b11899ae-84}{%
\ \ \ \ \ \ Neg\ {\isacharparenleft}Pre\ {\isadigit{1}}\ {\isacharbrackleft}Fun\ {\isadigit{0}}\ {\isacharbrackleft}{\isacharbrackright}{\isacharbrackright}{\isacharparenright}\isanewline
}
\DefineSnippet{Examples:lemma:b6645243b11899ae-85}{%
\ \ \ \ {\isacharbrackright}\isanewline
}
\DefineSnippet{Examples:lemma:b6645243b11899ae-86}{%
\ \ \ \ {\isacartoucheclose}\isanewline
}
\DefineSnippet{Examples:lemma:b6645243b11899ae-87}{%
\ \ \ \ \isacommand{using}\isamarkupfalse%
\ that\ \isacommand{by}\isamarkupfalse%
\ simp\isanewline
}
\DefineSnippet{Examples:lemma:b6645243b11899ae-88}{%
\ \ \isacommand{with}\isamarkupfalse%
\ Basic\ \isacommand{show}\isamarkupfalse%
\ {\isacharquery}thesis\isanewline
}
\DefineSnippet{Examples:lemma:b6645243b11899ae-89}{%
\ \ \ \ \isacommand{by}\isamarkupfalse%
\ simp\isanewline
}
\DefineSnippet{Examples:lemma:b6645243b11899ae-90}{%
\isacommand{qed}\isamarkupfalse%
\endisatagproof
{\isafoldproof}%
\isadelimproof
\endisadelimproof
}
\DefineSnippet{Examples:end:7158f6523c28304a-0}{%
\isadelimtheory
\endisadelimtheory
\isatagtheory
\isacommand{end}\isamarkupfalse%
\endisatagtheory
{\isafoldtheory}%
\isadelimtheory
\endisadelimtheory
}
\DefineSnippet{SeCaV:section:ce6e401f0587b57a-0}{%
\isadelimdocument
\endisadelimdocument
\isatagdocument
\isamarkupsection{Sequent Calculus Verifier (SeCaV)%
}
\isamarkuptrue%
\endisatagdocument
{\isafolddocument}%
\isadelimdocument
\endisadelimdocument
}
\DefineSnippet{SeCaV:theory:SeCaV-0}{%
\isadelimtheory
\endisadelimtheory
\isatagtheory
\isacommand{theory}\isamarkupfalse%
\ SeCaV\ \isakeyword{imports}\ Main\ \isakeyword{begin}%
\endisatagtheory
{\isafoldtheory}%
\isadelimtheory
\endisadelimtheory
}
\DefineSnippet{SeCaV:section:6d0d8914823af9b6-0}{%
\isadelimdocument
\endisadelimdocument
\isatagdocument
\isamarkupsection{Syntax: Terms / Formulas%
}
\isamarkuptrue%
\endisatagdocument
{\isafolddocument}%
\isadelimdocument
\endisadelimdocument
}
\DefineSnippet{SeCaV:datatype:tm-0}{%
\isacommand{datatype}\isamarkupfalse%
\ tm\ {\isacharequal}\ Fun\ nat\ {\isacartoucheopen}tm\ list{\isacartoucheclose}\ {\isacharbar}\ Var\ nat%
}
\DefineSnippet{SeCaV:datatype:fm-0}{%
\isacommand{datatype}\isamarkupfalse%
\ fm\ {\isacharequal}\ Pre\ nat\ {\isacartoucheopen}tm\ list{\isacartoucheclose}\ {\isacharbar}\ Imp\ fm\ fm\ {\isacharbar}\ Dis\ fm\ fm\ {\isacharbar}\ Con\ fm\ fm\ {\isacharbar}\ Exi\ fm\ {\isacharbar}\ Uni\ fm\ {\isacharbar}\ Neg\ fm%
}
\DefineSnippet{SeCaV:section:88c599726ee99150-0}{%
\isadelimdocument
\endisadelimdocument
\isatagdocument
\isamarkupsection{Semantics: Terms / Formulas%
}
\isamarkuptrue%
\endisatagdocument
{\isafolddocument}%
\isadelimdocument
\endisadelimdocument
}
\DefineSnippet{SeCaV:definition:shift-0}{%
\isacommand{definition}\isamarkupfalse%
\ {\isacartoucheopen}shift\ e\ v\ x\ {\isasymequiv}\ {\isasymlambda}n{\isachardot}\ if\ n\ {\isacharless}\ v\ then\ e\ n\ else\ if\ n\ {\isacharequal}\ v\ then\ x\ else\ e\ {\isacharparenleft}n\ {\isacharminus}\ {\isadigit{1}}{\isacharparenright}{\isacartoucheclose}%
}
\DefineSnippet{SeCaV:primrec:semantics-term-0}{%
\isacommand{primrec}\isamarkupfalse%
\ semantics{\isacharunderscore}term\ \isakeyword{and}\ semantics{\isacharunderscore}list\ \isakeyword{where}\isanewline
}
\DefineSnippet{SeCaV:primrec:semantics-term-1}{%
\ \ {\isacartoucheopen}semantics{\isacharunderscore}term\ e\ f\ {\isacharparenleft}Var\ n{\isacharparenright}\ {\isacharequal}\ e\ n{\isacartoucheclose}\ {\isacharbar}\isanewline
}
\DefineSnippet{SeCaV:primrec:semantics-term-2}{%
\ \ {\isacartoucheopen}semantics{\isacharunderscore}term\ e\ f\ {\isacharparenleft}Fun\ i\ l{\isacharparenright}\ {\isacharequal}\ f\ i\ {\isacharparenleft}semantics{\isacharunderscore}list\ e\ f\ l{\isacharparenright}{\isacartoucheclose}\ {\isacharbar}\isanewline
}
\DefineSnippet{SeCaV:primrec:semantics-term-3}{%
\ \ {\isacartoucheopen}semantics{\isacharunderscore}list\ e\ f\ {\isacharbrackleft}{\isacharbrackright}\ {\isacharequal}\ {\isacharbrackleft}{\isacharbrackright}{\isacartoucheclose}\ {\isacharbar}\isanewline
}
\DefineSnippet{SeCaV:primrec:semantics-term-4}{%
\ \ {\isacartoucheopen}semantics{\isacharunderscore}list\ e\ f\ {\isacharparenleft}t\ {\isacharhash}\ l{\isacharparenright}\ {\isacharequal}\ semantics{\isacharunderscore}term\ e\ f\ t\ {\isacharhash}\ semantics{\isacharunderscore}list\ e\ f\ l{\isacartoucheclose}%
}
\DefineSnippet{SeCaV:primrec:semantics-0}{%
\isacommand{primrec}\isamarkupfalse%
\ semantics\ \isakeyword{where}\isanewline
}
\DefineSnippet{SeCaV:primrec:semantics-1}{%
\ \ {\isacartoucheopen}semantics\ e\ f\ g\ {\isacharparenleft}Pre\ i\ l{\isacharparenright}\ {\isacharequal}\ g\ i\ {\isacharparenleft}semantics{\isacharunderscore}list\ e\ f\ l{\isacharparenright}{\isacartoucheclose}\ {\isacharbar}\isanewline
}
\DefineSnippet{SeCaV:primrec:semantics-2}{%
\ \ {\isacartoucheopen}semantics\ e\ f\ g\ {\isacharparenleft}Imp\ p\ q{\isacharparenright}\ {\isacharequal}\ {\isacharparenleft}semantics\ e\ f\ g\ p\ {\isasymlongrightarrow}\ semantics\ e\ f\ g\ q{\isacharparenright}{\isacartoucheclose}\ {\isacharbar}\isanewline
}
\DefineSnippet{SeCaV:primrec:semantics-3}{%
\ \ {\isacartoucheopen}semantics\ e\ f\ g\ {\isacharparenleft}Dis\ p\ q{\isacharparenright}\ {\isacharequal}\ {\isacharparenleft}semantics\ e\ f\ g\ p\ {\isasymor}\ semantics\ e\ f\ g\ q{\isacharparenright}{\isacartoucheclose}\ {\isacharbar}\isanewline
}
\DefineSnippet{SeCaV:primrec:semantics-4}{%
\ \ {\isacartoucheopen}semantics\ e\ f\ g\ {\isacharparenleft}Con\ p\ q{\isacharparenright}\ {\isacharequal}\ {\isacharparenleft}semantics\ e\ f\ g\ p\ {\isasymand}\ semantics\ e\ f\ g\ q{\isacharparenright}{\isacartoucheclose}\ {\isacharbar}\isanewline
}
\DefineSnippet{SeCaV:primrec:semantics-5}{%
\ \ {\isacartoucheopen}semantics\ e\ f\ g\ {\isacharparenleft}Exi\ p{\isacharparenright}\ {\isacharequal}\ {\isacharparenleft}{\isasymexists}x{\isachardot}\ semantics\ {\isacharparenleft}shift\ e\ {\isadigit{0}}\ x{\isacharparenright}\ f\ g\ p{\isacharparenright}{\isacartoucheclose}\ {\isacharbar}\isanewline
}
\DefineSnippet{SeCaV:primrec:semantics-6}{%
\ \ {\isacartoucheopen}semantics\ e\ f\ g\ {\isacharparenleft}Uni\ p{\isacharparenright}\ {\isacharequal}\ {\isacharparenleft}{\isasymforall}x{\isachardot}\ semantics\ {\isacharparenleft}shift\ e\ {\isadigit{0}}\ x{\isacharparenright}\ f\ g\ p{\isacharparenright}{\isacartoucheclose}\ {\isacharbar}\isanewline
}
\DefineSnippet{SeCaV:primrec:semantics-7}{%
\ \ {\isacartoucheopen}semantics\ e\ f\ g\ {\isacharparenleft}Neg\ p{\isacharparenright}\ {\isacharequal}\ {\isacharparenleft}{\isasymnot}\ semantics\ e\ f\ g\ p{\isacharparenright}{\isacartoucheclose}
}
\DefineSnippet{SeCaV:corollary:59241be5b3b14054-0}{%
\isacommand{corollary}\isamarkupfalse%
\ {\isacartoucheopen}semantics\ e\ f\ g\ {\isacharparenleft}Imp\ {\isacharparenleft}Pre\ {\isadigit{0}}\ {\isacharbrackleft}{\isacharbrackright}{\isacharparenright}\ {\isacharparenleft}Pre\ {\isadigit{0}}\ {\isacharbrackleft}{\isacharbrackright}{\isacharparenright}{\isacharparenright}{\isacartoucheclose}\isanewline
}
\DefineSnippet{SeCaV:corollary:59241be5b3b14054-1}{%
\isadelimproof
\ \ %
\endisadelimproof
\isatagproof
\isacommand{by}\isamarkupfalse%
\ simp%
\endisatagproof
{\isafoldproof}%
\isadelimproof
\endisadelimproof
}
\DefineSnippet{SeCaV:lemma:9b0d6477a67c94eb-0}{%
\isacommand{lemma}\isamarkupfalse%
\ {\isacartoucheopen}{\isasymnot}\ semantics\ e\ f\ g\ {\isacharparenleft}Neg\ {\isacharparenleft}Imp\ {\isacharparenleft}Pre\ {\isadigit{0}}\ {\isacharbrackleft}{\isacharbrackright}{\isacharparenright}\ {\isacharparenleft}Pre\ {\isadigit{0}}\ {\isacharbrackleft}{\isacharbrackright}{\isacharparenright}{\isacharparenright}{\isacharparenright}{\isacartoucheclose}\isanewline
}
\DefineSnippet{SeCaV:lemma:9b0d6477a67c94eb-1}{%
\isadelimproof
\ \ %
\endisadelimproof
\isatagproof
\isacommand{by}\isamarkupfalse%
\ simp%
\endisatagproof
{\isafoldproof}%
\isadelimproof
\endisadelimproof
}
\DefineSnippet{SeCaV:section:ea29346afd68a871-0}{%
\isadelimdocument
\endisadelimdocument
\isatagdocument
\isamarkupsection{Auxiliary Functions%
}
\isamarkuptrue%
\endisatagdocument
{\isafolddocument}%
\isadelimdocument
\endisadelimdocument
}
\DefineSnippet{SeCaV:primrec:new-term-0}{%
\isacommand{primrec}\isamarkupfalse%
\ new{\isacharunderscore}term\ \isakeyword{and}\ new{\isacharunderscore}list\ \isakeyword{where}\isanewline
}
\DefineSnippet{SeCaV:primrec:new-term-1}{%
\ \ {\isacartoucheopen}new{\isacharunderscore}term\ c\ {\isacharparenleft}Var\ n{\isacharparenright}\ {\isacharequal}\ True{\isacartoucheclose}\ {\isacharbar}\isanewline
}
\DefineSnippet{SeCaV:primrec:new-term-2}{%
\ \ {\isacartoucheopen}new{\isacharunderscore}term\ c\ {\isacharparenleft}Fun\ i\ l{\isacharparenright}\ {\isacharequal}\ {\isacharparenleft}if\ i\ {\isacharequal}\ c\ then\ False\ else\ new{\isacharunderscore}list\ c\ l{\isacharparenright}{\isacartoucheclose}\ {\isacharbar}\isanewline
}
\DefineSnippet{SeCaV:primrec:new-term-3}{%
\ \ {\isacartoucheopen}new{\isacharunderscore}list\ c\ {\isacharbrackleft}{\isacharbrackright}\ {\isacharequal}\ True{\isacartoucheclose}\ {\isacharbar}\isanewline
}
\DefineSnippet{SeCaV:primrec:new-term-4}{%
\ \ {\isacartoucheopen}new{\isacharunderscore}list\ c\ {\isacharparenleft}t\ {\isacharhash}\ l{\isacharparenright}\ {\isacharequal}\ {\isacharparenleft}if\ new{\isacharunderscore}term\ c\ t\ then\ new{\isacharunderscore}list\ c\ l\ else\ False{\isacharparenright}{\isacartoucheclose}%
}
\DefineSnippet{SeCaV:primrec:new-0}{%
\isacommand{primrec}\isamarkupfalse%
\ new\ \isakeyword{where}\isanewline
}
\DefineSnippet{SeCaV:primrec:new-1}{%
\ \ {\isacartoucheopen}new\ c\ {\isacharparenleft}Pre\ i\ l{\isacharparenright}\ {\isacharequal}\ new{\isacharunderscore}list\ c\ l{\isacartoucheclose}\ {\isacharbar}\isanewline
}
\DefineSnippet{SeCaV:primrec:new-2}{%
\ \ {\isacartoucheopen}new\ c\ {\isacharparenleft}Imp\ p\ q{\isacharparenright}\ {\isacharequal}\ {\isacharparenleft}if\ new\ c\ p\ then\ new\ c\ q\ else\ False{\isacharparenright}{\isacartoucheclose}\ {\isacharbar}\isanewline
}
\DefineSnippet{SeCaV:primrec:new-3}{%
\ \ {\isacartoucheopen}new\ c\ {\isacharparenleft}Dis\ p\ q{\isacharparenright}\ {\isacharequal}\ {\isacharparenleft}if\ new\ c\ p\ then\ new\ c\ q\ else\ False{\isacharparenright}{\isacartoucheclose}\ {\isacharbar}\isanewline
}
\DefineSnippet{SeCaV:primrec:new-4}{%
\ \ {\isacartoucheopen}new\ c\ {\isacharparenleft}Con\ p\ q{\isacharparenright}\ {\isacharequal}\ {\isacharparenleft}if\ new\ c\ p\ then\ new\ c\ q\ else\ False{\isacharparenright}{\isacartoucheclose}\ {\isacharbar}\isanewline
}
\DefineSnippet{SeCaV:primrec:new-5}{%
\ \ {\isacartoucheopen}new\ c\ {\isacharparenleft}Exi\ p{\isacharparenright}\ {\isacharequal}\ new\ c\ p{\isacartoucheclose}\ {\isacharbar}\isanewline
}
\DefineSnippet{SeCaV:primrec:new-6}{%
\ \ {\isacartoucheopen}new\ c\ {\isacharparenleft}Uni\ p{\isacharparenright}\ {\isacharequal}\ new\ c\ p{\isacartoucheclose}\ {\isacharbar}\isanewline
}
\DefineSnippet{SeCaV:primrec:new-7}{%
\ \ {\isacartoucheopen}new\ c\ {\isacharparenleft}Neg\ p{\isacharparenright}\ {\isacharequal}\ new\ c\ p{\isacartoucheclose}%
}
\DefineSnippet{SeCaV:primrec:news-0}{%
\isacommand{primrec}\isamarkupfalse%
\ news\ \isakeyword{where}\isanewline
}
\DefineSnippet{SeCaV:primrec:news-1}{%
\ \ {\isacartoucheopen}news\ c\ {\isacharbrackleft}{\isacharbrackright}\ {\isacharequal}\ True{\isacartoucheclose}\ {\isacharbar}\isanewline
}
\DefineSnippet{SeCaV:primrec:news-2}{%
\ \ {\isacartoucheopen}news\ c\ {\isacharparenleft}p\ {\isacharhash}\ z{\isacharparenright}\ {\isacharequal}\ {\isacharparenleft}if\ new\ c\ p\ then\ news\ c\ z\ else\ False{\isacharparenright}{\isacartoucheclose}%
}
\DefineSnippet{SeCaV:primrec:inc-term-0}{%
\isacommand{primrec}\isamarkupfalse%
\ inc{\isacharunderscore}term\ \isakeyword{and}\ inc{\isacharunderscore}list\ \isakeyword{where}\isanewline
}
\DefineSnippet{SeCaV:primrec:inc-term-1}{%
\ \ {\isacartoucheopen}inc{\isacharunderscore}term\ {\isacharparenleft}Var\ n{\isacharparenright}\ {\isacharequal}\ Var\ {\isacharparenleft}n\ {\isacharplus}\ {\isadigit{1}}{\isacharparenright}{\isacartoucheclose}\ {\isacharbar}\isanewline
}
\DefineSnippet{SeCaV:primrec:inc-term-2}{%
\ \ {\isacartoucheopen}inc{\isacharunderscore}term\ {\isacharparenleft}Fun\ i\ l{\isacharparenright}\ {\isacharequal}\ Fun\ i\ {\isacharparenleft}inc{\isacharunderscore}list\ l{\isacharparenright}{\isacartoucheclose}\ {\isacharbar}\isanewline
}
\DefineSnippet{SeCaV:primrec:inc-term-3}{%
\ \ {\isacartoucheopen}inc{\isacharunderscore}list\ {\isacharbrackleft}{\isacharbrackright}\ {\isacharequal}\ {\isacharbrackleft}{\isacharbrackright}{\isacartoucheclose}\ {\isacharbar}\isanewline
}
\DefineSnippet{SeCaV:primrec:inc-term-4}{%
\ \ {\isacartoucheopen}inc{\isacharunderscore}list\ {\isacharparenleft}t\ {\isacharhash}\ l{\isacharparenright}\ {\isacharequal}\ inc{\isacharunderscore}term\ t\ {\isacharhash}\ inc{\isacharunderscore}list\ l{\isacartoucheclose}%
}
\DefineSnippet{SeCaV:primrec:sub-term-0}{%
\isacommand{primrec}\isamarkupfalse%
\ sub{\isacharunderscore}term\ \isakeyword{and}\ sub{\isacharunderscore}list\ \isakeyword{where}\isanewline
}
\DefineSnippet{SeCaV:primrec:sub-term-1}{%
\ \ {\isacartoucheopen}sub{\isacharunderscore}term\ v\ s\ {\isacharparenleft}Var\ n{\isacharparenright}\ {\isacharequal}\ {\isacharparenleft}if\ n\ {\isacharless}\ v\ then\ Var\ n\ else\ if\ n\ {\isacharequal}\ v\ then\ s\ else\ Var\ {\isacharparenleft}n\ {\isacharminus}\ {\isadigit{1}}{\isacharparenright}{\isacharparenright}{\isacartoucheclose}\ {\isacharbar}\isanewline
}
\DefineSnippet{SeCaV:primrec:sub-term-2}{%
\ \ {\isacartoucheopen}sub{\isacharunderscore}term\ v\ s\ {\isacharparenleft}Fun\ i\ l{\isacharparenright}\ {\isacharequal}\ Fun\ i\ {\isacharparenleft}sub{\isacharunderscore}list\ v\ s\ l{\isacharparenright}{\isacartoucheclose}\ {\isacharbar}\isanewline
}
\DefineSnippet{SeCaV:primrec:sub-term-3}{%
\ \ {\isacartoucheopen}sub{\isacharunderscore}list\ v\ s\ {\isacharbrackleft}{\isacharbrackright}\ {\isacharequal}\ {\isacharbrackleft}{\isacharbrackright}{\isacartoucheclose}\ {\isacharbar}\isanewline
}
\DefineSnippet{SeCaV:primrec:sub-term-4}{%
\ \ {\isacartoucheopen}sub{\isacharunderscore}list\ v\ s\ {\isacharparenleft}t\ {\isacharhash}\ l{\isacharparenright}\ {\isacharequal}\ sub{\isacharunderscore}term\ v\ s\ t\ {\isacharhash}\ sub{\isacharunderscore}list\ v\ s\ l{\isacartoucheclose}%
}
\DefineSnippet{SeCaV:primrec:sub-0}{%
\isacommand{primrec}\isamarkupfalse%
\ sub\ \isakeyword{where}\isanewline
}
\DefineSnippet{SeCaV:primrec:sub-1}{%
\ \ {\isacartoucheopen}sub\ v\ s\ {\isacharparenleft}Pre\ i\ l{\isacharparenright}\ {\isacharequal}\ Pre\ i\ {\isacharparenleft}sub{\isacharunderscore}list\ v\ s\ l{\isacharparenright}{\isacartoucheclose}\ {\isacharbar}\isanewline
}
\DefineSnippet{SeCaV:primrec:sub-2}{%
\ \ {\isacartoucheopen}sub\ v\ s\ {\isacharparenleft}Imp\ p\ q{\isacharparenright}\ {\isacharequal}\ Imp\ {\isacharparenleft}sub\ v\ s\ p{\isacharparenright}\ {\isacharparenleft}sub\ v\ s\ q{\isacharparenright}{\isacartoucheclose}\ {\isacharbar}\isanewline
}
\DefineSnippet{SeCaV:primrec:sub-3}{%
\ \ {\isacartoucheopen}sub\ v\ s\ {\isacharparenleft}Dis\ p\ q{\isacharparenright}\ {\isacharequal}\ Dis\ {\isacharparenleft}sub\ v\ s\ p{\isacharparenright}\ {\isacharparenleft}sub\ v\ s\ q{\isacharparenright}{\isacartoucheclose}\ {\isacharbar}\isanewline
}
\DefineSnippet{SeCaV:primrec:sub-4}{%
\ \ {\isacartoucheopen}sub\ v\ s\ {\isacharparenleft}Con\ p\ q{\isacharparenright}\ {\isacharequal}\ Con\ {\isacharparenleft}sub\ v\ s\ p{\isacharparenright}\ {\isacharparenleft}sub\ v\ s\ q{\isacharparenright}{\isacartoucheclose}\ {\isacharbar}\isanewline
}
\DefineSnippet{SeCaV:primrec:sub-5}{%
\ \ {\isacartoucheopen}sub\ v\ s\ {\isacharparenleft}Exi\ p{\isacharparenright}\ {\isacharequal}\ Exi\ {\isacharparenleft}sub\ {\isacharparenleft}v\ {\isacharplus}\ {\isadigit{1}}{\isacharparenright}\ {\isacharparenleft}inc{\isacharunderscore}term\ s{\isacharparenright}\ p{\isacharparenright}{\isacartoucheclose}\ {\isacharbar}\isanewline
}
\DefineSnippet{SeCaV:primrec:sub-6}{%
\ \ {\isacartoucheopen}sub\ v\ s\ {\isacharparenleft}Uni\ p{\isacharparenright}\ {\isacharequal}\ Uni\ {\isacharparenleft}sub\ {\isacharparenleft}v\ {\isacharplus}\ {\isadigit{1}}{\isacharparenright}\ {\isacharparenleft}inc{\isacharunderscore}term\ s{\isacharparenright}\ p{\isacharparenright}{\isacartoucheclose}\ {\isacharbar}\isanewline
}
\DefineSnippet{SeCaV:primrec:sub-7}{%
\ \ {\isacartoucheopen}sub\ v\ s\ {\isacharparenleft}Neg\ p{\isacharparenright}\ {\isacharequal}\ Neg\ {\isacharparenleft}sub\ v\ s\ p{\isacharparenright}{\isacartoucheclose}%
}
\DefineSnippet{SeCaV:primrec:member-0}{%
\isacommand{primrec}\isamarkupfalse%
\ member\ \isakeyword{where}\isanewline
}
\DefineSnippet{SeCaV:primrec:member-1}{%
\ \ {\isacartoucheopen}member\ p\ {\isacharbrackleft}{\isacharbrackright}\ {\isacharequal}\ False{\isacartoucheclose}\ {\isacharbar}\isanewline
}
\DefineSnippet{SeCaV:primrec:member-2}{%
\ \ {\isacartoucheopen}member\ p\ {\isacharparenleft}q\ {\isacharhash}\ z{\isacharparenright}\ {\isacharequal}\ {\isacharparenleft}if\ p\ {\isacharequal}\ q\ then\ True\ else\ member\ p\ z{\isacharparenright}{\isacartoucheclose}%
}
\DefineSnippet{SeCaV:primrec:ext-0}{%
\isacommand{primrec}\isamarkupfalse%
\ ext\ \isakeyword{where}\isanewline
}
\DefineSnippet{SeCaV:primrec:ext-1}{%
\ \ {\isacartoucheopen}ext\ y\ {\isacharbrackleft}{\isacharbrackright}\ {\isacharequal}\ True{\isacartoucheclose}\ {\isacharbar}\isanewline
}
\DefineSnippet{SeCaV:primrec:ext-2}{%
\ \ {\isacartoucheopen}ext\ y\ {\isacharparenleft}p\ {\isacharhash}\ z{\isacharparenright}\ {\isacharequal}\ {\isacharparenleft}if\ member\ p\ y\ then\ ext\ y\ z\ else\ False{\isacharparenright}{\isacartoucheclose}
}
\DefineSnippet{SeCaV:lemma:member-0}{%
\isacommand{lemma}\isamarkupfalse%
\ member\ {\isacharbrackleft}iff{\isacharbrackright}{\isacharcolon}\ {\isacartoucheopen}member\ p\ z\ {\isasymlongleftrightarrow}\ p\ {\isasymin}\ set\ z{\isacartoucheclose}\isanewline
}
\DefineSnippet{SeCaV:lemma:member-1}{%
\isadelimproof
\ \ %
\endisadelimproof
\isatagproof
\isacommand{by}\isamarkupfalse%
\ {\isacharparenleft}induct\ z{\isacharparenright}\ simp{\isacharunderscore}all%
\endisatagproof
{\isafoldproof}%
\isadelimproof
\endisadelimproof
}
\DefineSnippet{SeCaV:lemma:ext-0}{%
\isacommand{lemma}\isamarkupfalse%
\ ext\ {\isacharbrackleft}iff{\isacharbrackright}{\isacharcolon}\ {\isacartoucheopen}ext\ y\ z\ {\isasymlongleftrightarrow}\ set\ z\ {\isasymsubseteq}\ set\ y{\isacartoucheclose}\isanewline
}
\DefineSnippet{SeCaV:lemma:ext-1}{%
\isadelimproof
\ \ %
\endisadelimproof
\isatagproof
\isacommand{by}\isamarkupfalse%
\ {\isacharparenleft}induct\ z{\isacharparenright}\ simp{\isacharunderscore}all%
\endisatagproof
{\isafoldproof}%
\isadelimproof
\endisadelimproof
}
\DefineSnippet{SeCaV:section:1d297c1e41f06c2e-0}{%
\isadelimdocument
\endisadelimdocument
\isatagdocument
\isamarkupsection{Sequent Calculus%
}
\isamarkuptrue%
\endisatagdocument
{\isafolddocument}%
\isadelimdocument
\endisadelimdocument
}
\DefineSnippet{SeCaV:inductive:sequent-calculus-0}{%
\isacommand{inductive}\isamarkupfalse%
\ sequent{\isacharunderscore}calculus\ {\isacharparenleft}{\isacartoucheopen}{\isasymtturnstile}\ {\isacharunderscore}{\isacartoucheclose}\ {\isadigit{0}}{\isacharparenright}\ \isakeyword{where}\isanewline
}
\DefineSnippet{SeCaV:inductive:sequent-calculus-1}{%
\ \ {\isacartoucheopen}{\isasymtturnstile}\ p\ {\isacharhash}\ z{\isacartoucheclose}\ \isakeyword{if}\ {\isacartoucheopen}member\ {\isacharparenleft}Neg\ p{\isacharparenright}\ z{\isacartoucheclose}\ {\isacharbar}\isanewline
}
\DefineSnippet{SeCaV:inductive:sequent-calculus-2}{%
\ \ {\isacartoucheopen}{\isasymtturnstile}\ Dis\ p\ q\ {\isacharhash}\ z{\isacartoucheclose}\ \isakeyword{if}\ {\isacartoucheopen}{\isasymtturnstile}\ p\ {\isacharhash}\ q\ {\isacharhash}\ z{\isacartoucheclose}\ {\isacharbar}\isanewline
}
\DefineSnippet{SeCaV:inductive:sequent-calculus-3}{%
\ \ {\isacartoucheopen}{\isasymtturnstile}\ Imp\ p\ q\ {\isacharhash}\ z{\isacartoucheclose}\ \isakeyword{if}\ {\isacartoucheopen}{\isasymtturnstile}\ Neg\ p\ {\isacharhash}\ q\ {\isacharhash}\ z{\isacartoucheclose}\ {\isacharbar}\isanewline
}
\DefineSnippet{SeCaV:inductive:sequent-calculus-4}{%
\ \ {\isacartoucheopen}{\isasymtturnstile}\ Neg\ {\isacharparenleft}Con\ p\ q{\isacharparenright}\ {\isacharhash}\ z{\isacartoucheclose}\ \isakeyword{if}\ {\isacartoucheopen}{\isasymtturnstile}\ Neg\ p\ {\isacharhash}\ Neg\ q\ {\isacharhash}\ z{\isacartoucheclose}\ {\isacharbar}\isanewline
}
\DefineSnippet{SeCaV:inductive:sequent-calculus-5}{%
\ \ {\isacartoucheopen}{\isasymtturnstile}\ Con\ p\ q\ {\isacharhash}\ z{\isacartoucheclose}\ \isakeyword{if}\ {\isacartoucheopen}{\isasymtturnstile}\ p\ {\isacharhash}\ z{\isacartoucheclose}\ \isakeyword{and}\ {\isacartoucheopen}{\isasymtturnstile}\ q\ {\isacharhash}\ z{\isacartoucheclose}\ {\isacharbar}\isanewline
}
\DefineSnippet{SeCaV:inductive:sequent-calculus-6}{%
\ \ {\isacartoucheopen}{\isasymtturnstile}\ Neg\ {\isacharparenleft}Imp\ p\ q{\isacharparenright}\ {\isacharhash}\ z{\isacartoucheclose}\ \isakeyword{if}\ {\isacartoucheopen}{\isasymtturnstile}\ p\ {\isacharhash}\ z{\isacartoucheclose}\ \isakeyword{and}\ {\isacartoucheopen}{\isasymtturnstile}\ Neg\ q\ {\isacharhash}\ z{\isacartoucheclose}\ {\isacharbar}\isanewline
}
\DefineSnippet{SeCaV:inductive:sequent-calculus-7}{%
\ \ {\isacartoucheopen}{\isasymtturnstile}\ Neg\ {\isacharparenleft}Dis\ p\ q{\isacharparenright}\ {\isacharhash}\ z{\isacartoucheclose}\ \isakeyword{if}\ {\isacartoucheopen}{\isasymtturnstile}\ Neg\ p\ {\isacharhash}\ z{\isacartoucheclose}\ \isakeyword{and}\ {\isacartoucheopen}{\isasymtturnstile}\ Neg\ q\ {\isacharhash}\ z{\isacartoucheclose}\ {\isacharbar}\isanewline
}
\DefineSnippet{SeCaV:inductive:sequent-calculus-8}{%
\ \ {\isacartoucheopen}{\isasymtturnstile}\ Exi\ p\ {\isacharhash}\ z{\isacartoucheclose}\ \isakeyword{if}\ {\isacartoucheopen}{\isasymtturnstile}\ sub\ {\isadigit{0}}\ t\ p\ {\isacharhash}\ z{\isacartoucheclose}\ {\isacharbar}\isanewline
}
\DefineSnippet{SeCaV:inductive:sequent-calculus-9}{%
\ \ {\isacartoucheopen}{\isasymtturnstile}\ Neg\ {\isacharparenleft}Uni\ p{\isacharparenright}\ {\isacharhash}\ z{\isacartoucheclose}\ \isakeyword{if}\ {\isacartoucheopen}{\isasymtturnstile}\ Neg\ {\isacharparenleft}sub\ {\isadigit{0}}\ t\ p{\isacharparenright}\ {\isacharhash}\ z{\isacartoucheclose}\ {\isacharbar}\isanewline
}
\DefineSnippet{SeCaV:inductive:sequent-calculus-10}{%
\ \ {\isacartoucheopen}{\isasymtturnstile}\ Uni\ p\ {\isacharhash}\ z{\isacartoucheclose}\ \isakeyword{if}\ {\isacartoucheopen}{\isasymtturnstile}\ sub\ {\isadigit{0}}\ {\isacharparenleft}Fun\ i\ {\isacharbrackleft}{\isacharbrackright}{\isacharparenright}\ p\ {\isacharhash}\ z{\isacartoucheclose}\ \isakeyword{and}\ {\isacartoucheopen}news\ i\ {\isacharparenleft}p\ {\isacharhash}\ z{\isacharparenright}{\isacartoucheclose}\ {\isacharbar}\isanewline
}
\DefineSnippet{SeCaV:inductive:sequent-calculus-11}{%
\ \ {\isacartoucheopen}{\isasymtturnstile}\ Neg\ {\isacharparenleft}Exi\ p{\isacharparenright}\ {\isacharhash}\ z{\isacartoucheclose}\ \isakeyword{if}\ {\isacartoucheopen}{\isasymtturnstile}\ Neg\ {\isacharparenleft}sub\ {\isadigit{0}}\ {\isacharparenleft}Fun\ i\ {\isacharbrackleft}{\isacharbrackright}{\isacharparenright}\ p{\isacharparenright}\ {\isacharhash}\ z{\isacartoucheclose}\ \isakeyword{and}\ {\isacartoucheopen}news\ i\ {\isacharparenleft}p\ {\isacharhash}\ z{\isacharparenright}{\isacartoucheclose}\ {\isacharbar}\isanewline
}
\DefineSnippet{SeCaV:inductive:sequent-calculus-12}{%
\ \ {\isacartoucheopen}{\isasymtturnstile}\ Neg\ {\isacharparenleft}Neg\ p{\isacharparenright}\ {\isacharhash}\ z{\isacartoucheclose}\ \isakeyword{if}\ {\isacartoucheopen}{\isasymtturnstile}\ p\ {\isacharhash}\ z{\isacartoucheclose}\ {\isacharbar}\isanewline
}
\DefineSnippet{SeCaV:inductive:sequent-calculus-13}{%
\ \ {\isacartoucheopen}{\isasymtturnstile}\ y{\isacartoucheclose}\ \isakeyword{if}\ {\isacartoucheopen}{\isasymtturnstile}\ z{\isacartoucheclose}\ \isakeyword{and}\ {\isacartoucheopen}ext\ y\ z{\isacartoucheclose}
}
\DefineSnippet{SeCaV:corollary:ef996ea6607316d-0}{%
\isacommand{corollary}\isamarkupfalse%
\ {\isacartoucheopen}{\isasymtturnstile}\ {\isacharbrackleft}Imp\ {\isacharparenleft}Pre\ {\isadigit{0}}\ {\isacharbrackleft}{\isacharbrackright}{\isacharparenright}\ {\isacharparenleft}Pre\ {\isadigit{0}}\ {\isacharbrackleft}{\isacharbrackright}{\isacharparenright}{\isacharbrackright}{\isacartoucheclose}\isanewline
}
\DefineSnippet{SeCaV:corollary:ef996ea6607316d-1}{%
\isadelimproof
\ \ %
\endisadelimproof
\isatagproof
\isacommand{using}\isamarkupfalse%
\ sequent{\isacharunderscore}calculus{\isachardot}intros{\isacharparenleft}{\isadigit{1}}{\isacharcomma}{\isadigit{3}}{\isacharcomma}{\isadigit{1}}{\isadigit{3}}{\isacharparenright}\ ext{\isachardot}simps\ member{\isachardot}simps{\isacharparenleft}{\isadigit{2}}{\isacharparenright}\ \isacommand{by}\isamarkupfalse%
\ metis%
\endisatagproof
{\isafoldproof}%
\isadelimproof
\endisadelimproof
}
\DefineSnippet{SeCaV:section:af1d5a2120eaa125-0}{%
\isadelimdocument
\endisadelimdocument
\isatagdocument
\isamarkupsection{Shorthands%
}
\isamarkuptrue%
\endisatagdocument
{\isafolddocument}%
\isadelimdocument
\endisadelimdocument
}
\DefineSnippet{SeCaV:lemmas:e5474352ca974094-0}{%
\isacommand{lemmas}\isamarkupfalse%
\ Basic\ {\isacharequal}\ sequent{\isacharunderscore}calculus{\isachardot}intros{\isacharparenleft}{\isadigit{1}}{\isacharparenright}%
}
\DefineSnippet{SeCaV:lemmas:7e6a02dbce13a23-0}{%
\isacommand{lemmas}\isamarkupfalse%
\ AlphaDis\ {\isacharequal}\ sequent{\isacharunderscore}calculus{\isachardot}intros{\isacharparenleft}{\isadigit{2}}{\isacharparenright}%
}
\DefineSnippet{SeCaV:lemmas:85bb465df0ebb680-0}{%
\isacommand{lemmas}\isamarkupfalse%
\ AlphaImp\ {\isacharequal}\ sequent{\isacharunderscore}calculus{\isachardot}intros{\isacharparenleft}{\isadigit{3}}{\isacharparenright}%
}
\DefineSnippet{SeCaV:lemmas:5c5f47dbaf9595b1-0}{%
\isacommand{lemmas}\isamarkupfalse%
\ AlphaCon\ {\isacharequal}\ sequent{\isacharunderscore}calculus{\isachardot}intros{\isacharparenleft}{\isadigit{4}}{\isacharparenright}%
}
\DefineSnippet{SeCaV:lemmas:cdd7e652d04bb81-0}{%
\isacommand{lemmas}\isamarkupfalse%
\ BetaCon\ {\isacharequal}\ sequent{\isacharunderscore}calculus{\isachardot}intros{\isacharparenleft}{\isadigit{5}}{\isacharparenright}%
}
\DefineSnippet{SeCaV:lemmas:24295edb476bf25a-0}{%
\isacommand{lemmas}\isamarkupfalse%
\ BetaImp\ {\isacharequal}\ sequent{\isacharunderscore}calculus{\isachardot}intros{\isacharparenleft}{\isadigit{6}}{\isacharparenright}%
}
\DefineSnippet{SeCaV:lemmas:af2f78a020951f89-0}{%
\isacommand{lemmas}\isamarkupfalse%
\ BetaDis\ {\isacharequal}\ sequent{\isacharunderscore}calculus{\isachardot}intros{\isacharparenleft}{\isadigit{7}}{\isacharparenright}%
}
\DefineSnippet{SeCaV:lemmas:fe1aaf282412e19e-0}{%
\isacommand{lemmas}\isamarkupfalse%
\ GammaExi\ {\isacharequal}\ sequent{\isacharunderscore}calculus{\isachardot}intros{\isacharparenleft}{\isadigit{8}}{\isacharparenright}%
}
\DefineSnippet{SeCaV:lemmas:276b55af33a9003-0}{%
\isacommand{lemmas}\isamarkupfalse%
\ GammaUni\ {\isacharequal}\ sequent{\isacharunderscore}calculus{\isachardot}intros{\isacharparenleft}{\isadigit{9}}{\isacharparenright}%
}
\DefineSnippet{SeCaV:lemmas:63b3254f43f8acb5-0}{%
\isacommand{lemmas}\isamarkupfalse%
\ DeltaUni\ {\isacharequal}\ sequent{\isacharunderscore}calculus{\isachardot}intros{\isacharparenleft}{\isadigit{1}}{\isadigit{0}}{\isacharparenright}%
}
\DefineSnippet{SeCaV:lemmas:53ee18320bdfe6fd-0}{%
\isacommand{lemmas}\isamarkupfalse%
\ DeltaExi\ {\isacharequal}\ sequent{\isacharunderscore}calculus{\isachardot}intros{\isacharparenleft}{\isadigit{1}}{\isadigit{1}}{\isacharparenright}%
}
\DefineSnippet{SeCaV:lemmas:f64e2b1ffce72652-0}{%
\isacommand{lemmas}\isamarkupfalse%
\ Neg\ {\isacharequal}\ sequent{\isacharunderscore}calculus{\isachardot}intros{\isacharparenleft}{\isadigit{1}}{\isadigit{2}}{\isacharparenright}%
}
\DefineSnippet{SeCaV:lemmas:ce17fe0131442de5-0}{%
\isacommand{lemmas}\isamarkupfalse%
\ Ext\ {\isacharequal}\ sequent{\isacharunderscore}calculus{\isachardot}intros{\isacharparenleft}{\isadigit{1}}{\isadigit{3}}{\isacharparenright}\isanewline
}
\DefineSnippet{SeCaV:lemmas:ce17fe0131442de5-1}{%
\isanewline
}
\DefineSnippet{SeCaV:lemmas:ce17fe0131442de5-2}{%
\isamarkupcmt{Test%
}%
}
\DefineSnippet{SeCaV:lemma:3e54878179f6bbc8-0}{%
\isacommand{lemma}\isamarkupfalse%
\ {\isacartoucheopen}{\isasymtturnstile}\isanewline
}
\DefineSnippet{SeCaV:lemma:3e54878179f6bbc8-1}{%
\ \ {\isacharbrackleft}\isanewline
}
\DefineSnippet{SeCaV:lemma:3e54878179f6bbc8-2}{%
\ \ \ \ Imp\ {\isacharparenleft}Pre\ {\isadigit{0}}\ {\isacharbrackleft}{\isacharbrackright}{\isacharparenright}\ {\isacharparenleft}Pre\ {\isadigit{0}}\ {\isacharbrackleft}{\isacharbrackright}{\isacharparenright}\isanewline
}
\DefineSnippet{SeCaV:lemma:3e54878179f6bbc8-3}{%
\ \ {\isacharbrackright}\isanewline
}
\DefineSnippet{SeCaV:lemma:3e54878179f6bbc8-4}{%
\ \ {\isacartoucheclose}\isanewline
}
\DefineSnippet{SeCaV:lemma:3e54878179f6bbc8-5}{%
\isadelimproof
\endisadelimproof
\isatagproof
\isacommand{proof}\isamarkupfalse%
\ {\isacharminus}\isanewline
}
\DefineSnippet{SeCaV:lemma:3e54878179f6bbc8-6}{%
\ \ \isacommand{from}\isamarkupfalse%
\ AlphaImp\ \isacommand{have}\isamarkupfalse%
\ {\isacharquery}thesis\ \isakeyword{if}\ {\isacartoucheopen}{\isasymtturnstile}\isanewline
}
\DefineSnippet{SeCaV:lemma:3e54878179f6bbc8-7}{%
\ \ \ \ {\isacharbrackleft}\isanewline
}
\DefineSnippet{SeCaV:lemma:3e54878179f6bbc8-8}{%
\ \ \ \ \ \ Neg\ {\isacharparenleft}Pre\ {\isadigit{0}}\ {\isacharbrackleft}{\isacharbrackright}{\isacharparenright}{\isacharcomma}\isanewline
}
\DefineSnippet{SeCaV:lemma:3e54878179f6bbc8-9}{%
\ \ \ \ \ \ Pre\ {\isadigit{0}}\ {\isacharbrackleft}{\isacharbrackright}\isanewline
}
\DefineSnippet{SeCaV:lemma:3e54878179f6bbc8-10}{%
\ \ \ \ {\isacharbrackright}\isanewline
}
\DefineSnippet{SeCaV:lemma:3e54878179f6bbc8-11}{%
\ \ \ \ {\isacartoucheclose}\isanewline
}
\DefineSnippet{SeCaV:lemma:3e54878179f6bbc8-12}{%
\ \ \ \ \isacommand{using}\isamarkupfalse%
\ that\ \isacommand{by}\isamarkupfalse%
\ simp\isanewline
}
\DefineSnippet{SeCaV:lemma:3e54878179f6bbc8-13}{%
\ \ \isacommand{with}\isamarkupfalse%
\ Ext\ \isacommand{have}\isamarkupfalse%
\ {\isacharquery}thesis\ \isakeyword{if}\ {\isacartoucheopen}{\isasymtturnstile}\isanewline
}
\DefineSnippet{SeCaV:lemma:3e54878179f6bbc8-14}{%
\ \ \ \ {\isacharbrackleft}\isanewline
}
\DefineSnippet{SeCaV:lemma:3e54878179f6bbc8-15}{%
\ \ \ \ \ \ Pre\ {\isadigit{0}}\ {\isacharbrackleft}{\isacharbrackright}{\isacharcomma}\isanewline
}
\DefineSnippet{SeCaV:lemma:3e54878179f6bbc8-16}{%
\ \ \ \ \ \ Neg\ {\isacharparenleft}Pre\ {\isadigit{0}}\ {\isacharbrackleft}{\isacharbrackright}{\isacharparenright}\isanewline
}
\DefineSnippet{SeCaV:lemma:3e54878179f6bbc8-17}{%
\ \ \ \ {\isacharbrackright}\isanewline
}
\DefineSnippet{SeCaV:lemma:3e54878179f6bbc8-18}{%
\ \ \ \ {\isacartoucheclose}\isanewline
}
\DefineSnippet{SeCaV:lemma:3e54878179f6bbc8-19}{%
\ \ \ \ \isacommand{using}\isamarkupfalse%
\ that\ \isacommand{by}\isamarkupfalse%
\ simp\isanewline
}
\DefineSnippet{SeCaV:lemma:3e54878179f6bbc8-20}{%
\ \ \isacommand{with}\isamarkupfalse%
\ Basic\ \isacommand{show}\isamarkupfalse%
\ {\isacharquery}thesis\isanewline
}
\DefineSnippet{SeCaV:lemma:3e54878179f6bbc8-21}{%
\ \ \ \ \isacommand{by}\isamarkupfalse%
\ simp\isanewline
}
\DefineSnippet{SeCaV:lemma:3e54878179f6bbc8-22}{%
\isacommand{qed}\isamarkupfalse%
\endisatagproof
{\isafoldproof}%
\isadelimproof
\endisadelimproof
}
\DefineSnippet{SeCaV:section:a4a5b2adf708bccc-0}{%
\isadelimdocument
\endisadelimdocument
\isatagdocument
\isamarkupsection{Appendix: Soundness%
}
\isamarkuptrue%
\endisatagdocument
{\isafolddocument}%
\isadelimdocument
\endisadelimdocument
}
\DefineSnippet{SeCaV:subsection:74d3ee99fd128c95-0}{%
\isadelimdocument
\endisadelimdocument
\isatagdocument
\isamarkupsubsection{Increment Function%
}
\isamarkuptrue%
\endisatagdocument
{\isafolddocument}%
\isadelimdocument
\endisadelimdocument
}
\DefineSnippet{SeCaV:primrec:liftt-0}{%
\isacommand{primrec}\isamarkupfalse%
\ liftt\ {\isacharcolon}{\isacharcolon}\ {\isacartoucheopen}tm\ {\isasymRightarrow}\ tm{\isacartoucheclose}\ \isakeyword{and}\ liftts\ {\isacharcolon}{\isacharcolon}\ {\isacartoucheopen}tm\ list\ {\isasymRightarrow}\ tm\ list{\isacartoucheclose}\ \isakeyword{where}\isanewline
}
\DefineSnippet{SeCaV:primrec:liftt-1}{%
\ \ {\isacartoucheopen}liftt\ {\isacharparenleft}Var\ i{\isacharparenright}\ {\isacharequal}\ Var\ {\isacharparenleft}Suc\ i{\isacharparenright}{\isacartoucheclose}\ {\isacharbar}\isanewline
}
\DefineSnippet{SeCaV:primrec:liftt-2}{%
\ \ {\isacartoucheopen}liftt\ {\isacharparenleft}Fun\ a\ ts{\isacharparenright}\ {\isacharequal}\ Fun\ a\ {\isacharparenleft}liftts\ ts{\isacharparenright}{\isacartoucheclose}\ {\isacharbar}\isanewline
}
\DefineSnippet{SeCaV:primrec:liftt-3}{%
\ \ {\isacartoucheopen}liftts\ {\isacharbrackleft}{\isacharbrackright}\ {\isacharequal}\ {\isacharbrackleft}{\isacharbrackright}{\isacartoucheclose}\ {\isacharbar}\isanewline
}
\DefineSnippet{SeCaV:primrec:liftt-4}{%
\ \ {\isacartoucheopen}liftts\ {\isacharparenleft}t\ {\isacharhash}\ ts{\isacharparenright}\ {\isacharequal}\ liftt\ t\ {\isacharhash}\ liftts\ ts{\isacartoucheclose}%
}
\DefineSnippet{SeCaV:subsection:1367e7f593197bc-0}{%
\isadelimdocument
\endisadelimdocument
\isatagdocument
\isamarkupsubsection{Parameters: Terms%
}
\isamarkuptrue%
\endisatagdocument
{\isafolddocument}%
\isadelimdocument
\endisadelimdocument
}
\DefineSnippet{SeCaV:primrec:paramst-0}{%
\isacommand{primrec}\isamarkupfalse%
\ paramst\ {\isacharcolon}{\isacharcolon}\ {\isacartoucheopen}tm\ {\isasymRightarrow}\ nat\ set{\isacartoucheclose}\ \isakeyword{and}\ paramsts\ {\isacharcolon}{\isacharcolon}\ {\isacartoucheopen}tm\ list\ {\isasymRightarrow}\ nat\ set{\isacartoucheclose}\ \isakeyword{where}\isanewline
}
\DefineSnippet{SeCaV:primrec:paramst-1}{%
\ \ {\isacartoucheopen}paramst\ {\isacharparenleft}Var\ n{\isacharparenright}\ {\isacharequal}\ {\isacharbraceleft}{\isacharbraceright}{\isacartoucheclose}\ {\isacharbar}\isanewline
}
\DefineSnippet{SeCaV:primrec:paramst-2}{%
\ \ {\isacartoucheopen}paramst\ {\isacharparenleft}Fun\ a\ ts{\isacharparenright}\ {\isacharequal}\ {\isacharbraceleft}a{\isacharbraceright}\ {\isasymunion}\ paramsts\ ts{\isacartoucheclose}\ {\isacharbar}\isanewline
}
\DefineSnippet{SeCaV:primrec:paramst-3}{%
\ \ {\isacartoucheopen}paramsts\ {\isacharbrackleft}{\isacharbrackright}\ {\isacharequal}\ {\isacharbraceleft}{\isacharbraceright}{\isacartoucheclose}\ {\isacharbar}\isanewline
}
\DefineSnippet{SeCaV:primrec:paramst-4}{%
\ \ {\isacartoucheopen}paramsts\ {\isacharparenleft}t\ {\isacharhash}\ ts{\isacharparenright}\ {\isacharequal}\ {\isacharparenleft}paramst\ t\ {\isasymunion}\ paramsts\ ts{\isacharparenright}{\isacartoucheclose}%
}
\DefineSnippet{SeCaV:lemma:p0-0}{%
\isacommand{lemma}\isamarkupfalse%
\ p{\isadigit{0}}\ {\isacharbrackleft}simp{\isacharbrackright}{\isacharcolon}\ {\isacartoucheopen}paramsts\ ts\ {\isacharequal}\ {\isasymUnion}{\isacharparenleft}set\ {\isacharparenleft}map\ paramst\ ts{\isacharparenright}{\isacharparenright}{\isacartoucheclose}\isanewline
}
\DefineSnippet{SeCaV:lemma:p0-1}{%
\isadelimproof
\ \ %
\endisadelimproof
\isatagproof
\isacommand{by}\isamarkupfalse%
\ {\isacharparenleft}induct\ ts{\isacharparenright}\ simp{\isacharunderscore}all%
\endisatagproof
{\isafoldproof}%
\isadelimproof
\endisadelimproof
}
\DefineSnippet{SeCaV:primrec:paramst'-0}{%
\isacommand{primrec}\isamarkupfalse%
\ paramst{\isacharprime}\ {\isacharcolon}{\isacharcolon}\ {\isacartoucheopen}tm\ {\isasymRightarrow}\ nat\ set{\isacartoucheclose}\ \isakeyword{where}\isanewline
}
\DefineSnippet{SeCaV:primrec:paramst'-1}{%
\ \ {\isacartoucheopen}paramst{\isacharprime}\ {\isacharparenleft}Var\ n{\isacharparenright}\ {\isacharequal}\ {\isacharbraceleft}{\isacharbraceright}{\isacartoucheclose}\ {\isacharbar}\isanewline
}
\DefineSnippet{SeCaV:primrec:paramst'-2}{%
\ \ {\isacartoucheopen}paramst{\isacharprime}\ {\isacharparenleft}Fun\ a\ ts{\isacharparenright}\ {\isacharequal}\ {\isacharbraceleft}a{\isacharbraceright}\ {\isasymunion}\ {\isasymUnion}{\isacharparenleft}set\ {\isacharparenleft}map\ paramst{\isacharprime}\ ts{\isacharparenright}{\isacharparenright}{\isacartoucheclose}%
}
\DefineSnippet{SeCaV:lemma:p1-0}{%
\isacommand{lemma}\isamarkupfalse%
\ p{\isadigit{1}}\ {\isacharbrackleft}simp{\isacharbrackright}{\isacharcolon}\ {\isacartoucheopen}paramst{\isacharprime}\ t\ {\isacharequal}\ paramst\ t{\isacartoucheclose}\isanewline
}
\DefineSnippet{SeCaV:lemma:p1-1}{%
\isadelimproof
\ \ %
\endisadelimproof
\isatagproof
\isacommand{by}\isamarkupfalse%
\ {\isacharparenleft}induct\ t{\isacharparenright}\ simp{\isacharunderscore}all%
\endisatagproof
{\isafoldproof}%
\isadelimproof
\endisadelimproof
}
\DefineSnippet{SeCaV:subsection:35dad610c7fe7a6f-0}{%
\isadelimdocument
\endisadelimdocument
\isatagdocument
\isamarkupsubsection{Parameters: Formulas%
}
\isamarkuptrue%
\endisatagdocument
{\isafolddocument}%
\isadelimdocument
\endisadelimdocument
}
\DefineSnippet{SeCaV:primrec:params-0}{%
\isacommand{primrec}\isamarkupfalse%
\ params\ {\isacharcolon}{\isacharcolon}\ {\isacartoucheopen}fm\ {\isasymRightarrow}\ nat\ set{\isacartoucheclose}\ \isakeyword{where}\isanewline
}
\DefineSnippet{SeCaV:primrec:params-1}{%
\ \ {\isacartoucheopen}params\ {\isacharparenleft}Pre\ b\ ts{\isacharparenright}\ {\isacharequal}\ paramsts\ ts{\isacartoucheclose}\ {\isacharbar}\isanewline
}
\DefineSnippet{SeCaV:primrec:params-2}{%
\ \ {\isacartoucheopen}params\ {\isacharparenleft}Imp\ p\ q{\isacharparenright}\ {\isacharequal}\ params\ p\ {\isasymunion}\ params\ q{\isacartoucheclose}\ {\isacharbar}\isanewline
}
\DefineSnippet{SeCaV:primrec:params-3}{%
\ \ {\isacartoucheopen}params\ {\isacharparenleft}Dis\ p\ q{\isacharparenright}\ {\isacharequal}\ params\ p\ {\isasymunion}\ params\ q{\isacartoucheclose}\ {\isacharbar}\isanewline
}
\DefineSnippet{SeCaV:primrec:params-4}{%
\ \ {\isacartoucheopen}params\ {\isacharparenleft}Con\ p\ q{\isacharparenright}\ {\isacharequal}\ params\ p\ {\isasymunion}\ params\ q{\isacartoucheclose}\ {\isacharbar}\isanewline
}
\DefineSnippet{SeCaV:primrec:params-5}{%
\ \ {\isacartoucheopen}params\ {\isacharparenleft}Exi\ p{\isacharparenright}\ {\isacharequal}\ params\ p{\isacartoucheclose}\ {\isacharbar}\isanewline
}
\DefineSnippet{SeCaV:primrec:params-6}{%
\ \ {\isacartoucheopen}params\ {\isacharparenleft}Uni\ p{\isacharparenright}\ {\isacharequal}\ params\ p{\isacartoucheclose}\ {\isacharbar}\isanewline
}
\DefineSnippet{SeCaV:primrec:params-7}{%
\ \ {\isacartoucheopen}params\ {\isacharparenleft}Neg\ p{\isacharparenright}\ {\isacharequal}\ params\ p{\isacartoucheclose}%
}
\DefineSnippet{SeCaV:primrec:params'-0}{%
\isacommand{primrec}\isamarkupfalse%
\ params{\isacharprime}\ {\isacharcolon}{\isacharcolon}\ {\isacartoucheopen}fm\ {\isasymRightarrow}\ nat\ set{\isacartoucheclose}\ \isakeyword{where}\isanewline
}
\DefineSnippet{SeCaV:primrec:params'-1}{%
\ \ {\isacartoucheopen}params{\isacharprime}\ {\isacharparenleft}Pre\ b\ ts{\isacharparenright}\ {\isacharequal}\ {\isasymUnion}{\isacharparenleft}set\ {\isacharparenleft}map\ paramst{\isacharprime}\ ts{\isacharparenright}{\isacharparenright}{\isacartoucheclose}\ {\isacharbar}\isanewline
}
\DefineSnippet{SeCaV:primrec:params'-2}{%
\ \ {\isacartoucheopen}params{\isacharprime}\ {\isacharparenleft}Imp\ p\ q{\isacharparenright}\ {\isacharequal}\ params{\isacharprime}\ p\ {\isasymunion}\ params{\isacharprime}\ q{\isacartoucheclose}\ {\isacharbar}\isanewline
}
\DefineSnippet{SeCaV:primrec:params'-3}{%
\ \ {\isacartoucheopen}params{\isacharprime}\ {\isacharparenleft}Dis\ p\ q{\isacharparenright}\ {\isacharequal}\ params{\isacharprime}\ p\ {\isasymunion}\ params{\isacharprime}\ q{\isacartoucheclose}\ {\isacharbar}\isanewline
}
\DefineSnippet{SeCaV:primrec:params'-4}{%
\ \ {\isacartoucheopen}params{\isacharprime}\ {\isacharparenleft}Con\ p\ q{\isacharparenright}\ {\isacharequal}\ params{\isacharprime}\ p\ {\isasymunion}\ params{\isacharprime}\ q{\isacartoucheclose}\ {\isacharbar}\isanewline
}
\DefineSnippet{SeCaV:primrec:params'-5}{%
\ \ {\isacartoucheopen}params{\isacharprime}\ {\isacharparenleft}Exi\ p{\isacharparenright}\ {\isacharequal}\ params{\isacharprime}\ p{\isacartoucheclose}\ {\isacharbar}\isanewline
}
\DefineSnippet{SeCaV:primrec:params'-6}{%
\ \ {\isacartoucheopen}params{\isacharprime}\ {\isacharparenleft}Uni\ p{\isacharparenright}\ {\isacharequal}\ params{\isacharprime}\ p{\isacartoucheclose}\ {\isacharbar}\isanewline
}
\DefineSnippet{SeCaV:primrec:params'-7}{%
\ \ {\isacartoucheopen}params{\isacharprime}\ {\isacharparenleft}Neg\ p{\isacharparenright}\ {\isacharequal}\ params{\isacharprime}\ p{\isacartoucheclose}%
}
\DefineSnippet{SeCaV:lemma:p2-0}{%
\isacommand{lemma}\isamarkupfalse%
\ p{\isadigit{2}}\ {\isacharbrackleft}simp{\isacharbrackright}{\isacharcolon}\ {\isacartoucheopen}params{\isacharprime}\ p\ {\isacharequal}\ params\ p{\isacartoucheclose}\isanewline
}
\DefineSnippet{SeCaV:lemma:p2-1}{%
\isadelimproof
\ \ %
\endisadelimproof
\isatagproof
\isacommand{by}\isamarkupfalse%
\ {\isacharparenleft}induct\ p{\isacharparenright}\ simp{\isacharunderscore}all%
\endisatagproof
{\isafoldproof}%
\isadelimproof
\endisadelimproof
}
\DefineSnippet{SeCaV:fun:paramst''-0}{%
\isacommand{fun}\isamarkupfalse%
\ paramst{\isacharprime}{\isacharprime}\ {\isacharcolon}{\isacharcolon}\ {\isacartoucheopen}tm\ {\isasymRightarrow}\ nat\ set{\isacartoucheclose}\ \isakeyword{where}\isanewline
}
\DefineSnippet{SeCaV:fun:paramst''-1}{%
\ \ {\isacartoucheopen}paramst{\isacharprime}{\isacharprime}\ {\isacharparenleft}Var\ n{\isacharparenright}\ {\isacharequal}\ {\isacharbraceleft}{\isacharbraceright}{\isacartoucheclose}\ {\isacharbar}\isanewline
}
\DefineSnippet{SeCaV:fun:paramst''-2}{%
\ \ {\isacartoucheopen}paramst{\isacharprime}{\isacharprime}\ {\isacharparenleft}Fun\ a\ ts{\isacharparenright}\ {\isacharequal}\ {\isacharbraceleft}a{\isacharbraceright}\ {\isasymunion}\ {\isacharparenleft}{\isasymUnion}t\ {\isasymin}\ set\ ts{\isachardot}\ paramst{\isacharprime}{\isacharprime}\ t{\isacharparenright}{\isacartoucheclose}%
}
\DefineSnippet{SeCaV:lemma:p1'-0}{%
\isacommand{lemma}\isamarkupfalse%
\ p{\isadigit{1}}{\isacharprime}\ {\isacharbrackleft}simp{\isacharbrackright}{\isacharcolon}\ {\isacartoucheopen}paramst{\isacharprime}{\isacharprime}\ t\ {\isacharequal}\ paramst\ t{\isacartoucheclose}\isanewline
}
\DefineSnippet{SeCaV:lemma:p1'-1}{%
\isadelimproof
\ \ %
\endisadelimproof
\isatagproof
\isacommand{by}\isamarkupfalse%
\ {\isacharparenleft}induct\ t{\isacharparenright}\ simp{\isacharunderscore}all%
\endisatagproof
{\isafoldproof}%
\isadelimproof
\endisadelimproof
}
\DefineSnippet{SeCaV:fun:params''-0}{%
\isacommand{fun}\isamarkupfalse%
\ params{\isacharprime}{\isacharprime}\ {\isacharcolon}{\isacharcolon}\ {\isacartoucheopen}fm\ {\isasymRightarrow}\ nat\ set{\isacartoucheclose}\ \isakeyword{where}\isanewline
}
\DefineSnippet{SeCaV:fun:params''-1}{%
\ \ {\isacartoucheopen}params{\isacharprime}{\isacharprime}\ {\isacharparenleft}Pre\ b\ ts{\isacharparenright}\ {\isacharequal}\ {\isacharparenleft}{\isasymUnion}t\ {\isasymin}\ set\ ts{\isachardot}\ paramst{\isacharprime}{\isacharprime}\ t{\isacharparenright}{\isacartoucheclose}\ {\isacharbar}\isanewline
}
\DefineSnippet{SeCaV:fun:params''-2}{%
\ \ {\isacartoucheopen}params{\isacharprime}{\isacharprime}\ {\isacharparenleft}Imp\ p\ q{\isacharparenright}\ {\isacharequal}\ params{\isacharprime}{\isacharprime}\ p\ {\isasymunion}\ params{\isacharprime}{\isacharprime}\ q{\isacartoucheclose}\ {\isacharbar}\isanewline
}
\DefineSnippet{SeCaV:fun:params''-3}{%
\ \ {\isacartoucheopen}params{\isacharprime}{\isacharprime}\ {\isacharparenleft}Dis\ p\ q{\isacharparenright}\ {\isacharequal}\ params{\isacharprime}{\isacharprime}\ p\ {\isasymunion}\ params{\isacharprime}{\isacharprime}\ q{\isacartoucheclose}\ {\isacharbar}\isanewline
}
\DefineSnippet{SeCaV:fun:params''-4}{%
\ \ {\isacartoucheopen}params{\isacharprime}{\isacharprime}\ {\isacharparenleft}Con\ p\ q{\isacharparenright}\ {\isacharequal}\ params{\isacharprime}{\isacharprime}\ p\ {\isasymunion}\ params{\isacharprime}{\isacharprime}\ q{\isacartoucheclose}\ {\isacharbar}\isanewline
}
\DefineSnippet{SeCaV:fun:params''-5}{%
\ \ {\isacartoucheopen}params{\isacharprime}{\isacharprime}\ {\isacharparenleft}Exi\ p{\isacharparenright}\ {\isacharequal}\ params{\isacharprime}{\isacharprime}\ p{\isacartoucheclose}\ {\isacharbar}\isanewline
}
\DefineSnippet{SeCaV:fun:params''-6}{%
\ \ {\isacartoucheopen}params{\isacharprime}{\isacharprime}\ {\isacharparenleft}Uni\ p{\isacharparenright}\ {\isacharequal}\ params{\isacharprime}{\isacharprime}\ p{\isacartoucheclose}\ {\isacharbar}\isanewline
}
\DefineSnippet{SeCaV:fun:params''-7}{%
\ \ {\isacartoucheopen}params{\isacharprime}{\isacharprime}\ {\isacharparenleft}Neg\ p{\isacharparenright}\ {\isacharequal}\ params{\isacharprime}{\isacharprime}\ p{\isacartoucheclose}%
}
\DefineSnippet{SeCaV:lemma:p2'-0}{%
\isacommand{lemma}\isamarkupfalse%
\ p{\isadigit{2}}{\isacharprime}\ {\isacharbrackleft}simp{\isacharbrackright}{\isacharcolon}\ {\isacartoucheopen}params{\isacharprime}{\isacharprime}\ p\ {\isacharequal}\ params\ p{\isacartoucheclose}\isanewline
}
\DefineSnippet{SeCaV:lemma:p2'-1}{%
\isadelimproof
\ \ %
\endisadelimproof
\isatagproof
\isacommand{by}\isamarkupfalse%
\ {\isacharparenleft}induct\ p{\isacharparenright}\ simp{\isacharunderscore}all%
\endisatagproof
{\isafoldproof}%
\isadelimproof
\endisadelimproof
}
\DefineSnippet{SeCaV:subsection:20fcbc69eab02ed4-0}{%
\isadelimdocument
\endisadelimdocument
\isatagdocument
\isamarkupsubsection{Update Lemmas%
}
\isamarkuptrue%
\endisatagdocument
{\isafolddocument}%
\isadelimdocument
\endisadelimdocument
}
\DefineSnippet{SeCaV:lemma:upd-lemma'-0}{%
\isacommand{lemma}\isamarkupfalse%
\ upd{\isacharunderscore}lemma{\isacharprime}\ {\isacharbrackleft}simp{\isacharbrackright}{\isacharcolon}\isanewline
}
\DefineSnippet{SeCaV:lemma:upd-lemma'-1}{%
\ \ {\isacartoucheopen}n\ {\isasymnotin}\ paramst\ t\ {\isasymLongrightarrow}\ semantics{\isacharunderscore}term\ e\ {\isacharparenleft}f{\isacharparenleft}n\ {\isacharcolon}{\isacharequal}\ z{\isacharparenright}{\isacharparenright}\ t\ {\isacharequal}\ semantics{\isacharunderscore}term\ e\ f\ t{\isacartoucheclose}\isanewline
}
\DefineSnippet{SeCaV:lemma:upd-lemma'-2}{%
\ \ {\isacartoucheopen}n\ {\isasymnotin}\ paramsts\ ts\ {\isasymLongrightarrow}\ semantics{\isacharunderscore}list\ e\ {\isacharparenleft}f{\isacharparenleft}n\ {\isacharcolon}{\isacharequal}\ z{\isacharparenright}{\isacharparenright}\ ts\ {\isacharequal}\ semantics{\isacharunderscore}list\ e\ f\ ts{\isacartoucheclose}\isanewline
}
\DefineSnippet{SeCaV:lemma:upd-lemma'-3}{%
\isadelimproof
\ \ %
\endisadelimproof
\isatagproof
\isacommand{by}\isamarkupfalse%
\ {\isacharparenleft}induct\ t\ \isakeyword{and}\ ts\ rule{\isacharcolon}\ paramst{\isachardot}induct\ paramsts{\isachardot}induct{\isacharparenright}\ auto%
\endisatagproof
{\isafoldproof}%
\isadelimproof
\endisadelimproof
}
\DefineSnippet{SeCaV:lemma:upd-lemma-0}{%
\isacommand{lemma}\isamarkupfalse%
\ upd{\isacharunderscore}lemma\ {\isacharbrackleft}iff{\isacharbrackright}{\isacharcolon}\ {\isacartoucheopen}n\ {\isasymnotin}\ params\ p\ {\isasymLongrightarrow}\ semantics\ e\ {\isacharparenleft}f{\isacharparenleft}n\ {\isacharcolon}{\isacharequal}\ z{\isacharparenright}{\isacharparenright}\ g\ p\ {\isasymlongleftrightarrow}\ semantics\ e\ f\ g\ p{\isacartoucheclose}\isanewline
}
\DefineSnippet{SeCaV:lemma:upd-lemma-1}{%
\isadelimproof
\ \ %
\endisadelimproof
\isatagproof
\isacommand{by}\isamarkupfalse%
\ {\isacharparenleft}induct\ p\ arbitrary{\isacharcolon}\ e{\isacharparenright}\ simp{\isacharunderscore}all%
\endisatagproof
{\isafoldproof}%
\isadelimproof
\endisadelimproof
}
\DefineSnippet{SeCaV:subsection:96afc51e0bef4324-0}{%
\isadelimdocument
\endisadelimdocument
\isatagdocument
\isamarkupsubsection{Substitution%
}
\isamarkuptrue%
\endisatagdocument
{\isafolddocument}%
\isadelimdocument
\endisadelimdocument
}
\DefineSnippet{SeCaV:primrec:substt-0}{%
\isacommand{primrec}\isamarkupfalse%
\ substt\ {\isacharcolon}{\isacharcolon}\ {\isacartoucheopen}tm\ {\isasymRightarrow}\ tm\ {\isasymRightarrow}\ nat\ {\isasymRightarrow}\ tm{\isacartoucheclose}\ \isakeyword{and}\ substts\ {\isacharcolon}{\isacharcolon}\ {\isacartoucheopen}tm\ list\ {\isasymRightarrow}\ tm\ {\isasymRightarrow}\ nat\ {\isasymRightarrow}\ tm\ list{\isacartoucheclose}\ \isakeyword{where}\isanewline
}
\DefineSnippet{SeCaV:primrec:substt-1}{%
\ \ {\isacartoucheopen}substt\ {\isacharparenleft}Var\ i{\isacharparenright}\ s\ k\ {\isacharequal}\ {\isacharparenleft}if\ k\ {\isacharless}\ i\ then\ Var\ {\isacharparenleft}i\ {\isacharminus}\ {\isadigit{1}}{\isacharparenright}\ else\ if\ i\ {\isacharequal}\ k\ then\ s\ else\ Var\ i{\isacharparenright}{\isacartoucheclose}\ {\isacharbar}\isanewline
}
\DefineSnippet{SeCaV:primrec:substt-2}{%
\ \ {\isacartoucheopen}substt\ {\isacharparenleft}Fun\ a\ ts{\isacharparenright}\ s\ k\ {\isacharequal}\ Fun\ a\ {\isacharparenleft}substts\ ts\ s\ k{\isacharparenright}{\isacartoucheclose}\ {\isacharbar}\isanewline
}
\DefineSnippet{SeCaV:primrec:substt-3}{%
\ \ {\isacartoucheopen}substts\ {\isacharbrackleft}{\isacharbrackright}\ s\ k\ {\isacharequal}\ {\isacharbrackleft}{\isacharbrackright}{\isacartoucheclose}\ {\isacharbar}\isanewline
}
\DefineSnippet{SeCaV:primrec:substt-4}{%
\ \ {\isacartoucheopen}substts\ {\isacharparenleft}t\ {\isacharhash}\ ts{\isacharparenright}\ s\ k\ {\isacharequal}\ substt\ t\ s\ k\ {\isacharhash}\ substts\ ts\ s\ k{\isacartoucheclose}%
}
\DefineSnippet{SeCaV:primrec:subst-0}{%
\isacommand{primrec}\isamarkupfalse%
\ subst\ {\isacharcolon}{\isacharcolon}\ {\isacartoucheopen}fm\ {\isasymRightarrow}\ tm\ {\isasymRightarrow}\ nat\ {\isasymRightarrow}\ fm{\isacartoucheclose}\ \isakeyword{where}\isanewline
}
\DefineSnippet{SeCaV:primrec:subst-1}{%
\ \ {\isacartoucheopen}subst\ {\isacharparenleft}Pre\ b\ ts{\isacharparenright}\ s\ k\ {\isacharequal}\ Pre\ b\ {\isacharparenleft}substts\ ts\ s\ k{\isacharparenright}{\isacartoucheclose}\ {\isacharbar}\isanewline
}
\DefineSnippet{SeCaV:primrec:subst-2}{%
\ \ {\isacartoucheopen}subst\ {\isacharparenleft}Imp\ p\ q{\isacharparenright}\ s\ k\ {\isacharequal}\ Imp\ {\isacharparenleft}subst\ p\ s\ k{\isacharparenright}\ {\isacharparenleft}subst\ q\ s\ k{\isacharparenright}{\isacartoucheclose}\ {\isacharbar}\isanewline
}
\DefineSnippet{SeCaV:primrec:subst-3}{%
\ \ {\isacartoucheopen}subst\ {\isacharparenleft}Dis\ p\ q{\isacharparenright}\ s\ k\ {\isacharequal}\ Dis\ {\isacharparenleft}subst\ p\ s\ k{\isacharparenright}\ {\isacharparenleft}subst\ q\ s\ k{\isacharparenright}{\isacartoucheclose}\ {\isacharbar}\isanewline
}
\DefineSnippet{SeCaV:primrec:subst-4}{%
\ \ {\isacartoucheopen}subst\ {\isacharparenleft}Con\ p\ q{\isacharparenright}\ s\ k\ {\isacharequal}\ Con\ {\isacharparenleft}subst\ p\ s\ k{\isacharparenright}\ {\isacharparenleft}subst\ q\ s\ k{\isacharparenright}{\isacartoucheclose}\ {\isacharbar}\isanewline
}
\DefineSnippet{SeCaV:primrec:subst-5}{%
\ \ {\isacartoucheopen}subst\ {\isacharparenleft}Exi\ p{\isacharparenright}\ s\ k\ {\isacharequal}\ Exi\ {\isacharparenleft}subst\ p\ {\isacharparenleft}liftt\ s{\isacharparenright}\ {\isacharparenleft}Suc\ k{\isacharparenright}{\isacharparenright}{\isacartoucheclose}\ {\isacharbar}\isanewline
}
\DefineSnippet{SeCaV:primrec:subst-6}{%
\ \ {\isacartoucheopen}subst\ {\isacharparenleft}Uni\ p{\isacharparenright}\ s\ k\ {\isacharequal}\ Uni\ {\isacharparenleft}subst\ p\ {\isacharparenleft}liftt\ s{\isacharparenright}\ {\isacharparenleft}Suc\ k{\isacharparenright}{\isacharparenright}{\isacartoucheclose}\ {\isacharbar}\isanewline
}
\DefineSnippet{SeCaV:primrec:subst-7}{%
\ \ {\isacartoucheopen}subst\ {\isacharparenleft}Neg\ p{\isacharparenright}\ s\ k\ {\isacharequal}\ Neg\ {\isacharparenleft}subst\ p\ s\ k{\isacharparenright}{\isacartoucheclose}%
}
\DefineSnippet{SeCaV:lemma:shift-eq-0}{%
\isacommand{lemma}\isamarkupfalse%
\ shift{\isacharunderscore}eq\ {\isacharbrackleft}simp{\isacharbrackright}{\isacharcolon}\ {\isacartoucheopen}i\ {\isacharequal}\ j\ {\isasymLongrightarrow}\ {\isacharparenleft}shift\ e\ i\ T{\isacharparenright}\ j\ {\isacharequal}\ T{\isacartoucheclose}\ \isakeyword{for}\ i\ {\isacharcolon}{\isacharcolon}\ nat\isanewline
}
\DefineSnippet{SeCaV:lemma:shift-eq-1}{%
\isadelimproof
\ \ %
\endisadelimproof
\isatagproof
\isacommand{unfolding}\isamarkupfalse%
\ shift{\isacharunderscore}def\ \isacommand{by}\isamarkupfalse%
\ simp%
\endisatagproof
{\isafoldproof}%
\isadelimproof
\endisadelimproof
}
\DefineSnippet{SeCaV:lemma:shift-gt-0}{%
\isacommand{lemma}\isamarkupfalse%
\ shift{\isacharunderscore}gt\ {\isacharbrackleft}simp{\isacharbrackright}{\isacharcolon}\ {\isacartoucheopen}j\ {\isacharless}\ i\ {\isasymLongrightarrow}\ {\isacharparenleft}shift\ e\ i\ T{\isacharparenright}\ j\ {\isacharequal}\ e\ j{\isacartoucheclose}\ \isakeyword{for}\ i\ {\isacharcolon}{\isacharcolon}\ nat\isanewline
}
\DefineSnippet{SeCaV:lemma:shift-gt-1}{%
\isadelimproof
\ \ %
\endisadelimproof
\isatagproof
\isacommand{unfolding}\isamarkupfalse%
\ shift{\isacharunderscore}def\ \isacommand{by}\isamarkupfalse%
\ simp%
\endisatagproof
{\isafoldproof}%
\isadelimproof
\endisadelimproof
}
\DefineSnippet{SeCaV:lemma:shift-lt-0}{%
\isacommand{lemma}\isamarkupfalse%
\ shift{\isacharunderscore}lt\ {\isacharbrackleft}simp{\isacharbrackright}{\isacharcolon}\ {\isacartoucheopen}i\ {\isacharless}\ j\ {\isasymLongrightarrow}\ {\isacharparenleft}shift\ e\ i\ T{\isacharparenright}\ j\ {\isacharequal}\ e\ {\isacharparenleft}j\ {\isacharminus}\ {\isadigit{1}}{\isacharparenright}{\isacartoucheclose}\ \isakeyword{for}\ i\ {\isacharcolon}{\isacharcolon}\ nat\isanewline
}
\DefineSnippet{SeCaV:lemma:shift-lt-1}{%
\isadelimproof
\ \ %
\endisadelimproof
\isatagproof
\isacommand{unfolding}\isamarkupfalse%
\ shift{\isacharunderscore}def\ \isacommand{by}\isamarkupfalse%
\ simp%
\endisatagproof
{\isafoldproof}%
\isadelimproof
\endisadelimproof
}
\DefineSnippet{SeCaV:lemma:shift-commute-0}{%
\isacommand{lemma}\isamarkupfalse%
\ shift{\isacharunderscore}commute\ {\isacharbrackleft}simp{\isacharbrackright}{\isacharcolon}\ {\isacartoucheopen}shift\ {\isacharparenleft}shift\ e\ i\ U{\isacharparenright}\ {\isadigit{0}}\ T\ {\isacharequal}\ shift\ {\isacharparenleft}shift\ e\ {\isadigit{0}}\ T{\isacharparenright}\ {\isacharparenleft}Suc\ i{\isacharparenright}\ U{\isacartoucheclose}\isanewline
}
\DefineSnippet{SeCaV:lemma:shift-commute-1}{%
\isadelimproof
\ \ %
\endisadelimproof
\isatagproof
\isacommand{unfolding}\isamarkupfalse%
\ shift{\isacharunderscore}def\ \isacommand{by}\isamarkupfalse%
\ force%
\endisatagproof
{\isafoldproof}%
\isadelimproof
\endisadelimproof
}
\DefineSnippet{SeCaV:lemma:subst-lemma'-0}{%
\isacommand{lemma}\isamarkupfalse%
\ subst{\isacharunderscore}lemma{\isacharprime}\ {\isacharbrackleft}simp{\isacharbrackright}{\isacharcolon}\isanewline
}
\DefineSnippet{SeCaV:lemma:subst-lemma'-1}{%
\ \ {\isacartoucheopen}semantics{\isacharunderscore}term\ e\ f\ {\isacharparenleft}substt\ t\ u\ i{\isacharparenright}\ {\isacharequal}\ semantics{\isacharunderscore}term\ {\isacharparenleft}shift\ e\ i\ {\isacharparenleft}semantics{\isacharunderscore}term\ e\ f\ u{\isacharparenright}{\isacharparenright}\ f\ t{\isacartoucheclose}\isanewline
}
\DefineSnippet{SeCaV:lemma:subst-lemma'-2}{%
\ \ {\isacartoucheopen}semantics{\isacharunderscore}list\ e\ f\ {\isacharparenleft}substts\ ts\ u\ i{\isacharparenright}\ {\isacharequal}\ semantics{\isacharunderscore}list\ {\isacharparenleft}shift\ e\ i\ {\isacharparenleft}semantics{\isacharunderscore}term\ e\ f\ u{\isacharparenright}{\isacharparenright}\ f\ ts{\isacartoucheclose}\isanewline
}
\DefineSnippet{SeCaV:lemma:subst-lemma'-3}{%
\isadelimproof
\ \ %
\endisadelimproof
\isatagproof
\isacommand{by}\isamarkupfalse%
\ {\isacharparenleft}induct\ t\ \isakeyword{and}\ ts\ rule{\isacharcolon}\ substt{\isachardot}induct\ substts{\isachardot}induct{\isacharparenright}\ simp{\isacharunderscore}all%
\endisatagproof
{\isafoldproof}%
\isadelimproof
\endisadelimproof
}
\DefineSnippet{SeCaV:lemma:lift-lemma-0}{%
\isacommand{lemma}\isamarkupfalse%
\ lift{\isacharunderscore}lemma\ {\isacharbrackleft}simp{\isacharbrackright}{\isacharcolon}\isanewline
}
\DefineSnippet{SeCaV:lemma:lift-lemma-1}{%
\ \ {\isacartoucheopen}semantics{\isacharunderscore}term\ {\isacharparenleft}shift\ e\ {\isadigit{0}}\ x{\isacharparenright}\ f\ {\isacharparenleft}liftt\ t{\isacharparenright}\ {\isacharequal}\ semantics{\isacharunderscore}term\ e\ f\ t{\isacartoucheclose}\isanewline
}
\DefineSnippet{SeCaV:lemma:lift-lemma-2}{%
\ \ {\isacartoucheopen}semantics{\isacharunderscore}list\ {\isacharparenleft}shift\ e\ {\isadigit{0}}\ x{\isacharparenright}\ f\ {\isacharparenleft}liftts\ ts{\isacharparenright}\ {\isacharequal}\ semantics{\isacharunderscore}list\ e\ f\ ts{\isacartoucheclose}\isanewline
}
\DefineSnippet{SeCaV:lemma:lift-lemma-3}{%
\isadelimproof
\ \ %
\endisadelimproof
\isatagproof
\isacommand{by}\isamarkupfalse%
\ {\isacharparenleft}induct\ t\ \isakeyword{and}\ ts\ rule{\isacharcolon}\ liftt{\isachardot}induct\ liftts{\isachardot}induct{\isacharparenright}\ simp{\isacharunderscore}all%
\endisatagproof
{\isafoldproof}%
\isadelimproof
\endisadelimproof
}
\DefineSnippet{SeCaV:lemma:subst-lemma-0}{%
\isacommand{lemma}\isamarkupfalse%
\ subst{\isacharunderscore}lemma\ {\isacharbrackleft}iff{\isacharbrackright}{\isacharcolon}\isanewline
}
\DefineSnippet{SeCaV:lemma:subst-lemma-1}{%
\ \ {\isacartoucheopen}semantics\ e\ f\ g\ {\isacharparenleft}subst\ a\ t\ i{\isacharparenright}\ {\isasymlongleftrightarrow}\ semantics\ {\isacharparenleft}shift\ e\ i\ {\isacharparenleft}semantics{\isacharunderscore}term\ e\ f\ t{\isacharparenright}{\isacharparenright}\ f\ g\ a{\isacartoucheclose}\isanewline
}
\DefineSnippet{SeCaV:lemma:subst-lemma-2}{%
\isadelimproof
\ \ %
\endisadelimproof
\isatagproof
\isacommand{by}\isamarkupfalse%
\ {\isacharparenleft}induct\ a\ arbitrary{\isacharcolon}\ e\ i\ t{\isacharparenright}\ simp{\isacharunderscore}all%
\endisatagproof
{\isafoldproof}%
\isadelimproof
\endisadelimproof
}
\DefineSnippet{SeCaV:subsection:5aa80acc4d0933ce-0}{%
\isadelimdocument
\endisadelimdocument
\isatagdocument
\isamarkupsubsection{Auxiliary Lemmas%
}
\isamarkuptrue%
\endisatagdocument
{\isafolddocument}%
\isadelimdocument
\endisadelimdocument
}
\DefineSnippet{SeCaV:lemma:s1-0}{%
\isacommand{lemma}\isamarkupfalse%
\ s{\isadigit{1}}\ {\isacharbrackleft}iff{\isacharbrackright}{\isacharcolon}\ {\isacartoucheopen}new{\isacharunderscore}term\ c\ t\ {\isasymlongleftrightarrow}\ {\isacharparenleft}c\ {\isasymnotin}\ paramst\ t{\isacharparenright}{\isacartoucheclose}\ {\isacartoucheopen}new{\isacharunderscore}list\ c\ l\ {\isasymlongleftrightarrow}\ {\isacharparenleft}c\ {\isasymnotin}\ paramsts\ l{\isacharparenright}{\isacartoucheclose}\isanewline
}
\DefineSnippet{SeCaV:lemma:s1-1}{%
\isadelimproof
\ \ %
\endisadelimproof
\isatagproof
\isacommand{by}\isamarkupfalse%
\ {\isacharparenleft}induct\ t\ \isakeyword{and}\ l\ rule{\isacharcolon}\ new{\isacharunderscore}term{\isachardot}induct\ new{\isacharunderscore}list{\isachardot}induct{\isacharparenright}\ simp{\isacharunderscore}all%
\endisatagproof
{\isafoldproof}%
\isadelimproof
\endisadelimproof
}
\DefineSnippet{SeCaV:lemma:s2-0}{%
\isacommand{lemma}\isamarkupfalse%
\ s{\isadigit{2}}\ {\isacharbrackleft}iff{\isacharbrackright}{\isacharcolon}\ {\isacartoucheopen}new\ c\ p\ {\isasymlongleftrightarrow}\ {\isacharparenleft}c\ {\isasymnotin}\ params\ p{\isacharparenright}{\isacartoucheclose}\isanewline
}
\DefineSnippet{SeCaV:lemma:s2-1}{%
\isadelimproof
\ \ %
\endisadelimproof
\isatagproof
\isacommand{by}\isamarkupfalse%
\ {\isacharparenleft}induct\ p{\isacharparenright}\ simp{\isacharunderscore}all%
\endisatagproof
{\isafoldproof}%
\isadelimproof
\endisadelimproof
}
\DefineSnippet{SeCaV:lemma:s3-0}{%
\isacommand{lemma}\isamarkupfalse%
\ s{\isadigit{3}}\ {\isacharbrackleft}iff{\isacharbrackright}{\isacharcolon}\ {\isacartoucheopen}news\ c\ z\ {\isasymlongleftrightarrow}\ list{\isacharunderscore}all\ {\isacharparenleft}{\isasymlambda}p{\isachardot}\ c\ {\isasymnotin}\ params\ p{\isacharparenright}\ z{\isacartoucheclose}\isanewline
}
\DefineSnippet{SeCaV:lemma:s3-1}{%
\isadelimproof
\ \ %
\endisadelimproof
\isatagproof
\isacommand{by}\isamarkupfalse%
\ {\isacharparenleft}induct\ z{\isacharparenright}\ simp{\isacharunderscore}all%
\endisatagproof
{\isafoldproof}%
\isadelimproof
\endisadelimproof
}
\DefineSnippet{SeCaV:lemma:s4-0}{%
\isacommand{lemma}\isamarkupfalse%
\ s{\isadigit{4}}\ {\isacharbrackleft}simp{\isacharbrackright}{\isacharcolon}\ {\isacartoucheopen}inc{\isacharunderscore}term\ t\ {\isacharequal}\ liftt\ t{\isacartoucheclose}\ {\isacartoucheopen}inc{\isacharunderscore}list\ l\ {\isacharequal}\ liftts\ l{\isacartoucheclose}\isanewline
}
\DefineSnippet{SeCaV:lemma:s4-1}{%
\isadelimproof
\ \ %
\endisadelimproof
\isatagproof
\isacommand{by}\isamarkupfalse%
\ {\isacharparenleft}induct\ t\ \isakeyword{and}\ l\ rule{\isacharcolon}\ inc{\isacharunderscore}term{\isachardot}induct\ inc{\isacharunderscore}list{\isachardot}induct{\isacharparenright}\ simp{\isacharunderscore}all%
\endisatagproof
{\isafoldproof}%
\isadelimproof
\endisadelimproof
}
\DefineSnippet{SeCaV:lemma:s5-0}{%
\isacommand{lemma}\isamarkupfalse%
\ s{\isadigit{5}}\ {\isacharbrackleft}simp{\isacharbrackright}{\isacharcolon}\ {\isacartoucheopen}sub{\isacharunderscore}term\ v\ s\ t\ {\isacharequal}\ substt\ t\ s\ v{\isacartoucheclose}\ {\isacartoucheopen}sub{\isacharunderscore}list\ v\ s\ l\ {\isacharequal}\ substts\ l\ s\ v{\isacartoucheclose}\isanewline
}
\DefineSnippet{SeCaV:lemma:s5-1}{%
\isadelimproof
\ \ %
\endisadelimproof
\isatagproof
\isacommand{by}\isamarkupfalse%
\ {\isacharparenleft}induct\ t\ \isakeyword{and}\ l\ rule{\isacharcolon}\ inc{\isacharunderscore}term{\isachardot}induct\ inc{\isacharunderscore}list{\isachardot}induct{\isacharparenright}\ simp{\isacharunderscore}all%
\endisatagproof
{\isafoldproof}%
\isadelimproof
\endisadelimproof
}
\DefineSnippet{SeCaV:lemma:s6-0}{%
\isacommand{lemma}\isamarkupfalse%
\ s{\isadigit{6}}\ {\isacharbrackleft}simp{\isacharbrackright}{\isacharcolon}\ {\isacartoucheopen}sub\ v\ s\ p\ {\isacharequal}\ subst\ p\ s\ v{\isacartoucheclose}\isanewline
}
\DefineSnippet{SeCaV:lemma:s6-1}{%
\isadelimproof
\ \ %
\endisadelimproof
\isatagproof
\isacommand{by}\isamarkupfalse%
\ {\isacharparenleft}induct\ p\ arbitrary{\isacharcolon}\ v\ s{\isacharparenright}\ simp{\isacharunderscore}all%
\endisatagproof
{\isafoldproof}%
\isadelimproof
\endisadelimproof
}
\DefineSnippet{SeCaV:subsection:d167292d2ed64437-0}{%
\isadelimdocument
\endisadelimdocument
\isatagdocument
\isamarkupsubsection{Soundness%
}
\isamarkuptrue%
\endisatagdocument
{\isafolddocument}%
\isadelimdocument
\endisadelimdocument
}
\DefineSnippet{SeCaV:theorem:sound-0}{%
\isacommand{theorem}\isamarkupfalse%
\ sound{\isacharcolon}\ {\isacartoucheopen}{\isasymtturnstile}\ z\ {\isasymLongrightarrow}\ {\isasymexists}p\ {\isasymin}\ set\ z{\isachardot}\ semantics\ e\ f\ g\ p{\isacartoucheclose}\isanewline
}
\DefineSnippet{SeCaV:theorem:sound-1}{%
\isadelimproof
\endisadelimproof
\isatagproof
\isacommand{proof}\isamarkupfalse%
\ {\isacharparenleft}induct\ arbitrary{\isacharcolon}\ f\ rule{\isacharcolon}\ sequent{\isacharunderscore}calculus{\isachardot}induct{\isacharparenright}\isanewline
}
\DefineSnippet{SeCaV:theorem:sound-2}{%
\ \ \isacommand{case}\isamarkupfalse%
\ {\isacharparenleft}{\isadigit{1}}{\isadigit{0}}\ i\ p\ z{\isacharparenright}\isanewline
}
\DefineSnippet{SeCaV:theorem:sound-3}{%
\ \ \isacommand{then}\isamarkupfalse%
\ \isacommand{show}\isamarkupfalse%
\ {\isacharquery}case\isanewline
}
\DefineSnippet{SeCaV:theorem:sound-4}{%
\ \ \isacommand{proof}\isamarkupfalse%
\ {\isacharparenleft}cases\ {\isacartoucheopen}{\isasymforall}x{\isachardot}\ semantics\ e\ {\isacharparenleft}f{\isacharparenleft}i\ {\isacharcolon}{\isacharequal}\ {\isasymlambda}{\isacharunderscore}{\isachardot}\ x{\isacharparenright}{\isacharparenright}\ g\ {\isacharparenleft}sub\ {\isadigit{0}}\ {\isacharparenleft}Fun\ i\ {\isacharbrackleft}{\isacharbrackright}{\isacharparenright}\ p{\isacharparenright}{\isacartoucheclose}{\isacharparenright}\isanewline
}
\DefineSnippet{SeCaV:theorem:sound-5}{%
\ \ \ \ \isacommand{case}\isamarkupfalse%
\ False\isanewline
}
\DefineSnippet{SeCaV:theorem:sound-6}{%
\ \ \ \ \isacommand{moreover}\isamarkupfalse%
\ \isacommand{have}\isamarkupfalse%
\ {\isacartoucheopen}list{\isacharunderscore}all\ {\isacharparenleft}{\isasymlambda}p{\isachardot}\ i\ {\isasymnotin}\ params\ p{\isacharparenright}\ z{\isacartoucheclose}\isanewline
}
\DefineSnippet{SeCaV:theorem:sound-7}{%
\ \ \ \ \ \ \isacommand{using}\isamarkupfalse%
\ {\isadigit{1}}{\isadigit{0}}\ \isacommand{by}\isamarkupfalse%
\ simp\isanewline
}
\DefineSnippet{SeCaV:theorem:sound-8}{%
\ \ \ \ \isacommand{ultimately}\isamarkupfalse%
\ \isacommand{show}\isamarkupfalse%
\ {\isacharquery}thesis\isanewline
}
\DefineSnippet{SeCaV:theorem:sound-9}{%
\ \ \ \ \ \ \isacommand{using}\isamarkupfalse%
\ {\isadigit{1}}{\isadigit{0}}\ Ball{\isacharunderscore}set\ insert{\isacharunderscore}iff\ list{\isachardot}set{\isacharparenleft}{\isadigit{2}}{\isacharparenright}\ upd{\isacharunderscore}lemma\ \isacommand{by}\isamarkupfalse%
\ metis\isanewline
}
\DefineSnippet{SeCaV:theorem:sound-10}{%
\ \ \isacommand{qed}\isamarkupfalse%
\ simp\isanewline
}
\DefineSnippet{SeCaV:theorem:sound-11}{%
\isacommand{next}\isamarkupfalse%
\isanewline
}
\DefineSnippet{SeCaV:theorem:sound-12}{%
\ \ \isacommand{case}\isamarkupfalse%
\ {\isacharparenleft}{\isadigit{1}}{\isadigit{1}}\ i\ p\ z{\isacharparenright}\isanewline
}
\DefineSnippet{SeCaV:theorem:sound-13}{%
\ \ \isacommand{then}\isamarkupfalse%
\ \isacommand{show}\isamarkupfalse%
\ {\isacharquery}case\isanewline
}
\DefineSnippet{SeCaV:theorem:sound-14}{%
\ \ \isacommand{proof}\isamarkupfalse%
\ {\isacharparenleft}cases\ {\isacartoucheopen}{\isasymforall}x{\isachardot}\ semantics\ e\ {\isacharparenleft}f{\isacharparenleft}i\ {\isacharcolon}{\isacharequal}\ {\isasymlambda}{\isacharunderscore}{\isachardot}\ x{\isacharparenright}{\isacharparenright}\ g\ {\isacharparenleft}Neg\ {\isacharparenleft}sub\ {\isadigit{0}}\ {\isacharparenleft}Fun\ i\ {\isacharbrackleft}{\isacharbrackright}{\isacharparenright}\ p{\isacharparenright}{\isacharparenright}{\isacartoucheclose}{\isacharparenright}\isanewline
}
\DefineSnippet{SeCaV:theorem:sound-15}{%
\ \ \ \ \isacommand{case}\isamarkupfalse%
\ False\isanewline
}
\DefineSnippet{SeCaV:theorem:sound-16}{%
\ \ \ \ \isacommand{moreover}\isamarkupfalse%
\ \isacommand{have}\isamarkupfalse%
\ {\isacartoucheopen}list{\isacharunderscore}all\ {\isacharparenleft}{\isasymlambda}p{\isachardot}\ i\ {\isasymnotin}\ params\ p{\isacharparenright}\ z{\isacartoucheclose}\isanewline
}
\DefineSnippet{SeCaV:theorem:sound-17}{%
\ \ \ \ \ \ \isacommand{using}\isamarkupfalse%
\ {\isadigit{1}}{\isadigit{1}}\ \isacommand{by}\isamarkupfalse%
\ simp\isanewline
}
\DefineSnippet{SeCaV:theorem:sound-18}{%
\ \ \ \ \isacommand{ultimately}\isamarkupfalse%
\ \isacommand{show}\isamarkupfalse%
\ {\isacharquery}thesis\isanewline
}
\DefineSnippet{SeCaV:theorem:sound-19}{%
\ \ \ \ \ \ \isacommand{using}\isamarkupfalse%
\ {\isadigit{1}}{\isadigit{1}}\ Ball{\isacharunderscore}set\ insert{\isacharunderscore}iff\ list{\isachardot}set{\isacharparenleft}{\isadigit{2}}{\isacharparenright}\ upd{\isacharunderscore}lemma\ \isacommand{by}\isamarkupfalse%
\ metis\isanewline
}
\DefineSnippet{SeCaV:theorem:sound-20}{%
\ \ \isacommand{qed}\isamarkupfalse%
\ simp\isanewline
}
\DefineSnippet{SeCaV:theorem:sound-21}{%
\isacommand{qed}\isamarkupfalse%
\ force{\isacharplus}%
\endisatagproof
{\isafoldproof}%
\isadelimproof
\endisadelimproof
}
\DefineSnippet{SeCaV:corollary:1dfb4481fef79b20-0}{%
\isacommand{corollary}\isamarkupfalse%
\ {\isacartoucheopen}{\isasymtturnstile}\ z\ {\isasymLongrightarrow}\ {\isasymexists}p{\isachardot}\ member\ p\ z\ {\isasymand}\ semantics\ e\ f\ g\ p{\isacartoucheclose}\isanewline
}
\DefineSnippet{SeCaV:corollary:1dfb4481fef79b20-1}{%
\isadelimproof
\ \ %
\endisadelimproof
\isatagproof
\isacommand{using}\isamarkupfalse%
\ sound\ \isacommand{by}\isamarkupfalse%
\ force%
\endisatagproof
{\isafoldproof}%
\isadelimproof
\endisadelimproof
}
\DefineSnippet{SeCaV:corollary:c38238813c5dc8d-0}{%
\isacommand{corollary}\isamarkupfalse%
\ {\isacartoucheopen}{\isasymtturnstile}\ {\isacharbrackleft}p{\isacharbrackright}\ {\isasymLongrightarrow}\ semantics\ e\ f\ g\ p{\isacartoucheclose}\isanewline
}
\DefineSnippet{SeCaV:corollary:c38238813c5dc8d-1}{%
\isadelimproof
\ \ %
\endisadelimproof
\isatagproof
\isacommand{using}\isamarkupfalse%
\ sound\ \isacommand{by}\isamarkupfalse%
\ force%
\endisatagproof
{\isafoldproof}%
\isadelimproof
\endisadelimproof
}
\DefineSnippet{SeCaV:corollary:4c6098ac2a6f2d11-0}{%
\isacommand{corollary}\isamarkupfalse%
\ {\isacartoucheopen}{\isasymnot}\ {\isacharparenleft}{\isasymtturnstile}\ {\isacharbrackleft}{\isacharbrackright}{\isacharparenright}{\isacartoucheclose}\isanewline
}
\DefineSnippet{SeCaV:corollary:4c6098ac2a6f2d11-1}{%
\isadelimproof
\ \ %
\endisadelimproof
\isatagproof
\isacommand{using}\isamarkupfalse%
\ sound\ \isacommand{by}\isamarkupfalse%
\ force%
\endisatagproof
{\isafoldproof}%
\isadelimproof
\endisadelimproof
}
\DefineSnippet{SeCaV:section:333af9ea13740c84-0}{%
\isadelimdocument
\endisadelimdocument
\isatagdocument
\isamarkupsection{Reference%
}
\isamarkuptrue%
\endisatagdocument
{\isafolddocument}%
\isadelimdocument
\endisadelimdocument
}
\DefineSnippet{SeCaV:text:acab9fd4d2f857cd-0}{%
\begin{isamarkuptext}%
Mordechai Ben-Ari (Springer 2012): Mathematical Logic for Computer Science (Third Edition)%
\end{isamarkuptext}\isamarkuptrue%
}
\DefineSnippet{SeCaV:end:7158f6523c28304a-0}{%
\isadelimtheory
\endisadelimtheory
\isatagtheory
\isacommand{end}\isamarkupfalse%
\endisatagtheory
{\isafoldtheory}%
\isadelimtheory
\endisadelimtheory
}

\newcommand{\Snippet}[1]{{%
  \newcount\i
  \i=0
  \loop
    \csname snippet--#1-\the\i\endcsname
    \advance \i 1
  \ifcsname snippet--#1-\the\i\endcsname
  \repeat
}}

\newcommand{\SnippetPart}[3]{{%
  \newcount\i
  \i=#1
  \loop
    \ifnum \i=#2
      \renewcommand{\isanewline}{}%
    \fi
    \csname snippet--#3-\the\i\endcsname
    \advance \i 1
    \ifnum \i>#2 {}
    \else \repeat
}}

\newcommand{\Rule}{\isa}

\def\isacartoucheopen{\isatext{\raise.3ex\hbox{$\scriptscriptstyle\langle\,\,\,$}}}%
\def\isacartoucheclose{\isatext{\raise.3ex\hbox{$\scriptscriptstyle\,\,\,\rangle$}}}%

\begin{document}

\maketitle

\begin{abstract}
We describe SeCaV, a sequent calculus verifier for first-order logic in Isabelle/HOL, and the SeCaV Unshortener, an online tool that expands succinct derivations into the full SeCaV syntax.
We leverage the power of Isabelle/HOL as a proof checker for our SeCaV derivations.
The interactive features of Isabelle/HOL make our system transparent.
For instance, the user can simply click on a side condition to see its exact definition.
Our formalized soundness and completeness proofs pertain exactly to the calculus as exposed to the user and not just to some model of our tool.
Users can also write their derivations in the SeCaV Unshortener, which provides a lighter syntax, and expand them for later verification.
We have used both tools in our teaching.
\end{abstract}

\section{Introduction}%
\label{sec:introduction}

Classical first-order logic plays an important role in mathematical logic and often occupies a central part in textbooks and courses on the subject.
The sequent calculus is used to exemplify formal deduction and to show theoretical results in proof theory.
It is instructive to write out concrete derivations in the calculus to get a feel for the rules and the method of reasoning.
While such derivations can be done with pen and paper and checked for mistakes by human eyes, we argue that there is benefit in computer assistance.

To this end, we introduce SeCaV, a sequent calculus verifier built on top of Isabelle/HOL.
SeCaV presents everything within the same unified system: the syntax of formulas, the proof rules, their side conditions, and the way derivations are written.
Moreover, it provides immediate feedback to the user on the correctness of their derivations.
This empowers the users when learning to write derivations and gives them an independence that is harder to achieve without computer assistance.
We recall Nipkow~\cite{Nipkow12} on the analogy between proof assistants and video games and especially the benefits of immediate feedback:

\begin{quote}
\begin{it}
    This is in contrast to the usual system of homework that is graded by a teaching assistant and returned a week later, long after the student struggled with it, and at a time when the course has moved on.
    This delay significantly reduces the impact that any feedback scribbled on the homework may have.
\end{it}
\end{quote}

In this paper we include the teaching and learning aspects only as background motivation, since this is a system description and we have discussed the teaching and learning aspects elsewhere~\cite{ThEdu19, ThEdu20}.
A detailed description of the soundness and completeness results, which does not overlap with the present paper, can be found in a separate paper by From, Schlichtkrull and Villadsen~\cite{CILC}.

Our main focus, then, is on the definition of the system itself and especially the benefits of building it on top of Isabelle/HOL.
While many tools are implemented independently and perhaps modeled in a proof assistant, we aim to show the benefits of working entirely within Isabelle/HOL.

A completely novel development is the SeCaV Unshortener, a web application that allows experienced users to forgo full verification and in return write more succinct derivations.
Such a derivation is automatically expanded into the full SeCaV syntax which can then be verified for correctness.

\begin{figure}[t]
\includegraphics[width=0.99\textwidth]{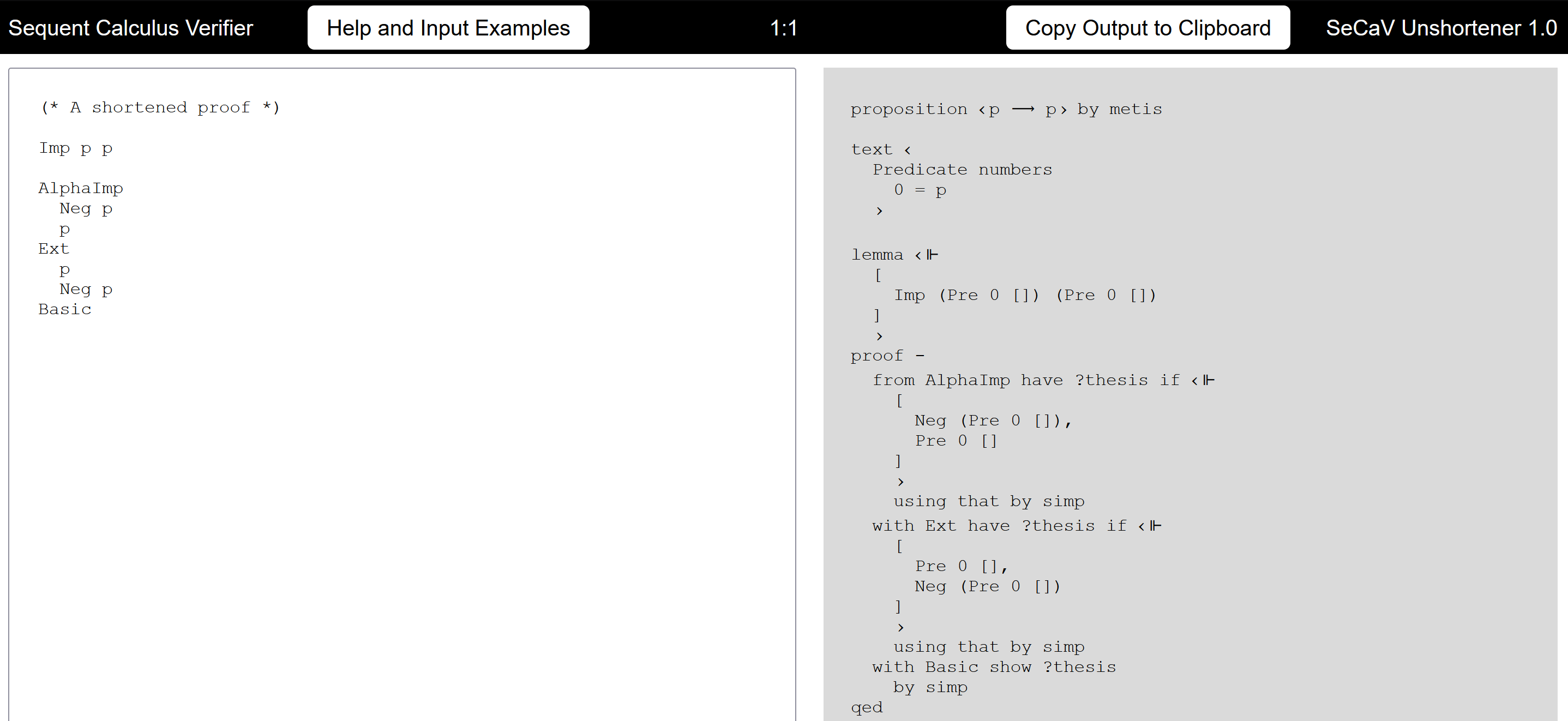}
\caption{The SeCaV Unshortener generating the example in \cref{fig:isa-gui}.}%
\label{fig:unshortener}
\end{figure}

The SeCaV Unshortener, shown in \cref{fig:unshortener}, allows proofs to be written in a much more compact syntax, cf.~\cref{sec:unshortener}.

We used the SeCaV system in our BSc course ``Logical Systems and Logic Programming'' in the fall of 2020.
71 students took the 2-hour exam where the exercises in SeCaV were worth 20\% of the grade.
We also used the SeCaV system in our MSc course ``Automated Reasoning'' in the spring of 2021, mainly in order to bridge the gap between our micro provers for propositional logic and our Natural Deduction Assistant (NaDeA), cf.~\cref{sec:related}.
Here, the students were also introduced to the recent SeCaV Unshortener.
34 students took the 2-hour exam where the exercises in SeCaV were worth 25\% of the grade.
Our experiences teaching these courses are detailed in a separate paper by Villadsen and Jacobsen~\cite{FMTea}.

The SeCaV system is available online (tested with Isabelle2020 and Isabelle2021):

\begin{center}
    \url{https://github.com/logic-tools/secav}
\end{center}

\noindent
The two relevant files are \texttt{SeCaV.thy}, which defines the sequent calculus and proves soundness, and \texttt{Sequent\_Calculus\_Verifier.thy}, which builds on our existing work~\cite{FOL-Seq-Calc1-AFP} to prove completeness (cf.~\cref{sec:secav}).

The SeCaV Unshortener 1.0 is available online (tested using the Chrome, Edge, Firefox and Safari browsers):

\begin{center}
    \url{https://secav.compute.dtu.dk/}
\end{center}

\noindent
Version 1.0 is fully functional and has an online tutorial with examples.
The online tutorial can be used to learn how to actually use the SeCaV system, while the present paper is a description of the system.

We continue by discussing existing work (\cref{sec:related}) before introducing our system and explaining a few design choices via a number of examples (\cref{sec:examples}).
We then introduce SeCaV formally, explain our design considerations further, outline the soundness and completeness results, and emphasize the benefits of the Isabelle/HOL integration (\cref{sec:secav}).
Next, we give an overview of the SeCaV Unshortener (\cref{sec:unshortener}) before concluding (\cref{sec:conclusion}).

We hope to convince the reader of the utility of SeCaV and its unshortener.
In combination, we have both an online system where students can easily practice derivations in sequent calculus and a transparent implementation of a proof checker for such derivations.
The transparency means that the full implementation is accessible, from the specification of the syntax as simple datatypes to the definition of the side conditions as functional programs.
Experienced students can even explore topics like soundness and completeness because Isabelle/HOL allows us to formalize semantics as well.
In summary, we believe that SeCaV, its implementation in Isabelle/HOL and its unshortener provide an excellent set of tools for introducing the topic of sequent calculus to students.

\section{Related Work}%
\label{sec:related}

There are many tools for sequent calculus, both online and offline.
We shall see that, while SeCaV defines one logic and one calculus, the transparency of the system makes the idea of extending or deviating from the system tangible.
An exploration of how proof assistants in general make the different layers and elements of logic easier to distinguish has previously been conducted by From, Villadsen and Blackburn~\cite{ThEdu20}.

The web application Logitext (\url{http://logitext.mit.edu}) allows users to derive a sequent by clicking the connective they want to apply a rule to.
As such, the rules are almost hidden away from the user who simply sees the appearance of new sub-derivations.
Sequoia~\cite{ReisNH20} allows users to input their own rules in a \LaTeX{} format and build derivations from them.
It also checks certain meta-theoretical properties of the stated calculus.
The online application was unavailable at the time of writing.
The \url{Carnap.io} site~\cite{Carnap17} allows users to specify their own logic as well as proof system, but in Haskell, which is then compiled to a web application.
The offline Sequent Calculus Trainer~\cite{SCT17} guides the user away from dead ends by alerting them if the current sequent is determined to be unprovable.
AXolotl~\cite{Axolotl19,CernaSSWB20} is an Android app that supports sequent calculus derivations in a classical notation.
It is designed to facilitate self-study.
Unlike SeCaV, none of these tools provide any formal guarantees of their correctness.
Each of them is a bespoke application in a regular programming language.

The Incredible Proof Machine~\cite{Breitner16,Incredible-Proof-Machine-AFP} is ``an interactive visual theorem prover which represents proofs as port graphs.''
It distinguishes itself by having a model of this proof representation formalized in Isabelle/HOL and shown to be as strong as natural deduction.
Unlike SeCaV, the formalized metatheoretical results only apply to a model of the system instead of to the actual implementation.

Our Natural Deduction Assistant (NaDeA)~\cite{NaDeA18} presents natural deduction in a more traditional style.
Its metatheory is formalized in Isabelle/HOL and the web application supports exporting proofs that can be verified in Isabelle/HOL, alleviating the problem of potential bugs in the online tool.

Our Students' Proof Assistant~\cite{SPA18} exists entirely inside Isabelle/HOL, where it defines a proof assistant within the proof assistant.
This helps make proof assistants and their design concrete, but makes the proving experience less natural than using the outer proof assistant directly as done in SeCaV.

Finally, we mention our micro provers for propositional logic~\cite{Villadsen20} whose formalized soundness and completeness results take up only a few dozen lines of Isabelle/HOL.
They are based on sequent calculus and can work as a first example in a course, before the full power of first-order logic and SeCaV is introduced.

\section{Design Choices and Examples}%
\label{sec:examples}
Before we go into the details of our proof system, we will explain some of our design choices by simple examples, which will hopefully also give some intuition about how our system works and how it is used in practice.

We use a simple programming-like syntax for formulas in SeCaV and abbreviate it further in the SeCaV Unshortener.
Both the syntax for SeCaV formulas and the abbreviated version are plain text formats, which makes them easy to write and avoids conflicts with logical operators in Isabelle/HOL.

\Cref{fig:ex1-isa} gives an example derivation of the formula \( p(a, b) \lor \neg p(a, b) \).
The formula is stated on line~3 as the sole member of the one-sided sequent spanning lines~2--4.
Recall that such a sequent is understood as a disjunction of formulas.
On line~3, the disjunction \( \lor \) is written using the constructor \isa{Dis} applied to two arguments separated by a space.
For predicates, the constructor \isa{Pre} takes a list of terms as arguments.
Here we use \isa{Fun\ 0\ \isacharbrackleft{}\isacharbrackright{}} and \isa{Fun\ 1\ \isacharbrackleft{}\isacharbrackright{}}, two function symbols taking no arguments, to represent the constants informally called \( a \) and \( b \).
We make this syntax precise in \cref{sec:secav}.
Note that both predicates and functions take their arguments as a list, which makes it easy to write simple functions that manipulate them by pattern matching.

The first rule application in \cref{fig:ex1-isa} occurs on line~7.
We will explain our proof rules later, but the interested reader may already now consult \cref{fig:rules-isa} to see their definitions.
We apply the \isa{AlphaDis} rule backwards, stating that our goal follows from the sequent listed on lines~9--10.
This sequent fits the shape of our \isa{Basic} axiom since it starts with a formula that also occurs negated.
Lines~14--15 finish the derivation based on this.
Note that our proof rules work only on the first element of each sequent, which makes the definition of the deductive system by pattern matching on the list representing the sequent simple.
The only exception is the structural rule \isa{Ext}, which works on the entire sequent.
We have chosen this representation over the common choice of representing sequents as sets to ensure that our definitions of e.g.\ side-conditions can be simple functional programs which users can actually execute to examine in detail why each step of their proof holds.

In the above we focused on the things essential to a human reader: the goal, the rules and their resulting sequents.
The remaining lines and keywords are for the benefit of Isabelle/HOL: they fit our derivations into the Isar syntax~\cite{Wenzel2007} giving us all the verification benefits of Isabelle/HOL.
If a rule is applied in a wrong manner, the editor tells us!

Our calculus is designed such that this boilerplate is predictable.
With the exception of two rules that require clarification when more than one variable is in play, we have yet to encounter a rule application that cannot be justified by Isabelle/HOL's simplifier.
This predictability means that we can write down only the essential parts of the derivation and then generate the boilerplate with the SeCaV Unshortener.
\Cref{fig:ex1-uns} contains an example of this, namely the same derivation as \cref{fig:ex1-isa}.
It starts with the goal formula on line~1, then the first rule application on line~3, followed by the resulting sequent, which spans lines~4--5, and finally line~6 finishes the derivation.
In fact, the SeCaV Unshortener produced the Isabelle/HOL code in \cref{fig:ex1-isa} automatically from this representation.
For brevity, we will generally favor the short representation.

\begin{figure}[tb]
\begin{internallinenumbers}[1]
\begin{isabelle}
    \Snippet{Examples:lemma:67c91b7751fada70}
\end{isabelle}
\end{internallinenumbers}
\caption{A sample SeCaV derivation in Isabelle/HOL.}%
\label{fig:ex1-isa}
\end{figure}

\begin{figure}[tb]
\bigskip 
\begin{internallinenumbers}[1]
\begin{verbatim}
Dis p[a, b] (Neg p[a, b])

AlphaDis
  p[a, b]
  Neg p[a, b]
Basic
\end{verbatim}
\end{internallinenumbers}
\caption{The sample SeCaV derivation in \cref{fig:ex1-isa} written in the syntax of the SeCaV Unshortener.}%
\label{fig:ex1-uns}
\end{figure}

At this point one might ask why we have not used Isabelle as a logical framework to implement SeCaV such that we can write proofs more directly instead of having to embed our deductive system within Isabelle/HOL.
First and foremost, doing so would prevent us from formalizing soundness and completeness of the SeCaV system directly, since we would then no longer have a formal metalanguage to work in.
Additionally, the implementation of such a system would be much more complex and include much more code to interface with the existing Isabelle system, obscuring the simplicity of the SeCaV system.
We would like students to see that implementing a basic formal deductive system and proving it sound and complete is not as daunting as it may seem when looking at full-fledged, and thus very complicated, proof assistants such as Isabelle/HOL or Coq.
This is of course a trade-off between ease of use for people who exclusively use the system to prove formulas and ease of understanding for people who would also like to understand why and how the system works.
Since our students are using the system for both purposes, we cannot stray too far towards either side of this balance.

\subsection{Instantiating Quantifiers}

Consider the additional example in \cref{fig:ex2}, which contains a derivation of:
\[ (\forall x.\, \forall y.\, p(x, y)) \rightarrow p(a, a) \]

We write the formulas using de~Bruijn indices to match the Isabelle/HOL formalization.
In the example, the variable~1 is bound by the outermost quantifier and 0 by the innermost.
The rule application on line~3 is propositional and straightforward: the implication holds if either the antecedent does not or the consequent does.
Consider instead line~6 where several things occur.
First, the \Rule{GammaUni} rule allows us to derive a negated, universally quantified formula from an example.
Applied backwards, we can insert any term for the bound variable while eliminating the quantifier.
We do so, replacing the bound variable with the term \texttt{a}.
Second, the notation \texttt{[a]} becomes a hint to Isabelle/HOL that \texttt{a} is the term used to replace the bound variable.
This ensures that the simplifier can verify the correctness and is necessary when more than one variable occurs in the term (we omit it on line~9).
Line~12 applies the \Rule{Ext} rule that rearranges the sequent such that the \Rule{Basic} rule applies on line~15.
We insist that the entire sequent is written down after each rule application so it is possible to read each application without referring back to previous ones.

\begin{figure}[tb]
\begin{internallinenumbers}[1]
\begin{verbatim}
Imp (Uni (Uni (p[1, 0]))) p[a, a]

AlphaImp
  Neg (Uni (Uni p[1, 0]))
  p[a, a]
GammaUni[a]
  Neg (Uni p[a, 0])
  p[a, a]
GammaUni
  Neg p[a, a]
  p[a, a]
Ext
  p[a, a]
  Neg p[a, a]
Basic
\end{verbatim}
\end{internallinenumbers}
\caption{SeCaV Unshortener example with instantiation of quantifiers.}
\label{fig:ex2}
\end{figure}

\subsection{Branching Derivations}

As a final example consider the longer \cref{fig:ex3}, which includes branching rules.
Lines~1--24 proceed using rules similar to those we have already seen.
We cover them in detail in \cref{sec:secav}.
Line~25 applies the \Rule{BetaImp} rule, which relies on two sub-derivations.
The first sequent that needs to be derived is given on lines~26--28 and the second sequent on lines~30--32, with a plus symbol (\texttt{+}) separating the two.
The application of \Rule{Basic} on line~33 closes the first branch and the rest of the derivation concerns only the second one.

The order of the two branches is not important for the Isabelle/HOL verification and the subsequent rules can be applied to either of the branches or even both at the same time.
For the sake of human-readability, however, we suggest working on the first branch.

\Cref{fig:ex3} displays another feature of our calculus that is worth pointing out.
On line~37, the \Rule{Ext} rule is used not just to rearrange the sequent, but also to drop a formula that was only necessary on one branch.
As such, it can be used to tidy up sequents during a derivation.

\begin{figure}[p]
\begin{internallinenumbers}[1]
\begin{verbatim}
Imp (Uni (Imp p[0] q[0])) (Imp (Exi p[0]) (Exi q[0]))

AlphaImp
  Neg (Uni (Imp p[0] q[0]))
  Imp (Exi p[0]) (Exi q[0])
Ext
  Imp (Exi p[0]) (Exi q[0])
  Neg (Uni (Imp p[0] q[0]))
AlphaImp
  Neg (Exi p[0])
  Exi q[0]
  Neg (Uni (Imp p[0] q[0]))
DeltaExi
  Neg p[a]
  Exi q[0]
  Neg (Uni (Imp p[0] q[0]))
Ext
  Neg (Uni (Imp p[0] q[0]))
  Neg p[a]
  Exi q[0]
GammaUni
  Neg (Imp p[a] q[a])
  Neg p[a]
  Exi q[0]
BetaImp
  p[a]
  Neg p[a]
  Exi q[0]
+
  Neg q[a]
  Neg p[a]
  Exi q[0]
Basic
  Neg q[a]
  Neg p[a]
  Exi q[0]
Ext
  Exi q[0]
  Neg q[a]
GammaExi
  q[a]
  Neg q[a]
Basic
\end{verbatim}
\end{internallinenumbers}
\caption{SeCaV Unshortener example with a branching derivation.}%
\label{fig:ex3}
\end{figure}

\section{SeCaV}%
\label{sec:secav}

\begin{figure}[tb]
\includegraphics[width=0.95\textwidth]{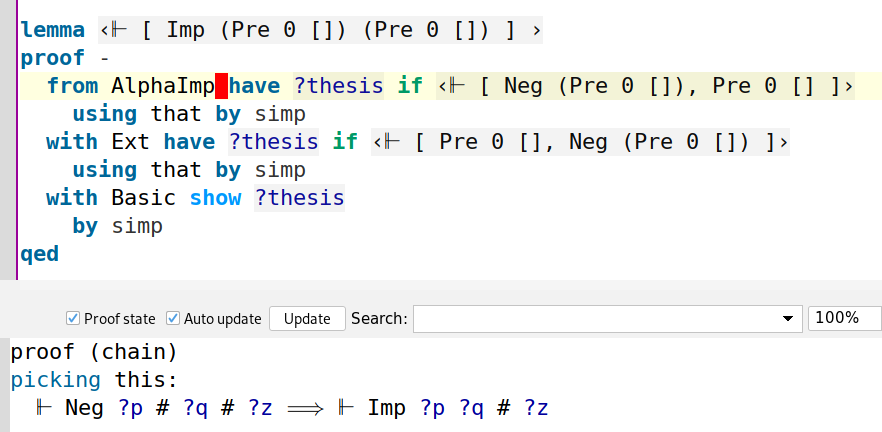}
\caption{A compressed SeCaV derivation in the Isabelle/jEdit editor. The ``Output'' panel at the bottom shows the rule under the cursor (\isa{AlphaImp}). Isabelle automatically instantiates each variable (\isa{?p, ?q, ?z}) appropriately.}%
\label{fig:isa-gui}
\end{figure}

\Cref{fig:isa-gui} shows a SeCaV derivation as it appears in the Isabelle/jEdit editor.
For brevity, we have removed a number of line breaks.
The SeCaV Unshortener allows the same proof to be written in a much more compact syntax, cf.~\cref{fig:unshortener}.

In this section we formally describe the SeCaV system: its syntax and semantics, proof rules, metatheory and Isabelle/HOL integration.

\subsection{Syntax and Semantics}

The syntax of terms and formulas in SeCaV is formally defined by the following two Isabelle/HOL datatype declarations:
\begin{isabelle}
    \Snippet{SeCaV:datatype:tm}
    
    \Snippet{SeCaV:datatype:fm}
\end{isabelle}

Terms are either functions identified by a natural number and applied to a list of terms, or variables identified by de Bruijn indices.
Formulas are either predicates, also identified by a natural number and applied to a list of terms, a connective applied to an appropriate number of formulas, or a quantifier.
This use of natural numbers as identifiers was chosen for simplicity, but we could also make the datatypes generic over the type of identifiers.
In the Unshortener (cf.~\ref{sec:unshortener}) we use strings of letters instead of numbers.

The embedding of SeCaV into Isabelle/HOL means that we do not need to write a parser for this syntax.
As seen in \cref{fig:isa-gui} we can write down a formula immediately.
We use a simple programming-like syntax here, but it is also possible to define a more regular infix syntax with various precedences and associativity.

To formalize metatheory, like the soundness and completeness of our proof system, it is essential to assign a meaning to our formulas.
We can do this because of our deep embedding of the syntax as a datatype.
While we could also use Isabelle as the generic proof assistant it is, define our logic in that style and still have it check our proofs, doing so would prevent us from formalizing our metatheory, and we would not even be able to prove soundness.

The following functions interpret terms and formulas into Isabelle/HOL's higher-order logic, given a variable assignment \isa{e}, a function denotation \isa{f} and a predicate denotation \isa{g}:
\begin{isabelle}
    \Snippet{SeCaV:primrec:semantics-term}
\end{isabelle}
\pagebreak
\begin{isabelle}
    \Snippet{SeCaV:primrec:semantics}
\end{isabelle}

We use the given connectives and quantifiers from Isabelle/HOL's higher-order logic to interpret our own, similarly to how semantics given using pen and paper uses natural language like ``and'' or ``there exists.''
The two quantifier clauses make the interpretation of de~Bruijn indices explicit.
When interpreting a quantifier, the function \isa{shift} adjusts the variable assignment \isa{e} so index 0 points at the newly quantified variable.
Its general definition is:
\begin{isabelle}
    \Snippet{SeCaV:definition:shift}
\end{isabelle}

This use matches the intuition that variable 0 is bound by the ``nearest'' quantifier.
Similarly, the existing indices are shifted by one since they appear one scope further out.
Keeping this definition explicit makes some of the proofs needed for soundness more manageable.

Both here and later we will prefer to define explicit but simple recursive functions instead of using higher-order library functions such as \isa{map} and \isa{list_all} in our definitions.
We find that, while these library functions make the definitions more succinct and understandable to experienced users of Isabelle/HOL, their use can easily confuse students who are not very good functional programmers.
Another issue specific to Isabelle/HOL is that the definition of e.g.\ the \isa{map} library function for lists is very cryptic, consisting only of the definition of the \isa{list} datatype and a specialization of a more general function.
Understanding how this works requires a deep dive into the internals of Isabelle/HOL, and since SeCaV is oriented towards beginners, we would like to keep this complexity out of sight.

\subsection{Substitution}

Returning to the topic of de~Bruijn indices we now cover how substitution is formalized, as we need it to specify our proof rules.
We include the definitions of such helper functions to make our presentation self-contained.
Our substitution function on formulas, \isa{sub}, is designed to be used whenever we instantiate a quantifier.
The application \isa{sub\ v\ s\ p} substitutes variable \isa{v} for the term \isa{s} in formula \isa{p}.
During this substitution, we ensure that no variable in \isa{s} gets bound by a quantifier in \isa{p}.
We define the function by structural recursion:
\begin{isabelle}
    \Snippet{SeCaV:primrec:sub}
\end{isabelle}

Only the predicate and quantifier cases are interesting; the rest simply apply the substitution to the sub-formulas.
In the predicate case we use the function \isa{sub-list} to apply the substitution across the list of argument terms.
It is defined mutually with \isa{sub-term}:
\begin{isabelle}
    \Snippet{SeCaV:primrec:sub-term}
\end{isabelle}

There are two cases for \isa{sub-term}.
At variables we leave smaller variables alone, substitute those matching the target index \isa{v}, and decrement larger variables to account for the instantiated quantifier whose scope is now gone.
At function symbols, we simply apply the substitution across the arguments.

Returning to \isa{sub}, in the quantifier cases we increment \isa{v} to account for the quantifier whose scope we are now under.
For the same reason, we use the function \isa{inc-term} to increment the variables in \isa{s}.

It is defined mutually with \isa{inc-list}:
\begin{isabelle}
    \Snippet{SeCaV:primrec:inc-term}
\end{isabelle}

With these at hand, we can now turn to the proof system itself.

\subsection{Proof System}

Our sequent calculus is a one-sided system like System G by Ben-Ari~\cite{BenAri2001}, which inspired it.
A one-sided system has a couple of advantages.
First, it can be explained and understood as simply meta-notation for a disjunction between formulas.
Second, it can be formalized as a single list of formulas, in turn reducing the syntactic burden of writing down a sequent.
Consider the SeCaV Unshortener syntax in e.g.~\cref{fig:ex2}.
Rule applications and sequents alternate throughout the derivation, with no need for a special symbol to distinguish a left- and right-hand side of the sequent as would be needed in a two-sided system \cite[p.~69]{BenAri2001}.

\begin{figure}[tb]
\begin{minipage}{.15\textwidth}
\begin{isabelle}
    \phantom{-}

    Basic
    
    AlphaDis
    
    AlphaImp
    
    AlphaCon
    
    BetaCon
    
    BetaImp
    
    BetaDis
    
    GammaExi
    
    GammaUni
    
    DeltaUni
    
    DeltaExi
    
    NegNeg
    
    Ext
\end{isabelle}
\end{minipage}%
\begin{minipage}{.85\textwidth}
\begin{isabelle}
    \Snippet{SeCaV:inductive:sequent-calculus}
\end{isabelle}
\end{minipage}
\caption{SeCaV proof rules in Isabelle/HOL with associated names inserted manually to the left.}%
\label{fig:rules-isa}
\end{figure}

\Cref{fig:rules-isa} contains our proof rules.
We use Smullyan's uniform notation~\cite{Smullyan1995} for the names, designating whether they branch (\( \beta \)) or not (\( \alpha \)) and whether the quantifiers can be built from any term (\( \gamma \)) or only a fresh witness (\( \delta \)).
Notice that each rule is actually a schema: the symbols \( p \) and \( q \) etc. are not concrete formulas but metavariables that can be instantiated with any type-correct value.

\subsubsection{Proof Rules}

As mentioned our sequents are lists of formulas, which means that they are ordered.
In Isabelle/HOL, the symbol \isacharhash{} separates the head and tail of a list.
All our rules except \Rule{Ext} use this notation to replace the head of the list.

\Cref{fig:rules-isa} begins with the only axiom, \Rule{Basic}, which states that a sequent with some formula \isa{p} at the head and \isa{Neg\ p} somewhere in the tail can be derived.
That is, we can derive the sequent \isa{\isasymtturnstile{} p \isacharhash{} z}, \isakeyword{if} we can demonstrate that \isa{Neg\ p} is a member of \isa{z}, i.e.~\isa{member \isacharparenleft{}Neg p\isacharparenright{} z}.

The function \isa{member} is defined in SeCaV as a simple primitive recursive function on lists for users to inspect (or even run):
\begin{isabelle}
    \Snippet{SeCaV:primrec:member}
\end{isabelle}

Since a sequent is understood as a disjunction, those of the \Rule{Basic} shape are clearly valid.
To derive a sequent that contains both some \isa{p} and a corresponding \isa{Neg\ p} but not necessarily in the order dictated by \Rule{Basic}, the final rule in \cref{fig:rules-isa}, \Rule{Ext}, can be used.
As we have seen in \cref{sec:examples}, it allows one to rearrange the formulas of a sequent or to drop formulas.
It should be read as follows: if we can derive a sequent \isa{z} and the sequent \isa{y} is an \emph{ext}ension of \isa{z}, then we are allowed to derive \isa{y} itself.
The function \isa{ext} builds on \isa{member} to check that the formulas in \isa{y} constitute a superset of those in \isa{z}:
\begin{isabelle}
    \Snippet{SeCaV:primrec:ext}
\end{isabelle}

\paragraph{Alpha Rules}

After \Rule{Basic} follow three \( \alpha \)-rules that rely on just one sub-derivation.
The \Rule{AlphaDis} rule moves the connective from the object language into the metalanguage, simply removing the connective and adding the two disjuncts to the sequent.
To show that an implication \isa{Imp\ p\ q} holds, we can either falsify the antecedent, \isa{Neg\ p}, or show the conclusion, \isa{q}, so the rule \Rule{AlphaImp} replaces an implication with exactly those formulas.
Finally, \Rule{AlphaCon} states that to show \isa{Neg\ \isacharparenleft{}Con\ p\ q\isacharparenright{}} we can falsify either \isa{p} or \isa{q}.
The pen-ultimate rule \Rule{NegNeg} also relies on just one sub-derivation, so we include it here.
It introduces a double negation.

\paragraph{Beta Rules}

After the first \( \alpha \)-rules follow three \( \beta \)-rules that make the derivation branch.
A conjunction only holds if both conjuncts do, so the \Rule{BetaCon} rule adds each to separate sub-derivations.
The \Rule{BetaImp} rule works on a negated implication, \isa{Neg\ \isacharparenleft{}Imp\ p\ q\isacharparenright{}}, and states that we must both prove \isa{p} and falsify \isa{q}.
Finally, \Rule{BetaDis} replaces \isa{Neg\ \isacharparenleft{}Dis\ p\ q\isacharparenright{}} with both \isa{Neg\ p} and \isa{Neg\ q} on separate branches, as both \isa{p} and \isa{q} must be falsified for their disjunction to be falsified.

\paragraph{Gamma Rules}

The \( \gamma \)-rules apply to formulas that are effectively existentially quantified.
Such formulas can be built from any witnessing term.
The next rule exemplifies this: \Rule{GammaExi} derives the sequent \isa{Exi\ p \isacharhash{} z} from \isa{sub\ 0\ t\ p \isacharhash{} z}.
In the sub-derivation, we have the formula \isa{p} with its outermost variable instantiated with the term \isa{t} using the \isa{sub} function.
This term, \isa{t}, witnesses the existence, so in the conclusion we quantify over the variable, giving \isa{Exi\ p}, instead of substituting it.

The \Rule{GammaUni} rule applies when the head of the sequent is \isa{Neg\ \isacharparenleft{}Uni p\isacharparenright{}} for some formula \isa{p}.
In the sub-derivation, the head is replaced by \isa{Neg\ \isacharparenleft{}sub\ 0\ t\ p\isacharparenright{}} since falsifying \isa{p} instantiated with any term \isa{t} is enough to falsify \isa{Uni\ p}.

\paragraph{Delta Rules}

The two \( \delta \)-rules apply to formulas that are effectively universally quantified.
To prove a universal quantifier, we cannot abstract over just any term like with \( \gamma \)-rules.
Instead, the term must be arbitrary, i.e.~\isa{new} to the sequent as formalized below, so that any other term could stand in its place.
Sometimes this is called \emph{fresh} rather than \emph{new}.

The \Rule{DeltaUni} rule allows the derivation of \isa{Uni\ p \isacharhash{} z} if we can derive \isa{sub\ 0\ \isacharparenleft{}Fun\ i\ \isacharbrackleft{}\isacharbrackright{}\isacharparenright{} p \isacharhash{} z} where the name \isa{i} does not occur in either \isa{p} or \isa{z}, as checked by \isa{news\ i\ \isacharparenleft{}p \isacharhash{} z\isacharparenright{}}.
The \Rule{DeltaExi} rule is similar but applies when the head of the sequent is \isa{Neg \isacharparenleft{}Exi\ p\isacharparenright{}}.

We define newness similarly to the other side conditions.
The function \isa{new} defines what it means for a function symbol \isa{c} to be new to a formula:
\begin{isabelle}
    \Snippet{SeCaV:primrec:new}
\end{isabelle}

Only the predicate case is interesting; the rest simply consider sub-formulas.
The function \isa{new-list} checks whether \isa{c} is new to a list of terms.
It is defined mutually with \isa{new-term}:
\begin{isabelle}
    \Snippet{SeCaV:primrec:new-term}
\end{isabelle}

If the term is a variable then the function symbol \isa{c} is obviously new.
Otherwise the term is a function application and we check whether the two function symbols coincide.
If they do, \isa{c} is not new, but even if they do not, \isa{c} still has to be new to the arguments of the function, which we check with \isa{new-list}.

The entry point to these functions is \isa{news}, which checks whether the function symbol \isa{c} is new to the given sequent:
\begin{isabelle}
    \Snippet{SeCaV:primrec:news}
\end{isabelle}

\subsubsection{Rule Design}

After seeing how the proof rules work, we want to point out several choices in their design.

While sequents are sometimes unordered (cf.~Ben-Ari~\cite{BenAri2001}, Nipkow and Michaelis~\cite{MichaelisN2017}) ours do have an order.
Where Ben-Ari underlines the formula in a sequent that the next rule applies to, our rules always work on the first one.
This simplification has several benefits: (i) it makes the formalization simpler to state and the success of the verification easier to predict, (ii) it reduces the notational burden in the SeCaV Unshortener syntax and (iii) it provides a straight-forward recipe for new users to get started: ``simply look at the first formula and see if any rules apply.''
Of course, the recipe in (iii) may result in derivations that are longer than necessary and because of our \( \gamma \)-rules the recipe may even be insufficient for certain formulas, but such formulas are also out of reach if the user starts out overwhelmed and never gets going.
The simplification forces us to include a structural rule like \Rule{Ext}.
This is the price of separating concerns, but as we have seen, \Rule{Ext} can also, for instance, be used to drop formulas on branches that do not need them.

Another point is that our rules do not just add to the sequent, but always replace the head of it in the sub-derivation(s).
Even the \( \gamma \)-rules do this, even though we may want to instantiate such formulas with several different terms (cf.~Ben-Ari~\cite{BenAri2001}).
Again we separate concerns: to apply a \( \gamma \)-rule twice, first duplicate the formula with \Rule{Ext} and then apply the rule.
This makes such duplication deliberate instead of an arbitrary feature of \( \gamma \)-rules that is sometimes useful and sometimes not.

Our last point relates to the benefit of specifying our system in Isabelle/HOL: every operation and side condition is explicitly spelled out and \emph{computational}.
It may seem obvious what membership in a sequent entails or what it means for a constant name to be \isa{new}, but something like substitution is notoriously tricky, no matter the representation.
In our system, these things are implemented by simple functional programs, accessible directly in the system.
They are not opaque pieces of natural language or hidden away in an implementation, but can be inspected by the user and even run on simple examples.

It speaks to the complexity of substitution that only rules that involve this operation can be hard to verify: \Rule{GammaUni} and \Rule{GammaExi}.
Otherwise, our definitions, like those of \isa{member} and \isa{ext}, play on the strengths of the Isabelle/HOL simplifier: they are simple functional programs that can be checked by rewriting.
Similarly, by letting our rules work on the first formula in the sequent, it becomes a simple problem to unify it with the current goal, solving the meta-variables to check if they match the stated sub-derivation.
These design choices make the system predictable to work with.

\subsection{Writing Proofs}

\begin{figure}[tb]
    \begin{isabelle}
        \input{template-proof}
    \end{isabelle}
    \caption{SeCaV derivation template. Branches are added using the \isacommand{and} keyword between sequents.}%
    \label{fig:template}
\end{figure}

As alluded to in \cref{sec:examples}, derivations in SeCaV follow a common template, which we have sketched in \cref{fig:template}.
Users of the system only need to worry about filling in the GOAL and a number of RULE applications with corresponding SUBGOALs.
Those curious about Isabelle/HOL can investigate the meaning of the remaining keywords if they want to.
Since derivations are entirely textual, it is easy to copy this template, or parts of it, from given examples or previous derivations.

\begin{figure}[tb]
\begin{verbatim}
Failed to finish proof:
goal (1 subgoal):
\end{verbatim}
\begin{internallinenumbers}[1]
\begin{isabelle}
    \input{example-error}
\end{isabelle}
\end{internallinenumbers}
\caption{Error message when \Rule{AlphaDis} is replaced by \Rule{AlphaImp} in \cref{fig:ex1-isa}.}%
\label{fig:example-error}
\end{figure}

In \cref{fig:example-error} we see the error message obtained when \Rule{AlphaDis} is replaced by \Rule{AlphaImp} in \cref{fig:ex1-isa}.
Isabelle/HOL will highlight the \isakeyword{by} keyword following the rule application, notifying the user that something is wrong.
The error is then displayed in the output panel when placing the cursor over the highlight.
Line 1 in \cref{fig:example-error} contains the rule being applied, lines 2--3 the stated subgoal and lines 4--5 the goal itself.
By inspection we see that since \isa{Imp} and \isa{Dis} do not match, the rule does not apply to produce the goal (the subgoal does not match either).

We obtain this error message completely for free by leveraging Isabelle/HOL as the platform for specifying our system.

We also inherit the interactive features of Isabelle/HOL.
Users can click a name and be taken to its definition, e.g.~that of \isa{sub} if in doubt about substitution.
Or when following the template, they can put their cursor on an applied rule and see its definition in the output panel as in \cref{fig:isa-gui}:
``This derivation uses \Rule{AlphaImp}, how does that look again? Oh, right: \isa{
\isasymtturnstile Neg\ ?p {\isacharhash} ?q {\isacharhash} ?z {\isasymLongrightarrow} \isasymtturnstile Imp\ ?p\ ?q {\isacharhash} ?z}.''

\subsection{Soundness and Completeness}

Having specified SeCaV in a proof assistant enables us to give certain guarantees about not just our calculus but its implementation as well.
The first and most obvious is soundness of the rules.
If we can derive a sequent, then for any interpretation, some formula in the sequent is satisfied:
\begin{isabelle}
    \SnippetPart00{SeCaV:theorem:sound}
\end{isabelle}

See the formalization for the proof which works by induction over the rules and using a substitution lemma.
We immediately obtain that if a derivable sequent contains just one formula then that formula must be valid:
\begin{isabelle}
    \SnippetPart00{SeCaV:corollary:c38238813c5dc8d}
\end{isabelle}

We build the completeness proof on existing work in the Archive of Formal Proofs, namely the entry ``A Sequent Calculus for First-Order Logic''~\cite{FOL-Seq-Calc1-AFP} (though this work also contains a soundness proof, we prefer to keep the soundness proof free of external dependencies since showing soundness is not a very difficult undertaking).
In less than a hundred lines of Isabelle/HOL, we relate our syntax, semantics, side conditions and operations to an existing sequent calculus formalization and show that derivations in that one (\isasymturnstile) can be translated into ours (\isasymtturnstile) (cf.~the formalization):
\begin{isabelle}
    \SnippetPart00{Sequent-Calculus-Verifier:lemma:sim}
\end{isabelle}

From these components, completeness follows straightforwardly:
\begin{isabelle}
    \SnippetPart00{Sequent-Calculus-Verifier:theorem:complete-sound}
\end{isabelle}

The symbol \isasymthen{} abbreviates validity in the universe of Herbrand terms.
Validity in just this universe is enough to show the existence of a derivation (a slightly stronger completeness result than assuming validity in all universes).
The soundness result, conversely, implies validity in any universe, as \isa{e}, \isa{f} and \isa{g} can be picked at will.

These aspects can be ignored when working with the system, but used to concretize discussions of soundness and completeness in a course.

\subsection{Isabelle/HOL Integration}
Next, we want to reiterate a few consequences of building our system on top of Isabelle/HOL.

In terms of infrastructure we are relieved from implementing a lot of work ourselves.
By giving two simple datatype declarations, in a syntax resembling BNF, we get to reuse Isabelle/HOL's parser when writing formulas in our object logic.
The same reusability applies to the proof system, both in its declarative specification using the \isacommand{inductive} command and when writing concrete derivations.
Given the declaration in \cref{fig:rules-isa}, we inherit proof checking completely for free: Isabelle/HOL verifies the correctness of derivations for us.

The use of Isabelle/HOL also means that we can reuse its mature graphical editor Isabelle/jEdit.
Besides regular editor features like undo, Isabelle/jEdit continually checks the correctness of what the user enters.
As seen, it produces decent errors that are displayed in the same window as the derivation and where the offending rule application is highlighted directly in the derivation.
Finally, all definitions used by the system can be inspected by using the editor to look them up within the same interface.

\section{SeCaV Unshortener}%
\label{sec:unshortener}

\begin{figure}[t]
\includegraphics[width=0.99\textwidth]{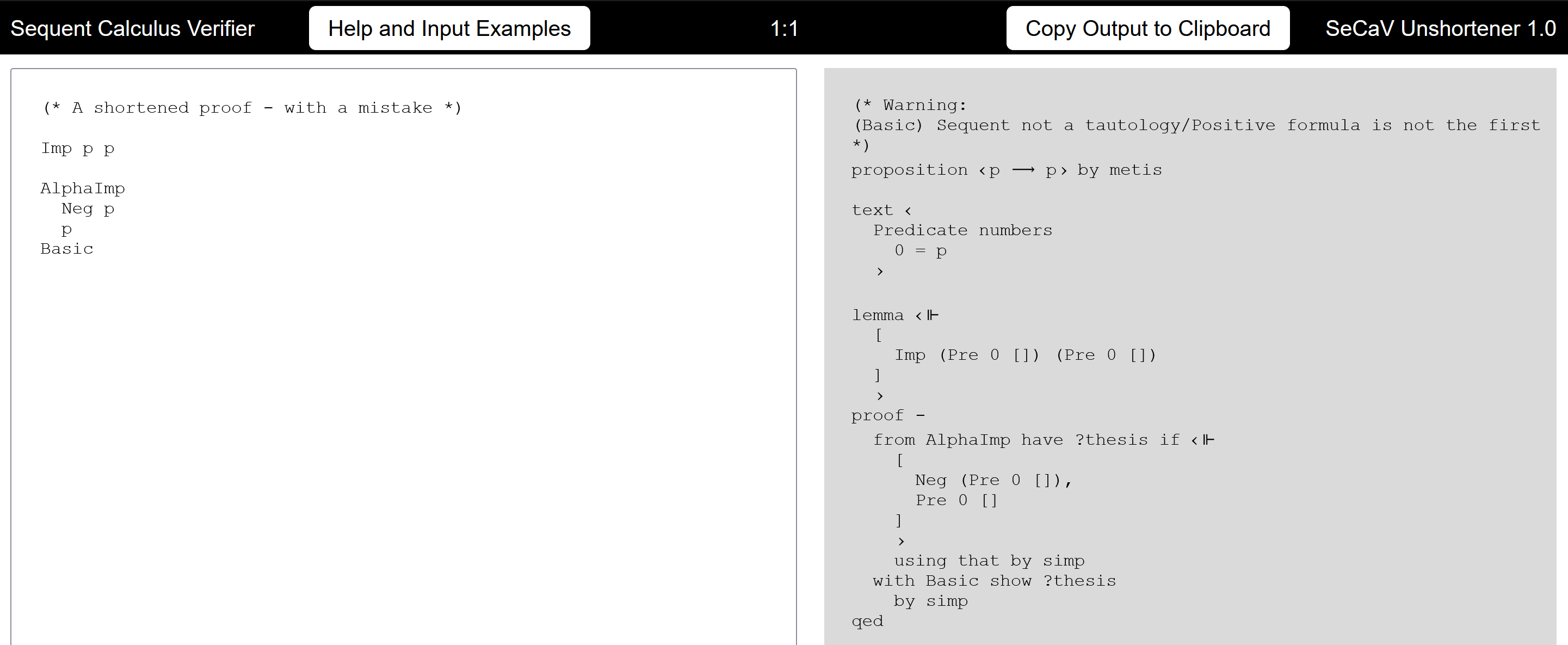}
\caption{The SeCaV Unshortener generating the example in \cref{fig:isa-gui} --- With a mistake.}%
\label{fig:unshortener-mistake}
\end{figure}

While the embedding of SeCaV into Isabelle/HOL makes it possible for users to get quick feedback on their proofs and provides us with an editor for free, actually writing out the proofs in the Isabelle/HOL syntax can become tedious.
To remedy this, we have introduced the SeCaV Unshortener, shown in \cref{fig:unshortener} and \cref{fig:unshortener-mistake}.
It allows proofs to be written in a much more compact syntax that resembles the style one might use when writing pen-and-paper proofs.
For instance, instead of \isa{Fun 0 []}, we can simply write \isa{a} and it will be ``unshortened'' for us.

The SeCaV Unshortener is a web application whose main page consists of two panes.
The first pane is a text area in which the user can write proofs in the compact SeCaV Unshortener syntax.
The second pane contains the result of ``unshortening'' the proofs written in the first pane into the full SeCaV syntax, ready to be copied into Isabelle/HOL for verification.
For each proof, the SeCaV Unshortener also generates a representation of the statement in usual logical syntax and a mapping from predicate and function names to the natural numbers used in the full SeCaV syntax.
The first of these is useful to detect misunderstandings in the statement to be proved, while the latter is needed to relate the actual proof to the representation in usual logical syntax, and may also be used to quickly detect typos in names.
The second pane reacts to changes in the first pane in real time, and will contain an error message if a proof is written using wrong syntax or if proof rules are applied in a wrong manner.

Along the top of the main page is a link to a page containing extensive help and a number of examples, an indication of the currently selected line and column in the first pane (for use with error messages), and a button that copies the unshortened proof to the clipboard.

The SeCaV Unshortener provides a canonical formatting of proofs and a lighter, more readable syntax at the expense of introducing another step in the proof procedure.
The SeCaV Unshortener does not verify the proofs entered into it to the same degree as Isabelle/HOL but it does add warnings when the proofs will most likely be rejected by Isabelle/HOL.
Examples of these warnings include forgetting to actually change the sequent after a rule application, changing the sequent in a way that is inconsistent with the latest rule application, misapplying the \Rule{Basic} rule and forgetting the side conditions on \( \delta \)-rules.

The SeCaV Unshortener is implemented in PureScript using the Concur web UI framework with a React backend.
The application is compiled down to a few JavaScript, HTML, and CSS files, which can be hosted on any web server or downloaded for local use.
The system basically consists of a parser for the Unshortener syntax and a generator for the full SeCaV syntax.
After parsing the user input, the abstract syntax tree is checked for errors and warnings are added to the generated SeCaV syntax.

\section{Conclusion}%
\label{sec:conclusion}
We have introduced SeCaV, a sequent calculus verifier built on top of Isabelle/HOL, and explained the syntax, semantics, and proof rules of the system.
SeCaV is designed to be easy to learn and understand for students, and is therefore implemented as a number of simple functional programs utilizing the interactive Isabelle/jEdit editor to allow inspection of every part of the system.
We have used Isabelle/HOL to prove soundness and completeness of the SeCaV calculus exactly as users work with it.

We have also introduced the SeCaV Unshortener, a web application that allows users of SeCaV to omit the boilerplate notation needed for the embedding in Isabelle/HOL.
We are currently (fall 2021) using the SeCaV system in our BSc course “Logical Systems and Logic Programming” and many of the 77 students prefer the SeCaV Unshortener with the lighter, more readable syntax.
We plan to release the teaching material as soon as possible.

\section*{Acknowledgements}
We thank Agnes Moesgård Eschen, Alexander Birch Jensen, Simon Tobias Lund, Emmanuel André Ryom and Anders Schlichtkrull for comments on a draft.
We also thank the anonymous reviewers, whose comments and suggestions have materially improved the text and caused us to reassess several aspects of our system.

\bibliographystyle{eptcs}
\bibliography{references}

\begin{thebibliography}{10}
\providecommand{\bibitemdeclare}[2]{}
\providecommand{\surnamestart}{}
\providecommand{\surnameend}{}
\providecommand{\urlprefix}{Available at }
\providecommand{\url}[1]{\texttt{#1}}
\providecommand{\href}[2]{\texttt{#2}}
\providecommand{\urlalt}[2]{\href{#1}{#2}}
\providecommand{\doi}[1]{doi:\urlalt{http://dx.doi.org/#1}{#1}}
\providecommand{\bibinfo}[2]{#2}

\bibitemdeclare{book}{BenAri2001}
\bibitem{BenAri2001}
\bibinfo{author}{Mordechai \surnamestart Ben{-}Ari\surnameend}
  (\bibinfo{year}{2012}): \emph{\bibinfo{title}{Mathematical Logic for Computer
  Science}}.
\newblock \bibinfo{publisher}{Springer}, \doi{10.1007/978-1-4471-4129-7}.

\bibitemdeclare{inproceedings}{Breitner16}
\bibitem{Breitner16}
\bibinfo{author}{Joachim \surnamestart Breitner\surnameend}
  (\bibinfo{year}{2016}): \emph{\bibinfo{title}{{Visual Theorem Proving with
  the Incredible Proof Machine}}}.
\newblock In \bibinfo{editor}{Jasmin~Christian \surnamestart
  Blanchette\surnameend} \& \bibinfo{editor}{Stephan \surnamestart
  Merz\surnameend}, editors: {\sl \bibinfo{booktitle}{Interactive Theorem
  Proving - 7th International Conference, {ITP} 2016, Nancy, France, August
  22-25, 2016, Proceedings}}, {\sl \bibinfo{series}{Lecture Notes in Computer
  Science}} \bibinfo{volume}{9807}, \bibinfo{publisher}{Springer}, pp.
  \bibinfo{pages}{123--139}, \doi{10.1007/978-3-319-43144-4\_8}.

\bibitemdeclare{article}{Incredible-Proof-Machine-AFP}
\bibitem{Incredible-Proof-Machine-AFP}
\bibinfo{author}{Joachim \surnamestart Breitner\surnameend} \&
  \bibinfo{author}{Denis \surnamestart Lohner\surnameend}
  (\bibinfo{year}{2016}): \emph{\bibinfo{title}{The meta theory of the
  Incredible Proof Machine}}.
\newblock {\sl \bibinfo{journal}{Archive of Formal Proofs}}.
\newblock
  \bibinfo{note}{\url{https://isa-afp.org/entries/Incredible_Proof_Machine.html},
  Formal proof development}.

\bibitemdeclare{inproceedings}{Axolotl19}
\bibitem{Axolotl19}
\bibinfo{author}{David~M. \surnamestart Cerna\surnameend},
  \bibinfo{author}{Rafael P.~D. \surnamestart Kiesel\surnameend} \&
  \bibinfo{author}{Alexandra \surnamestart Dzhiganskaya\surnameend}
  (\bibinfo{year}{2019}): \emph{\bibinfo{title}{A Mobile Application for
  Self-Guided Study of Formal Reasoning}}.
\newblock In \bibinfo{editor}{Pedro \surnamestart Quaresma\surnameend},
  \bibinfo{editor}{Walther \surnamestart Neuper\surnameend} \&
  \bibinfo{editor}{Jo{\~{a}}o \surnamestart Marcos\surnameend}, editors: {\sl
  \bibinfo{booktitle}{Proceedings 8th International Workshop on Theorem Proving
  Components for Educational Software, ThEdu@CADE 2019, Natal, Brazil, 25th
  August 2019}}, {\sl \bibinfo{series}{{EPTCS}}} \bibinfo{volume}{313}, pp.
  \bibinfo{pages}{35--53}, \doi{10.4204/EPTCS.313.3}.

\bibitemdeclare{inproceedings}{CernaSSWB20}
\bibitem{CernaSSWB20}
\bibinfo{author}{David~M. \surnamestart Cerna\surnameend},
  \bibinfo{author}{Martina \surnamestart Seidl\surnameend},
  \bibinfo{author}{Wolfgang \surnamestart Schreiner\surnameend},
  \bibinfo{author}{Wolfgang \surnamestart Windsteiger\surnameend} \&
  \bibinfo{author}{Armin \surnamestart Biere\surnameend}
  (\bibinfo{year}{2020}): \emph{\bibinfo{title}{Aiding an Introduction to
  Formal Reasoning Within a First-Year Logic Course for {CS} Majors Using a
  Mobile Self-Study App}}.
\newblock In \bibinfo{editor}{Michail~N. \surnamestart Giannakos\surnameend},
  \bibinfo{editor}{Guttorm \surnamestart Sindre\surnameend},
  \bibinfo{editor}{Andrew \surnamestart Luxton{-}Reilly\surnameend} \&
  \bibinfo{editor}{Monica \surnamestart Divitini\surnameend}, editors: {\sl
  \bibinfo{booktitle}{Proceedings of the 2020 {ACM} Conference on Innovation
  and Technology in Computer Science Education, ITiCSE 2020, Trondheim, Norway,
  June 15-19, 2020}}, \bibinfo{publisher}{{ACM}}, pp. \bibinfo{pages}{61--67},
  \doi{10.1145/3341525.3387409}.

\bibitemdeclare{inproceedings}{SCT17}
\bibitem{SCT17}
\bibinfo{author}{Arno \surnamestart Ehle\surnameend}, \bibinfo{author}{Norbert
  \surnamestart Hundeshagen\surnameend} \& \bibinfo{author}{Martin
  \surnamestart Lange\surnameend} (\bibinfo{year}{2017}):
  \emph{\bibinfo{title}{The Sequent Calculus Trainer with Automated Reasoning -
  Helping Students to Find Proofs}}.
\newblock In \bibinfo{editor}{Pedro \surnamestart Quaresma\surnameend} \&
  \bibinfo{editor}{Walther \surnamestart Neuper\surnameend}, editors: {\sl
  \bibinfo{booktitle}{Proceedings 6th International Workshop on Theorem proving
  components for Educational software, ThEdu@CADE 2017, Gothenburg, Sweden, 6
  August 2017}}, {\sl \bibinfo{series}{{EPTCS}}} \bibinfo{volume}{267}, pp.
  \bibinfo{pages}{19--37}, \doi{10.4204/EPTCS.267.2}.

\bibitemdeclare{inproceedings}{ThEdu19}
\bibitem{ThEdu19}
\bibinfo{author}{Asta~Halkj{\ae}r \surnamestart From\surnameend},
  \bibinfo{author}{Alexander~Birch \surnamestart Jensen\surnameend},
  \bibinfo{author}{Anders \surnamestart Schlichtkrull\surnameend} \&
  \bibinfo{author}{J{\o}rgen \surnamestart Villadsen\surnameend}
  (\bibinfo{year}{2019}): \emph{\bibinfo{title}{Teaching a Formalized Logical
  Calculus}}.
\newblock In \bibinfo{editor}{Pedro \surnamestart Quaresma\surnameend},
  \bibinfo{editor}{Walther \surnamestart Neuper\surnameend} \&
  \bibinfo{editor}{Jo{\~{a}}o \surnamestart Marcos\surnameend}, editors: {\sl
  \bibinfo{booktitle}{Proceedings 8th International Workshop on Theorem Proving
  Components for Educational Software, ThEdu@CADE 2019, Natal, Brazil, 25th
  August 2019}}, {\sl \bibinfo{series}{{EPTCS}}} \bibinfo{volume}{313}, pp.
  \bibinfo{pages}{73--92}, \doi{10.4204/EPTCS.313.5}.

\bibitemdeclare{inproceedings}{ThEdu20}
\bibitem{ThEdu20}
\bibinfo{author}{Asta~Halkj{\ae}r \surnamestart From\surnameend},
  \bibinfo{author}{J{\o}rgen \surnamestart Villadsen\surnameend} \&
  \bibinfo{author}{Patrick \surnamestart Blackburn\surnameend}
  (\bibinfo{year}{2020}): \emph{\bibinfo{title}{Isabelle/{HOL} as a
  Meta-Language for Teaching Logic}}.
\newblock In \bibinfo{editor}{Pedro \surnamestart Quaresma\surnameend},
  \bibinfo{editor}{Walther \surnamestart Neuper\surnameend} \&
  \bibinfo{editor}{Jo{\~{a}}o \surnamestart Marcos\surnameend}, editors: {\sl
  \bibinfo{booktitle}{Proceedings 9th International Workshop on Theorem Proving
  Components for Educational Software, ThEdu@IJCAR 2020, Paris, France, 29th
  June 2020}}, {\sl \bibinfo{series}{{EPTCS}}} \bibinfo{volume}{328}, pp.
  \bibinfo{pages}{18--34}, \doi{10.4204/EPTCS.328.2}.

\bibitemdeclare{article}{FOL-Seq-Calc1-AFP}
\bibitem{FOL-Seq-Calc1-AFP}
\bibinfo{author}{Asta~Halkjær \surnamestart From\surnameend}
  (\bibinfo{year}{2019}): \emph{\bibinfo{title}{A Sequent Calculus for
  First-Order Logic}}.
\newblock {\sl \bibinfo{journal}{Archive of Formal Proofs}}.
\newblock \bibinfo{note}{\url{https://isa-afp.org/entries/FOL_Seq_Calc1.html},
  Formal proof development}.

\bibitemdeclare{inproceedings}{CILC}
\bibitem{CILC}
\bibinfo{author}{Asta~Halkjær \surnamestart From\surnameend},
  \bibinfo{author}{Anders \surnamestart Schlichtkrull\surnameend} \&
  \bibinfo{author}{Jørgen \surnamestart Villadsen\surnameend}
  (\bibinfo{year}{2021}): \emph{\bibinfo{title}{A Sequent Calculus for
  First-Order Logic Formalized in Isabelle/{HOL}}}.
\newblock In \bibinfo{editor}{Stefania \surnamestart Monica\surnameend} \&
  \bibinfo{editor}{Federico \surnamestart Bergenti\surnameend}, editors: {\sl
  \bibinfo{booktitle}{Proceedings of the 36th Italian Conference on
  Computational Logic - {CILC} 2021, Parma, Italy, September 7-9, 2021}}, {\sl
  \bibinfo{series}{{CEUR} Workshop Proceedings}} \bibinfo{volume}{3002},
  \bibinfo{publisher}{CEUR-WS.org}, pp. \bibinfo{pages}{107--121}.
\newblock \urlprefix\url{http://ceur-ws.org/Vol-3002/paper7.pdf}.

\bibitemdeclare{inproceedings}{Carnap17}
\bibitem{Carnap17}
\bibinfo{author}{Graham \surnamestart Leach{-}Krouse\surnameend}
  (\bibinfo{year}{2017}): \emph{\bibinfo{title}{Carnap: An Open Framework for
  Formal Reasoning in the Browser}}.
\newblock In \bibinfo{editor}{Pedro \surnamestart Quaresma\surnameend} \&
  \bibinfo{editor}{Walther \surnamestart Neuper\surnameend}, editors: {\sl
  \bibinfo{booktitle}{Proceedings 6th International Workshop on Theorem proving
  components for Educational software, ThEdu@CADE 2017, Gothenburg, Sweden, 6
  August 2017}}, {\sl \bibinfo{series}{{EPTCS}}} \bibinfo{volume}{267}, pp.
  \bibinfo{pages}{70--88}, \doi{10.4204/EPTCS.267.5}.

\bibitemdeclare{inproceedings}{MichaelisN2017}
\bibitem{MichaelisN2017}
\bibinfo{author}{Julius \surnamestart Michaelis\surnameend} \&
  \bibinfo{author}{Tobias \surnamestart Nipkow\surnameend}
  (\bibinfo{year}{2018}): \emph{\bibinfo{title}{Formalized Proof Systems for
  Propositional Logic}}.
\newblock In \bibinfo{editor}{A.~\surnamestart Abel\surnameend},
  \bibinfo{editor}{F.~Nordvall \surnamestart Forsberg\surnameend} \&
  \bibinfo{editor}{A.~\surnamestart Kaposi\surnameend}, editors: {\sl
  \bibinfo{booktitle}{23rd International Conference on Types for Proofs and
  Programs (TYPES 2017)}}, {\sl \bibinfo{series}{LIPIcs}}
  \bibinfo{volume}{104}, \bibinfo{publisher}{Schloss Dagstuhl - Leibniz-Zentrum
  fuer Informatik}, pp. \bibinfo{pages}{6:1--6:16},
  \doi{10.4230/LIPIcs.TYPES.2017.5}.

\bibitemdeclare{inproceedings}{Nipkow12}
\bibitem{Nipkow12}
\bibinfo{author}{Tobias \surnamestart Nipkow\surnameend}
  (\bibinfo{year}{2012}): \emph{\bibinfo{title}{Teaching Semantics with a Proof
  Assistant: No More {LSD} Trip Proofs}}.
\newblock In \bibinfo{editor}{Viktor \surnamestart Kuncak\surnameend} \&
  \bibinfo{editor}{Andrey \surnamestart Rybalchenko\surnameend}, editors: {\sl
  \bibinfo{booktitle}{Verification, Model Checking, and Abstract
  Interpretation}}, \bibinfo{publisher}{Springer}, pp. \bibinfo{pages}{24--38},
  \doi{10.1007/978-3-642-27940-9_3}.

\bibitemdeclare{inproceedings}{ReisNH20}
\bibitem{ReisNH20}
\bibinfo{author}{Giselle \surnamestart Reis\surnameend}, \bibinfo{author}{Zan
  \surnamestart Naeem\surnameend} \& \bibinfo{author}{Mohammed \surnamestart
  Hashim\surnameend} (\bibinfo{year}{2020}): \emph{\bibinfo{title}{Sequoia: {A}
  Playground for Logicians - (System Description)}}.
\newblock In \bibinfo{editor}{Nicolas \surnamestart Peltier\surnameend} \&
  \bibinfo{editor}{Viorica \surnamestart Sofronie{-}Stokkermans\surnameend},
  editors: {\sl \bibinfo{booktitle}{Automated Reasoning - 10th International
  Joint Conference, {IJCAR} 2020, Paris, France, July 1-4, 2020, Proceedings,
  Part {II}}}, {\sl \bibinfo{series}{Lecture Notes in Computer Science}}
  \bibinfo{volume}{12167}, \bibinfo{publisher}{Springer}, pp.
  \bibinfo{pages}{480--488}, \doi{10.1007/978-3-030-51054-1_32}.

\bibitemdeclare{inproceedings}{SPA18}
\bibitem{SPA18}
\bibinfo{author}{Anders \surnamestart Schlichtkrull\surnameend},
  \bibinfo{author}{J{\o}rgen \surnamestart Villadsen\surnameend} \&
  \bibinfo{author}{Andreas~Halkj{\ae}r \surnamestart From\surnameend}
  (\bibinfo{year}{2018}): \emph{\bibinfo{title}{{Students' Proof Assistant
  (SPA)}}}.
\newblock In \bibinfo{editor}{Pedro \surnamestart Quaresma\surnameend} \&
  \bibinfo{editor}{Walther \surnamestart Neuper\surnameend}, editors: {\sl
  \bibinfo{booktitle}{Proceedings 7th International Workshop on Theorem proving
  components for Educational software, ThEdu@FLoC 2018, Oxford, United Kingdom,
  18 July 2018}}, {\sl \bibinfo{series}{{EPTCS}}} \bibinfo{volume}{290}, pp.
  \bibinfo{pages}{1--13}, \doi{10.4204/EPTCS.290.1}.

\bibitemdeclare{book}{Smullyan1995}
\bibitem{Smullyan1995}
\bibinfo{author}{Raymond~M. \surnamestart Smullyan\surnameend}
  (\bibinfo{year}{1995}): \emph{\bibinfo{title}{First-Order Logic}}.
\newblock \bibinfo{publisher}{Dover Publications}.

\bibitemdeclare{inproceedings}{Villadsen20}
\bibitem{Villadsen20}
\bibinfo{author}{J{\o}rgen \surnamestart Villadsen\surnameend}
  (\bibinfo{year}{2020}): \emph{\bibinfo{title}{{Tautology Checkers in Isabelle
  and Haskell}}}.
\newblock In \bibinfo{editor}{Francesco \surnamestart Calimeri\surnameend},
  \bibinfo{editor}{Simona \surnamestart Perri\surnameend} \&
  \bibinfo{editor}{Ester \surnamestart Zumpano\surnameend}, editors: {\sl
  \bibinfo{booktitle}{Proceedings of the 35th Italian Conference on
  Computational Logic - {CILC} 2020, Rende, Italy, October 13-15, 2020}}, {\sl
  \bibinfo{series}{{CEUR} Workshop Proceedings}} \bibinfo{volume}{2710},
  \bibinfo{publisher}{CEUR-WS.org}, pp. \bibinfo{pages}{327--341}.
\newblock \urlprefix\url{http://ceur-ws.org/Vol-2710/paper21.pdf}.

\bibitemdeclare{inproceedings}{NaDeA18}
\bibitem{NaDeA18}
\bibinfo{author}{J{\o}rgen \surnamestart Villadsen\surnameend},
  \bibinfo{author}{Andreas~Halkj{\ae}r \surnamestart From\surnameend} \&
  \bibinfo{author}{Anders \surnamestart Schlichtkrull\surnameend}
  (\bibinfo{year}{2018}): \emph{\bibinfo{title}{{Natural Deduction Assistant
  (NaDeA)}}}.
\newblock In \bibinfo{editor}{Pedro \surnamestart Quaresma\surnameend} \&
  \bibinfo{editor}{Walther \surnamestart Neuper\surnameend}, editors: {\sl
  \bibinfo{booktitle}{Proceedings 7th International Workshop on Theorem proving
  components for Educational software, ThEdu@FLoC 2018, Oxford, United Kingdom,
  18 July 2018}}, {\sl \bibinfo{series}{{EPTCS}}} \bibinfo{volume}{290}, pp.
  \bibinfo{pages}{14--29}, \doi{10.4204/EPTCS.290.2}.

\bibitemdeclare{inproceedings}{FMTea}
\bibitem{FMTea}
\bibinfo{author}{Jørgen \surnamestart Villadsen\surnameend} \&
  \bibinfo{author}{Frederik~Krogsdal \surnamestart Jacobsen\surnameend}
  (\bibinfo{year}{2021}): \emph{\bibinfo{title}{Using Isabelle in Two Courses
  on Logic and Automated Reasoning}}.
\newblock In \bibinfo{editor}{João~F. \surnamestart Ferreira\surnameend},
  \bibinfo{editor}{Alexandra \surnamestart Mendes\surnameend} \&
  \bibinfo{editor}{Claudio \surnamestart Menghi\surnameend}, editors: {\sl
  \bibinfo{booktitle}{Formal Methods Teaching}}, \bibinfo{publisher}{Springer
  International Publishing}, \bibinfo{address}{Cham}, pp.
  \bibinfo{pages}{117--132}, \doi{10.1007/978-3-030-91550-6_9}.

\bibitemdeclare{article}{Wenzel2007}
\bibitem{Wenzel2007}
\bibinfo{author}{Makarius \surnamestart Wenzel\surnameend}
  (\bibinfo{year}{2007}): \emph{\bibinfo{title}{Isabelle/{I}sar - a generic
  framework for human-readable proof documents}}.
\newblock {\sl \bibinfo{journal}{From Insight to Proof - Festschrift in Honour
  of Andrzej Trybulec, Studies in Logic, Grammar, and Rhetoric. University of
  Białystok}} \bibinfo{volume}{10}(\bibinfo{number}{23}), pp.
  \bibinfo{pages}{277--298}.

\end{thebibliography}

\end{document}